\documentclass[draftcls,onecolumn]{IEEEtran}
\usepackage{epsfig}
\usepackage{times}
\usepackage{float}
\usepackage{afterpage}
\usepackage{amsmath}
\usepackage{amstext}
\usepackage{amssymb,bm}
\usepackage{latexsym}
\usepackage{color}
\usepackage{array,algpseudocode}
\usepackage{makecell}
\usepackage{diagbox}
\usepackage{amsmath}
\usepackage{amsthm}
\usepackage{graphicx}
\usepackage[center]{caption}
\usepackage{pstricks}
\usepackage{subfigure}
\usepackage{booktabs}
\usepackage{enumerate}
\usepackage[ruled]{algorithm}
\usepackage{romannum}
\usepackage[normalem]{ulem}
\usepackage{url}
\newtheorem{theorem}{Theorem}
\newtheorem{lemma}{Lemma}

\newtheorem{mechanism}{Mechanism}

\newcommand{\Ksf}{\mathsf{K}} 
\newcommand{\Qsf}{\mathsf{Q}} 
\newcommand{\esf}{\mathsf{e}} 
\newcommand{\Nbf}{\mathbf{N}} 

\newcommand{\Ec}{\mathcal{E}} 
\newcommand{\Oc}{\mathcal{O}} 
\newcommand{\comment}[1]{}
\newcommand{\floor}[1]{\lfloor #1 \rfloor}
\newcommand{\ceil}[1]{\lceil #1 \rceil}
\let\svthefootnote\thefootnote
\newcommand\blankfootnote[1]{%
  \let\thefootnote\relax\footnotetext{#1}%
  \let\thefootnote\svthefootnote%
}

\allowdisplaybreaks

\begin{document}
%
\title{
Capacity and Stability Regions for Layered Packet Erasure Broadcast Channels with Feedback}
%
%
%

\author{Siyao Li, Daniela Tuninetti, Natasha Devroye and Hulya Seferoglu \\
University of Illinois at Chicago, Chicago, IL 60607, USA\\
Email: \{sli210, danielat, devroye, hulya\}@uic.edu}

\maketitle

\begin{abstract}
This paper focuses on the Layered Packet Erasure Broadcast Channel (LPE-BC) with Channel Output Feedback (COF) available at the transmitter. 
The LPE-BC is a high-SNR approximation of the fading Gaussian BC recently proposed by Tse and Yates, who characterized the capacity region for any number of users and any number of layers when there is no COF. 
This paper provides a comparative overview of this channel model along the following lines: First, inner and outer bounds to the  \emph{capacity region}  (set of achievable rates with backlogged arrivals)  are presented:  a) a new outer bound based on the idea of the physically degraded broadcast channel, and b) an inner bound of the LPE-BC with COF for the case of two users and any number of layers. 
Next, an inner bound on the  \emph{stability region} (set of exogenous arrival rates for which packet arrival queues are stable) for the same model is derived.  
The capacity region inner bound generalizes past results for the two-user erasure BC, which is a special case of the LPE-BC with COF with only one layer. 
 The novelty lies in the use of \emph{inter-user and inter-layer network coding} retransmissions (for those packets that have only been received by the unintended user), where each random linear combination may involve packets intended for any user originally sent on any of the layers.
For the case of $\Ksf=2$ users and $\Qsf\geq 1$ layers, the inner bounds to the capacity region and the stability region coincide; both strategically employ the novel retransmission protocol. For the case of $\Qsf=2$ layers, sufficient conditions are derived  by Fourier-Motzkin elimination for the inner bound on the stability region to coincide with the capacity outer bound, thus showing that in those cases the capacity and stability regions coincide.
\end{abstract}


\section{Introduction}
\label{sec:intro}
\blankfootnote{This work was in part supported by NSF award 1900911. This paper was presented in part at ICC 2019 and ITW 2019.} 
The Broadcast Channel (BC) has been extensively employed as a model for downlink communication systems. 
A class of channels that has received significant attention is erasure channels, where at each channel use a packet is sent, and the packet is either received or erased at each receiver. The erasure channel is used as  a model for lossy packet networks. Another class of channel particularly important in wireless communications is the Additive White Gaussian Noise fading BC (AWGN-BC), where the channel between the single transmitter or base-station sending signal $X$, and multiple users is modeled as $Y_i = h_i X + N_i$ for user $i$, where $N_i$ is the AWGN, and $h_i$ is the fading parameter, or Channel State Information (CSI).  
With Channel Output Feedback (COF), the signal at the receivers is fed back to the transmitter.
When the transmitter has 
independent messages to send to different subsets of users, the capacity region 
captures some of the tension seen in BCs: a single signal must be encoded such that when correlated versions of this signal are received at the users, each can extract their own intended message(s). 

The Layered Packet Erasure Broadcast Channel (LPE-BC) model proposed in~\cite{FadingBC} approximates the AWGN-BC without Channel State Information at the Transmitter (CSIT) in the high SNR regime and also generalizes the (single-layer) Binary Erasure Channel (BEC-BC).
In the LPE-BC, the base-station at each channel use sends a vector of inputs (or layers of packets). At each time, each receiver receives a random number of layers, and missing layers are said to be ``erased''. Erasures are correlated because when a layer is erased, all the layers with larger indices are also erased.

\subsection{Past Work}
{\bf{Capacity results}}. 
 The capacity region characterizes the largest set of simultaneously achievable message rates that can be reliably transmitted~\cite{book:ElGamal-Kim}. The capacity region assumes that all users, or nodes, have messages, or packets, to send at all times.  That is, the packet arrival queues are infinitely backlogged. 
While the capacity regions of the general BC remains unknown, 
it is known for the degraded BC, 
the BC with degraded message sets, 
the AWGN-BC without fading, 
and the AWGN-BC with fading known at the transmitter and the receivers~\cite{Cover:2006}. 
The capacity region of the BEC-BC without COF is known for any number of users (because the channel is stochasticaly degraded)~\cite{Cover:2006}. 
For the BEC-BC, the presence of COF allows the transmitter to know if a packet was erased or not at each receiver. 
COF information allows the sender to re-send certain packets, and may do so in a {\it network-coded fashion} (by sending linear combinations of packets intended for different users). 
 In~\cite{on-the-capacity-of-1-to-k}, the capacity region for 3-user BEC-BC with COF, as well as two types of symmetric $K$-user Packet Erasure Broadcast Channels (PEBCs) and spatially independent PEBCs with one-sided fairness constraints with COF, were derived. Similar results to~\cite{on-the-capacity-of-1-to-k} were also obtained in~\cite{multiuser-BEC-with-feedback}.
 The authors in \cite{FadingBC} determined the capacity region of the LPE-BC exactly and bounded that of the AWGN-BC to within a constant gap of approximately $6$ bits per channel use, regardless of the fading distribution.
The capacity of the AWGN-BC with COF is unknown, but it may be enlarged by feedback even in the non-fading regime~\cite{Ozarow:1984,4655434Bhaskaran},  in sharp contrast to memoryless point-to-point channels. However, feedback cannot enlarge the capacity of physically degraded BCs~\cite{ElGamal-BC}. 
A partial characterization of the capacity region of the two-user Gaussian fading BC was provided in~\cite{two-user-Gaussian-fading-BC}. The work in~\cite{achievable-throughput-of-a-multiantenna-Gaussian-BC} studied the achievable throughput of a multi-antenna Gaussian BC. 

{\bf{Stability results}}. 
The stability region assumes that packets arrive stochastically, and may be queued before transmission. A networked system is called stable if the packet queues are asymptotically finite, with finite packet  delays. In~\cite{dynamic-power-allocation}, the stability region is defined as the closure of the set of all arrival rate vectors that can be stably supported by the network. 
The exact characterization of the stability region of networks with bursty sources is known to be a difficult problem~\cite{Dimitriou}. This is due to the interaction of the queues, i.e., when the service rate of a queue depends on the state of the other queues. 
In~\cite{stable-scheduling} the authors presented a class of scheduling policies for an $\Ksf$-user broadcast channel and showed that the system is stable in the mean through the use of a Lyapunov argument. 
Sufficient conditions for stability in a broadcast setting were derived in~\cite{fundamental-stability}.
 For the degraded BCs,~\cite{stability-degraded-bc} presented an outer
bound to the stability region of message arrival rate vectors
achievable by the class of stationary scheduling policies and showed
that the stability region of information arrival rate vectors is the
information-theoretic capacity region under an asymptotic regime. 
The work in~\cite{duality-and-stability-regions} investigated stability regions of two-user Gaussian fading multiple access and broadcast networks with centralized scheduling under the assumption of infinitely backlogged users.
The stability properties of different transmission schemes with and without network coding over a BEC-BC with COF were evaluated in~\cite{BroadcastStability}. 
In~\cite{Sagduyu-unusual} the authors characterized the stability region of a 2-user BEC-BC with COF and constructed several algorithms that employ network coding of packets received at the un-intended receiver and stabilize the system. The conditions under which the capacity and stability regions coincide  are not known in general~\cite{Ephremides}. 



\subsection{Contributions}
All exact capacity results for the LPE-BC are without COF~\cite{FadingBC}, or for the single-layer case with COF and up to $\Ksf = 3$ users~\cite{on-the-capacity-of-1-to-k,multiuser-BEC-with-feedback,Sagduyu-unusual}. We study the capacity  and the stability regions of the (multi-layer) LPE-BC with COF, combine and extend the works in~\cite{FadingBC,on-the-capacity-of-1-to-k,multiuser-BEC-with-feedback,Sagduyu-unusual}. 

In this paper, \emph{rate region}  (with backlogged packet traffic) refers to an achievable message rate region, which can not be larger than the capacity region; similarly, \emph{arrival region}  (with stochastic packet traffic) refers to an achievable arrival rate region, which can not be larger than the stability region. Our main contributions are listed as follows.
\begin{enumerate}
\item 
{\bf{We provide a general outer bound to the capacity region for LPE-BC with COF for $\Ksf$ receivers ($\Ksf\geq 2$) and $\Qsf$ layers ($\Qsf\geq 1$).}} 
 The outer bound is easily characterized by augmenting the model to various degraded versions of the LPE-BC for which capacity is known.

\item {\bf{ We present an achievable rate region and an achievable arrival region for LPE-BC with COF for $\Ksf =2 $ users and $\Qsf$ layers ($\Qsf\geq 1$).}} 
 The achievable rate and arrival regions are obtained by using schemes that employ network coding across layers in case retransmissions are needed. 
 We show the correctness of our schemes by induction. 
 Our proof techniques here (for the case of any number of layers) differs from that of~\cite{Sagduyu-unusual} (for a single layer): we do not rely on a ``Markov chain''-argument as~\cite{Sagduyu-unusual} but rather on a ``concentration to the mean ''-argument explained in Appendix~\ref{sec:proof of time concentration} and~\ref{sec:multiple-layers}.
 

\item {\bf{Conditions are given under which these regions match for the case of $\Ksf=2 $ users and $\Qsf= 2$ layers. }} Hence, for such channels both the capacity region and the arrival region are fully characterized, and they coincide. 
 Consequently, our results highlight similarity between the capacity and stability regions, both measuring rates  albeit for different traffic scenario.

\end{enumerate}

\subsection{Paper Organization}
The rest of the paper is organized as follows. 
Section~\ref{sec:model} introduces the LPE-BC.
Section~\ref{sec:schemes} presents the information theoretic inner and outer bounds on the capacity region.
Section~\ref{subsec:stochastic} presents a queueing theoretic arrival rate rates region, an inner bound on the stability region of the LPE-BC with COF.
Section~\ref{subsec:optimality} gives sufficient conditions under which the inner bound to the stability region coincide with the outer bound to the capacity region.
Section~\ref{sec:numerical} illustrates the derived bounds by means of numerical examples. 
Section~\ref{sec:concl} concludes the paper.
Most of the proofs may be found in the Appendices. 

\section{System Model and Definitions} 
\label{sec:model}

\begin{figure}
  \centering
  \includegraphics[ height= 0.7\columnwidth]{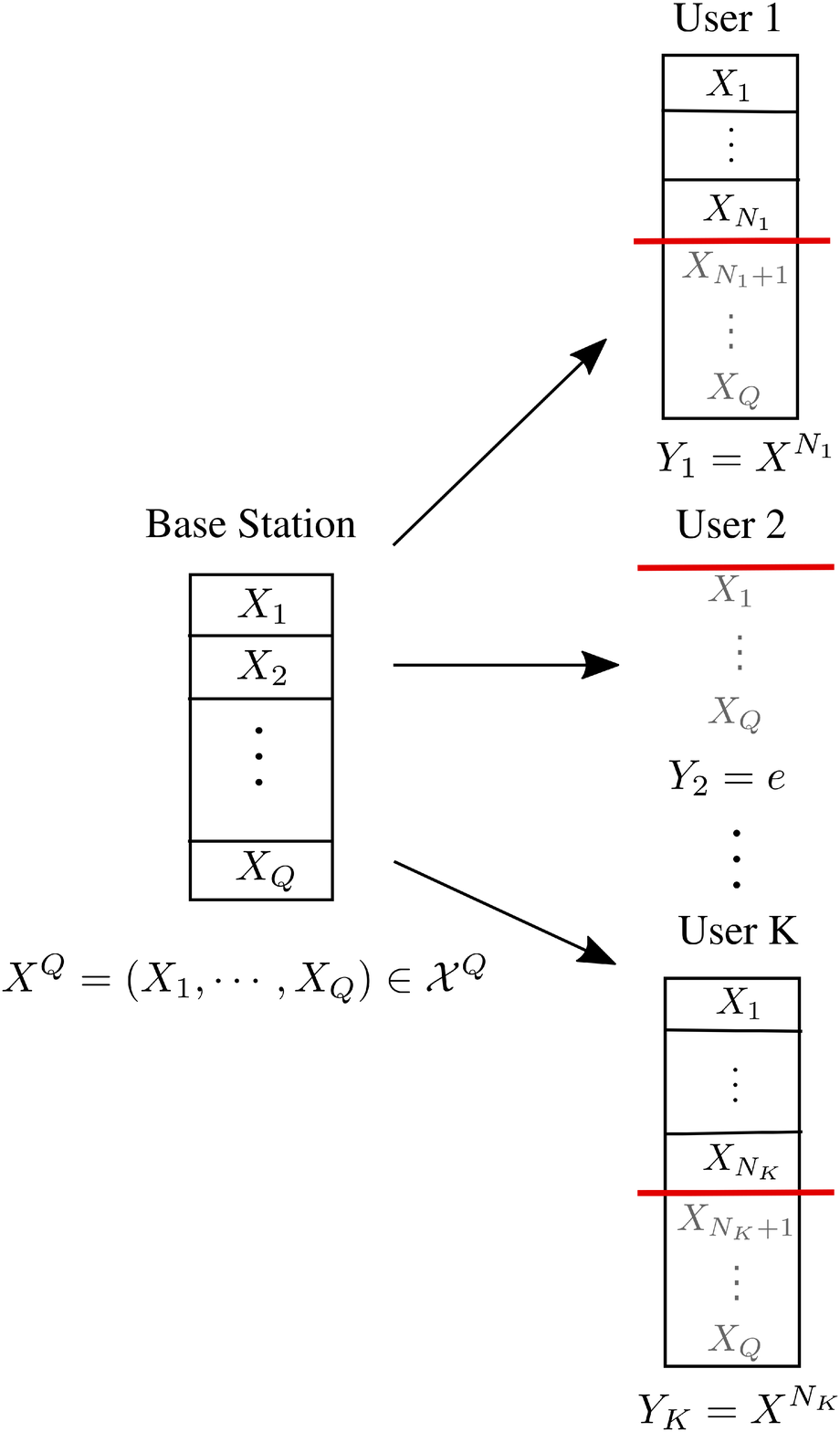}
  \caption{\small Illustration of the LPE-BC model.}
  \label{fig:system-model}
\end{figure}

\subsection{Notation}
For the rest of this paper, we use the following convention:
\begin{itemize}
\item We use capital letters for random variables (apart from rates which are denoted by capital $R$'s but are not random, to conform with standard information theoretic notation) and small letters for their realization.
\item We  shall also sometimes use  small letters to indicate expectation of the corresponding random variables.
\item We use$\floor{x}$ to indicate the greatest integer less than or equal to $x$ and $\ceil{x}$ to indicate the smallest integer greater than or equal to $x$.
\item We define $[x]^+ = \max\{ 0, x\}$ for $x\in \mathbb{R}$.
\item We consider a time-slotted system where slot $t \in \mathbb{N}$ denotes the slot index. 
\item We use  
$f(n) = o(g(n))$ to denote $\lim_{n\to \infty} f(n)/ g(n) = 0$. 
\item The notation $[n]$ for $n \in \mathbb{N}$ denotes the set $\{1,2,\cdots n\}$.
\item The notation $|A|$ denotes the cardinality of a set $A$. 
\item We use $\Ksf  \in \mathbb{N}$ to denote the  number of receivers (users).
\item We use $\Qsf  \in \mathbb{N}$ to denote the  number of layers.
\item For an integer $N$, the symbol $X^{N}$ indicates the length-$N$ vector $ (X_{1}, \ldots, X_{N})$. Also, we use $X_i^j$ to denote  $(X_i, \cdots, X_j)$.   
\item For the plots, the logarithms are in base 2, i.e., rates are expressed in bits/s/Hz.
\end{itemize}

\subsection{LPE-BC Model}
The LPE-BC illustrated in Fig.\ref{fig:system-model}, as originally proposed in~\cite{FadingBC}, consists of one transmitter (base-station) and $\Ksf$ receivers (users).
At each channel use ({\it slot}) the transmitter sends $\Qsf$ symbols ({\it packets / layers}), each symbol from an input alphabet $\mathcal{X}$, where $\mathcal{X}$ is assumed to be a discrete finite set; 
the input is denoted as $X^{\Qsf} := (X_{1}, \ldots, X_{\Qsf}) \in \mathcal{X}^{\Qsf}$.
The LPE-BC is characterized by the random vector ({\it channel state}) $N^\Ksf := (N_{1}, \ldots, N_{\Ksf})\in[0:\Qsf]^{\Ksf}$, 
where $N_{k}  \in \mathbb{N}$ denotes how many layers have been successfully received by user $k\in[\Ksf]$. 
The LPE-BC channel output for user $k\in[\Ksf]$ is $Y_{k} := X^{N_{k}} = (X_{1}, \ldots, X_{N_{k}})$ for $N_{k}>0$, that is, layers $(X_{N_{k}+1}, \ldots, X_{\Qsf})$ have been {\it erased} (i.e., all the layers below the red bar in Fig.\ref{fig:system-model}); if $N_{k}=0$ then all layers have been erased and we set $Y_{k}=\esf$ for some constant ``erasure'' symbol $\esf$, which is distinct from any other possible transmitted symbol. The channel state $\Nbf$ is assumed to be independent and identically distributed (i.i.d.) across time slots, that is, the channel is memoryless. In the LPE-BC, the erasures are correlated so as to capture the high SNR behavior of the fading AWGN-BC~\cite{FadingBC}.
The case $\Qsf=1$ and $\mathcal{X}=\text{GF}(2)$ is the well studied $\Ksf$-user BEC-BC~\cite{on-the-capacity-of-1-to-k,multiuser-BEC-with-feedback}.

There are two regions under consideration for this channel: 1) the information theoretic capacity region, and 2) the queueing theoretic stability region. The first region describes the transmission rates under which it is possible to transmit sets of messages (one for each user) so that all users receive the messages destined to them with probability of error vanishing in the blocklength. The second region describes the set of average arrival rates at which the packets may randomly arrive and be delivered to the destination without letting the queues blow up, i.e., the queues must remain stable, where we follow the definition of system stability from~\cite{Sagduyu-unusual}.
We formalize these notions next.

\subsection{Definitions in Information Theory: Capacity Region}
In this setup, the transmitter has $\Ksf$ queues of packets\footnote{In this paper, we use the term {\it{packet}} to describe an arriving unit of data in a communication network.}, one per receiver, and all the queues have infinitely many packets.
A code  $\mathcal{C}(|\mathcal{X}|^{n R_{1}}, $ $\ldots, |\mathcal{X}|^{n R_{\Ksf}}, n)$ for the LPE-BC is defined as follows.
The transmitter must convey $|\mathcal{X}|^{n R_{k}}$ (private) messages reliably to user $k\in[\Ksf]$ in $n$ channel uses. Note that the rate $R_{k}$ is measured in number of packets per channel use. Let $(W_{1},\ldots,W_{\Ksf})$ be the messages to be sent to the users. With COF, the encoding function at time $t$ is $X^{\Qsf}_{t}(W_{1},\ldots,W_{\Ksf}, N^{\Ksf, t-1}),$  where $N^{\Ksf, t-1} := (N_{1,1}, \cdots, N_{1,t-1}, \cdots, N_{\Ksf, 1}, \cdots, N_{\Ksf, t-1} ) \ t\in[n]$.
We assume that all receivers have full CSI, namely, by time $t=n$ they know $N^{\Ksf, n}$.
User $k\in[\Ksf]$ estimates $\widehat{W}_{k} = {\rm dec}_{k}(Y_{k}^{n}, N^{\Ksf, n} )$ for some decoding function ${\rm dec}_{k}$. The probability of error is 
$P_e^{(n)} := 1-\Pr[{\rm dec}_{k}(Y_{k}^{n}, N^{\Ksf, n})=W_{k}, \ \forall k\in[\Ksf]].$ 
The set of rates $(R_{1}, \ldots, R_{\Ksf})\in\mathbb{R}^{\Ksf}_{+}$ is said to be achievable if there exists a sequence of channel codes $\mathcal{C}(|\mathcal{X}|^{n R_{1}}, $ $\ldots, |\mathcal{X}|^{n R_{\Ksf}}, n), n= 1,2 \ldots,$ 
such that  $\lim_{n\to\infty} P_e^{(n)} = 0.$
The capacity region is the convex closure of  the set of all achievable rate vectors.

\subsection{Definitions in Queueing Theory: Stability Region}
In this setup, the transmitter maintains $\Ksf$ packet queues, one per receiver, and exogenous packets arrive randomly at each queue. 
Let $A_{u,t}$ be the packets that arrived at the beginning of slot $t\in\mathbb{N}$ and are intended for user $u\in[\Ksf]$. Let $\mathbf{A}_t := (A_{1,t},\ldots,A_{\Ksf,t})$ be the vector of exogenous arrivals, assumed to be i.i.d. over time, with average arrival rates $\lambda_u := \mathbb{E}[|A_{u,t}|], u\in[\Ksf]$.
Let $Q_{u,t}$ be the queue that contains the packets that still need to be transmitted to user $u\in[\Ksf]$ at slot $t\in\mathbb{N}$ (i.e., it includes the exogenous packets $A_{u,t}$, as well as those packets that were not yet delivered to user $u$ at previous slots, as described next).
With COF, the transmitter sends $X^{\Qsf}_{t}(\mathbf{Q}_t, N^{\Ksf, t-1} )$. 
User $u\in[\Ksf]$ applies decoding function ${\rm dec}_{k}(Y_{k}^{n}, N^{\Ksf, n} )$ that returns the packets that could be retrieved \emph{error-free} by using all channel outputs and all channel states available to it up to time $t\in\mathbb{N}$.
 A successfully received packet is removed from its queue; this can be tracked at the transmitter thanks to COF.
The evolution of the queue length over time is given by 
$ |Q_{u,t}| = \big[|Q_{u,t-1}|+ |A_{u,t}|- |{\rm dec}_{k}(Y_{k}^{t}, N^{\Ksf, t} ) |\big]^+, \ t\in\mathbb{N}, u\in[\Ksf],$ 
where $|\cdot|$ denotes the number of packets in the queue.
The \emph{stability region} is the  convex closure of the set of all arrival rate-tuples $(\lambda_{1}, \ldots, \lambda_{\Ksf})\in\mathbb{R}^{\Ksf}_{+}$ for which the process of queue lengths $\{ (|Q_{1,t}|,\ldots,|Q_{\Ksf,t}|) \}_{t\in\mathbb{N}}$ is stable\footnote{
From~\cite{Sagduyu-unusual}:
The process $\{\mathbf{X}_t\}_{t\in\mathbb{N}}$, where $\mathbf{X}_t := (X_{1,t},\ldots,X_{\Ksf,t})$, is \emph{stable} if 
the following holds at all points of continuity of some cumulative distribution function $F(\mathbf{x})$:
$\lim_{t\to\infty} \Pr[\mathbf{X}_t \leq \mathbf{x}] = F(\mathbf{x})$
and
$\lim_{\min(x_1,\ldots,x_\Ksf) \to \infty} F(\mathbf{x})= 1$,
where $\mathbf{x}:=(x_1,\ldots,x_\Ksf)$ and $\mathbf{X}_t \leq \mathbf{x}$ means coordinate-wise inequalities.
The process $\{\mathbf{X}_t\}_{t\in\mathbb{N}}$ is \emph{substable} if $\lim_{\min(x_1,\ldots,x_\Ksf) \to \infty} \lim\inf_{t\to\infty} \Pr[\mathbf{X}_t \leq \mathbf{x}] = 1$.
If the processes $\{ X_{i,t}\}_{t\in\mathbb{N}}$ are substable for all $i\in[\Ksf]$, then the process $\{\mathbf{X}_t\}_{t\in\mathbb{N}}$ is substable. 
In our case, $\{\mathbf{X}_t\}_{t\in\mathbb{N}}$ will represent the process of queue lengths.
}.

\section{Capacity Region} 
\label{sec:schemes}






%
In this section, we bound the capacity region of the LPE-BC with COF. 
%

In Section~\ref{sec:outer-bound}, we propose a new outer bound of the LPE-BC with COF. It is based on a channel enhancement that creates a degraded BC for which the capacity region is known. 
Then, in Section~\ref{capacity:inner-bounds}, we introduce a two-phase protocol and present a trivial inner bound. 
The novelty lies in the use of {\emph{inter-user $\&$ inter-layer network coding}} retransmissions in the achievable scheme. 

\subsection{Outer Bound of the LPE-BC with COF}
\label{sec:outer-bound}
Although COF does not increase the capacity of a memoryless point-to-point channels, it enlarges the capacity region of broadcast channels in general~\cite{Ozarow:1984,4655434Bhaskaran}. 
The following theorem characterizes the outer bound in the weighted sum rate form. 
\begin{theorem}[New outer bound]
\label{thm:COFnewOuter}
The capacity region of the LPE-BC with COF is contained in
\begin{align}
&\sum_{k\in[\Ksf]} \omega_{k} R_{k} 
  \leq 
  \sum_{q\in[\Qsf]} \max_{k\in[\Ksf]}\left( \omega_{\pi(k)} \Pr[\max(
 N_{\pi(k)}^{\pi(\Ksf)}
 ) \geq q] \right),
\label{eq:converse thm:COFnewOuter}
\end{align}
for all $(\omega_{1}, \ldots, \omega_{\Ksf})\in\mathbb{R}^{\Ksf}_{+}$  and for all permutations $\pi$ of $[\Ksf]$. 
\end{theorem}
\begin{IEEEproof}
\begin{subequations}
We enhance the original LPE-BC to a physically degraded LPE-BC by using a cooperation-based argument. Consider a permutation $\pi$ of $[\Ksf]$ and define 
\begin{align}
\widetilde{N}_{\pi(k)} := \max(N_{\pi(k)},N_{\pi(k+1)},\ldots,N_{\pi(\Ksf)}). \label{def:wideN}
\end{align}
Based on our system model, user $\pi(k)$ with channel state $\widetilde{N}_{\pi(k)}$ can receive all the packets received by user ${\pi(k)}, $ ${\pi(k+1)}, $ $\ldots, {\pi(\Ksf)}$.
Thus, the following Markov chains hold
\begin{align}
X^{\Qsf} \to X^{\widetilde{N}_{\pi(1)}} \to X^{\widetilde{N}_{\pi(2)}} \ldots \to X^{\widetilde{N}_{\pi(\Ksf)}},
\\
X^{\Qsf} \to X^{\widetilde{N}_{k}} \to X^{N_{k}}, \ \forall k\in[\Ksf].
\end{align}
\label{eq:enhancement thm:COFnewOuter}
That is, the BC with CSI $\widetilde{N}_{k}$ is physically degraded and its capacity is not enlarged by feedback~\cite{ElGamal-BC}. 
By~\cite{FadingBC}, 
the capacity region of the LPE-BC with no CSIT is characterized by 
\end{subequations}
\begin{align}
\sum_{k\in[\Ksf]} \omega_{k} R_{k} \leq \sum_{q \in[\Qsf]} \max_{u\in[\Ksf]}\left( \omega_{u} \Pr[\widetilde{N}_{u} \geq q] \right), 
\label{thm:Tse-Yates ITA-2011}
\end{align}
for all $(\omega_{1}, \ldots, \omega_{\Ksf})\in\mathbb{R}^{\Ksf}_{+}$. 
With $\widetilde{N}_{u}$ in~\eqref{def:wideN}, the region in~\eqref{thm:Tse-Yates ITA-2011} is the same as~\eqref{eq:converse thm:COFnewOuter}. 
\end{IEEEproof}


\subsection{Inner Bound of the LPE-BC with COF}
\label{capacity:inner-bounds}
We proposed several achievable schemes in~\cite{Siyao-Li}.
In this subsection, 
an achievable  two-phase protocol with the best performance will be introduced.  
 As a comparison point, we also present a trivial achievable inner bound, which is the simple extension of the single-layer case algorithm in~\cite{Sagduyu-unusual} to multiple layers where the layers operate independently. The analysis of the inner bound in this section is based on the law of large numbers. 
 We provide the detailed proof of Theorem~\ref{thm:Prototype} in Appendix~\ref{sec:multiple-layers}, which is an extension of the single-layer case proof in Appendix~\ref{sec:proof of time concentration} to multiple layers. 
Note that Appendix ~\ref{sec:proof of time concentration} is a new, alternative proof of the single-layer algorithm  in~\cite{Sagduyu-unusual}; we present it as our generalization to multiple layers follows this new proof closely.

\begin{table}
\caption{ Definition and initialization value of different notations. }
\centering
\begin{tabular}{ | p{2.9cm}  | p{6cm} |  p{2.5cm} | }
\hline
Notation & Definition & Initial value  \\  \hline
  $\Qsf_{\{u\},q} , u \in [2],   q\in[\Qsf] $      
  &   The queues store the packets destined to user $u$   assigned on layer $q$   
  & $k_{u,q}$ packets
   \\ \hline
   $\Qsf_{\{1,2\},q} , q\in[\Qsf] $       
   &  The queues store the packets destined to user $u$ but only received by the other user $\bar{u}, u\neq \bar{u}$ on layer $q$. 
   &    empty 
       \\ \hline
$K_{u,q}^{\text{(unc)}}(t)$       
&     The number of  uncoded packets not yet transmitted to user $u$  on layer $q$ at time $t$  
& $k_{u,q}$ 
 \\ \hline
  $K_{u,q}^{\text{(rem)}}(t)$     
  &    The number of packets not received by user $u$ but overheard by the other user on layer $q$  at time $t$
  & 0        \\ \hline
    $K_{u}^{\text{(rem)}}(t)$     
    &   The number of packets not received by user $u$ but overheard by the other user on all layers at time $t$   
    & 0       \\ \hline
     $K_{u,q}^{\text{(NC)}}(t)$     
     &   The number of coded packets received by user $u$ on layer $q$ at time $t$
     & 0          \\ \hline
     $K^\text{(rtx)}_{u}[j]$ 
     &  The number of packets destined to user~$u\in[2]$ but not yet successfully received by user~$u$ at the end of $j$-th sub-phase 
     & $k_u :=\sum_{q\in[\Qsf]}k_{u,q}$  
      \\ \hline 
     $T_{\pi(j)}^{\text{(unc)}}$   &  The number of slots needed for layer $\pi(j)$ to complete the $j$-th sub-phase &  $\infty$
     \\ \hline
     $T^{\text{(NC)}}$ & The number of slots for Phase2 & $\infty$ 
     \\ \hline
\end{tabular}

\label{tab-notations}
\end{table}

Our assumptions regarding the system are described as follows:
\begin{enumerate}
\item 
All terminals have enough storage to keep track of which packets have been sent and which have been successfully received (for whichever user(s) they have access to).
\item 
 When a network coded packet is sent (network coding to be defined soon) 
 the code (i.e., set of coefficients used for a linear combination) has been agreed upon in advance and is known to all terminals, i.e., every terminal knows the codebook. 
\end{enumerate}

 \begin{figure}
\vspace{-0.4cm}
  \centering
  \includegraphics[width=0.8\columnwidth]{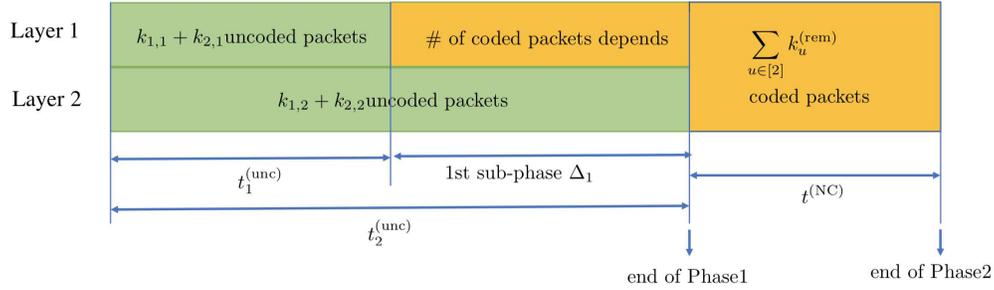}
  \caption{\small Two-phase protocol with two layers. }
  \label{fig:two-phase-demo}
\vspace*{-.4cm}
\end{figure}


The achievable regions for the LPE-BC with COF and $\Ksf = 2$ users will be of the form presented in  Theorem~\ref{thm:Prototype} next, which was inspired by~\cite{Sagduyu-unusual}.
We shall use the following nomenclature:
\begin{itemize}
\item An {\it uncoded packet} is packet that is sent by itself, i.e., not coded together with other packets, on some layer, which is also defined as the {\it native packet} in~\cite{XOR-in-the-air}.
\item An {\it overheard packet} is a packet that has not yet been delivered uncoded to the intended user but has  been successfully received at the non-intended user and that serves as the side information when we apply network coding. 
\item A {\it (network) coded packet} is a packet that is sent on some layer in a linear combination involving overheard packets from other users, which has been widely used in networking to improve the throughput of the communication systems. 
\end{itemize}

\begin{algorithm} 
$\textit{Input.}$ Initialize $Q_{\mathcal{S},q}, K_{u,q}^{\text{(unc)}}, K_{u,q}^{\text{(rem)}}, K_{u}^{\text{(rem)}},K_u^{\text{(NC)}}$ for all  $\mathcal{S} \subseteq \{1,2\}, q\in[\Qsf], u\in \mathcal{S}$ as in Table~\ref{tab-notations}.

$\textit{Output.}$ $T^{\text{(unc)}}, T^{\text{(NC)}}$. The number of slots to finish Phase1 and Phase2 respectively.
\begin{algorithmic}[1]
\State $t \gets 0;$
\State \textbf{while} $   \sum_{u\in[2]} \left( \sum_{q\in[\Qsf]} K_{u,q}^{\text{(unc)}}(t) + \sum_{u\in[2]}K_{u}^{\text{(rem)}}(t) \right) > 0$ \textbf{do}
\State \hspace{3mm} \textbf{if} $\left( \sum_{u\in[2]}  K_{u,q}^{\text{(unc)}}(t)  > 0, \forall q\in[\Qsf] \right)$ \textbf{then}  \hspace{12mm}$\triangleright$ Phase1
\State \hspace{6mm}  \textbf{for} $q \gets 1, \cdots, \Qsf $ \textbf{do}
\State \hspace{9mm} $A_{q} = \{u\}$ randomly based on the proportion of the number of  packets in $Q_{\{1\},q}$ and $Q_{\{2\},q}$.
\State  \hspace{6mm}  \textbf{end for}
\State \hspace{6mm} run  Algorithm~\ref{update1} based on the feedback from both users;
\State \hspace{3mm} \textbf{else if} $ \left(\exists q\in[\Qsf]: \sum_{u\in[2]}  K_{u,q}^{\text{(unc)}}(t) \leq 0 \right)$ \textbf{then}  \hspace{4mm}$\triangleright$ Sub-phases of Phase1
\State \hspace{6mm} $T_{q}^{\text{(unc)}} \gets t$;  \ 
\State \hspace{6mm}  \textbf{for} $q \gets 1, \cdots, \Qsf $ \textbf{do}
\State \hspace{9mm} \textbf{if } $ \sum_{u\in[2]}K_{u}^{\text{(rem)}} >0$  \textbf{then}
\State \hspace{12mm} $A_{q} = \{1,2\}$;
\State \hspace{12mm} layer $q$ transmits a random linear combination of all the packets in $\cup_{j\in[\Qsf]}  Q_{\{1,2\},j}$;
\State \hspace{9mm} \textbf{end if}
\State \hspace{6mm}\textbf{end for}

\State \hspace{3mm} \textbf{else} $\left( \sum_{u\in[2]}  K_{u,q}^{\text{(unc)}}(t)  \leq 0, \forall q\in[\Qsf] \right)$ \textbf{then}  \hspace{9mm}$\triangleright$ Phase2
\State \hspace{6mm} $T^{\text{(unc)}} \gets t$; 
\State \hspace{6mm} transmit a different random linear combination of all the packets in $\cup_{q\in[\Qsf]} Q_{\{1,2\},q}$ on each layer;
\State \hspace{6mm} run  Algorithm~\ref{update2} based on the feedback from both users;
\State \hspace{3mm} \textbf{end if}
\State \hspace{3mm} $t \gets t+1;$
\State \textbf{end while}
\State $T^{\text{(NC)}} \gets t - T^{\text{(unc)}}$
\end{algorithmic}
\caption{Two-phase Protocol }
 \label{prototype}
\end{algorithm} 


The definition and initial value of some notations used in this section are listed in Table~\ref{tab-notations}. 
Before the protocol description, we give a brief discussion of the underlying rationale. 
The idea is to have a protocol with two phases: Phase1 corresponds to uncoded transmission on some layers (and can be split in sub-phases), while Phase2 corresponds to network coded transmissions on all layers. Herein, user $u$ must decode exactly $k_u := \sum_{q\in[\Qsf]} k_{u,q}$ packets in $ \cup_{q\in[\Qsf]} Q_{{u},q}$ and the transmitter sends uncoded or coded packets to each user $u$. 
Hence, to show the correctness of the algorithm, each user $u$ must eventually receive $k_u$ linearly independent combinations of the packets in $ \cup_{q\in[\Qsf]} Q_{{u},q}$. 
\paragraph*{ Description of the Two-phase Protocol}
It  is succinctly described in pseudocode form in Algorithm~\ref{prototype}. 
 The two-layer case is demonstrated in Fig.~\ref{fig:two-phase-demo}.
The transmitter maintains a network of queues $Q_{\{1\},q}, Q_{\{2\},q}$ and $Q_{\{1,2\},q}, q\in[\Qsf]$, with  initial value listed in~Table~\ref{tab-notations}.

{\bf Phase1}. In each slot, we transmit $\Qsf$ uncoded packets, one from each layer, simultaneously according to a predetermined order  in line 5 of Algorithm~\ref{prototype} that is known to all users. $A_q \subseteq \{1,2\}$ indicates which user is served on layer $q\in[\Qsf]$, i.e., $A_q = \{u\}, u\in[2]$ means an uncoded packet for user $u$ is transmitted on layer $q$; $A_q = \{1,2\}$ means a coded packet is transmitted on layer $q$.  
Each receiver's output is fed back to the transmitter at the end of slot $t$ (ACK for received packet, NACK for erased packet), which implies that the transmitter knows $N^{2, t}$  
at the beginning of slot $t+1$. The entities $Q_{\mathcal{S},q}, K_{u,q}^{\text{(unc)}}, K_{u,q}^{\text{(rem)}}, K_{u}^{\text{(rem)}}, K_{u,q}^{\text{(NC)}}$ are dynamically updated based on the feedback from both users as follows: 
\begin{itemize}
\item If a transmitted packet from queue $Q_{\{u\},q}$ is erased at both users, it remains in queue $Q_{\{u\},q}$.
\item
If a transmitted packet from queue $Q_{\{u\},q}$ is received by at least one of the receivers, it is removed from $Q_{\{u\},q}$ and the counting number $K_{u,q}^{\text{(unc)}}$ is reduced by one.
\item
If a transmitted packet destined to user $u$ from queue $Q_{\{u\},q}$ is only received by user $\bar{u}, \bar{u}\neq u$, it is moved from queue $Q_{\{u\},q}$ to queue $Q_{\{1,2\},q}$ and the counting number $K_{u,q}^{\text{(unc)}}$ is reduced by one, $K_{u,q}^{\text{(rem)}}$ and $K_{u}^{\text{(rem)}}$ are increased by one.
\end{itemize}

\begin{algorithm} \caption{Update of Phase1 }\label{update1}
\begin{algorithmic}[1]
\State \textbf{for} $q \gets 1, \cdots, \Qsf$ \textbf{do} 
\State  \hspace{3mm} pick $A_q$ according to line 5 in Algorithm~\ref{prototype};
\State  \hspace{3mm} \textbf{if} $A_q = \{ u\}$ \textbf{then} 
\State  \hspace{6mm} \textbf{if} $ \max(N_{1},N_{2}) < q$ \textbf{then}
\State \hspace{9mm} retransmit that uncoded packet from $Q_{\{u\},q}$ on layer $q$; 
\State  \hspace{6mm} \textbf{else if} $\left( N_{u} \geq q, K_{u,q}^{\text{(unc)}}>0 \right)$  \textbf{then}
\State \hspace{9mm} remove that uncoded packet from $Q_{\{u\},q}$; 
\State \hspace{9mm} $K_{u,q}^{\text{(unc)}} \gets K_{u,q}^{\text{(unc)}} -1$;
\State  \hspace{6mm} \textbf{else if}  $\left( N_u < q, N_{\bar{u}} \geq q, K_{u,q}^{\text{(unc)}} >0, u\neq \bar{u} \right)$   \textbf{then}
\State \hspace{9mm} remove that uncoded packet from $Q_{\{u\},q}$ and move that uncoded packet to $Q_{\{1,2\},q}$; 
\State \hspace{9mm} $K_{u,q}^{\text{(unc)}} \gets K_{u,q}^{\text{(unc)}} -1$;
\State \hspace{9mm} $K_{u,q}^{\text{(rem)}} \gets K_{u,q}^{\text{(rem)}}+1$;
\State \hspace{9mm} $K_{u}^{\text{(rem)}} \gets K_{u}^{\text{(rem)}} + K_{u,q}^{\text{(rem)}}$;
\State \hspace{6mm}\textbf{end if}
\State  \hspace{3mm} \textbf{else if}  $A_{q} = \{1,2 \}$ \textbf{then}
\State  \hspace{6mm}    \textbf{if} $\left( N_u \geq q, K_{u}^{\text{(rem)}} >0, u\in[2] \right)$ \textbf{then}
\State   \hspace{9mm}  $K_{u}^{\text{(rem)}} \gets K_{u}^{\text{(rem)}} -1$;
\State   \hspace{9mm}  $K_{u,q}^{\text{(NC)}} \gets K_{u,q}^{\text{(NC)}} +1$;
\State  \hspace{6mm} \textbf{end if}
\State  \hspace{4mm}\textbf{end if}
\State \textbf{end for}
\end{algorithmic}
\end{algorithm}



\begin{algorithm}\caption{Update of Phase2 }\label{update2}
\begin{algorithmic}[1]
\State \textbf{for} $q \gets 1, \cdots, \Qsf$ \textbf{do} 
\State  \hspace{3mm} \textbf{ if} $ \left( N_{u,t}\geq q, K_{u}^{\text{(rem)}}>0, u\in[2] \right)$   \textbf{then}
\State \hspace{6mm} $K_{u}^{\text{(rem)}} \gets K_{u}^{\text{(rem)}}-1$;
\State \hspace{6mm} $\sum_{q\in[\Qsf]}K_{u,q}^{\text{(NC)}}\gets \sum_{q\in[\Qsf]}K_{u,q}^{\text{(NC)}}+1$
\State \hspace{3mm} \textbf{end if}
\State \textbf{end for}
\end{algorithmic}
\end{algorithm}


{\bf Sub-phases}. In line 9 of Algorithm~\ref{prototype}, we mean that $T^\text{\rm(unc)}_q$ records the first time $\sum_{u\in[2]}  K_{u,q}^{\text{(unc)}}(t) \leq 0$ and it will not be updated as time goes on. 
Thus, $T^\text{\rm(unc)}_q$ is the time at which layer $q$ finishes its all uncoded packets in queues $Q_{\{1\},q}$ and $Q_{\{2\},q}$. 
In line 17 of Algorithm~\ref{prototype}, we mean that $T^\text{\rm(unc)}$ records the first time 
$\sum_{u\in[2]} \sum_{q\in[\Qsf]} K_{u,q}^{\text{(unc)}}(t) \leq 0$ and it will not be updated as time goes on. 
Thus,  $T^\text{\rm(unc)}$ is the time at which all sub-phases complete and Phase 1 ends. 
Since the time needed for each layer to complete the transmission of its uncoded packets may be different, let $\pi$ be the permutation of $[\Qsf]$ such that  
\begin{align}
0\equiv T^\text{\rm(unc)}_{\pi(0)}\leq T^\text{\rm(unc)}_{\pi(1)} \leq T^\text{\rm(unc)}_{\pi(2)} \ldots \leq T^\text{\rm(unc)}_{\pi(\Qsf)} \equiv T^\text{\rm(unc)}.
\label{time-permutation}
\end{align} 
At time $T^\text{\rm(unc)}_{\pi(j)}$, the layers $\pi(1), \ldots, \pi(j)$ have finished their uncoded phase. 
Phase1 is hence composed of $\Qsf$ sub-phases. 
The slot $\left[T^\text{\rm(unc)}_{\pi(j-1)},T^\text{\rm(unc)}_{\pi(j)}\right), \ j\in[\Qsf],$ is where of the $j$-th sub-phase of Phase1 takes place. 
There are $\Qsf!$ possible configurations of sub-phases, one for each  permutation of $[\Qsf]$.
Once layer $q$ has finished sending its uncoded packets at time $T^\text{\rm(unc)}_{q}$, we send linear combinations of {\it all} overheard packets on {\it all} layers up to the current time slot (note: each layer gets a linearly independent linear combination); we refer to this scheme as an {\it  inter-user  \& inter-layer network coding scheme}. 
 The coded packets transmitted during the sub-phases in slot $t\in[T_{\pi(1)}^{\text{(unc)}}, T^{\text{(unc)}}]$ can be written as 
 \begin{align}
s = \sum_{q\in[\Qsf]} \sum_{p\in Q_{\{1,2\},q}(t)} a_s(p)p, \label{coded-packet-mech3}
\end{align}
where $a_s(p)$ is the random encoding coefficient of packet $p$
over the finite field input alphabet $\mathcal{X}$ of dimension $\sum_{q\in[\Qsf]}| Q_{\{1,2\},q}(t)|$ and  it is generated by a random number generation algorithm known a priori to both users.
New packets may be added to $Q_{\{1,2\},q}$ after each slot and the packet $s$ in~\eqref{coded-packet-mech3}  is the linear combinations of packets from dynamic queues $Q_{\{1,2\},q}(t)$. We use $Q_{\{1,2\},q}(t)$ to emphasize that the queue will be updated as time goes on. 
 During the sub-phases, some of  $Q_{\{1,2\},q}$ finishes updating, i.e., the uncoded phase of layer $q$ is done and no more packets will be added to $Q_{\{1,2\},q}$, while some of $Q_{\{1,2\},j}$ are still updating, i.e., the uncoded phase of layer $j$ is not done yet and new overheard uncoded packets may be added to $Q_{\{1,2\},j}$ later. 
 
{\bf Phase2}. In each slot, 
the coded packet $s$ transmitted in line 18 of Algorithm~\ref{prototype} 
 is the linear combination of all the overheard packets from all $Q_{\{1,2\},q}, q\in[\Qsf]$ in the transmitter and can be expressed as
\begin{align}
s = \sum_{q\in[\Qsf]} \sum_{p\in Q_{\{1,2\},q}(T^{\text{(unc)}})} a_s(p)p. 
\label{packet:phase2}
\end{align}
 We use $K_{u,q}^{\text{(NC)}}$ to track the number of coded packets received by user $u$ on layer $q$ in line 18 of Algorithm~\ref{update1}. 
 Once a coded packet is received by user $u$, $K_{u}^{\text{(rem)}} $ is reduced by one and 
 $\sum_{q\in[\Qsf]}K_{u,q}^{\text{(NC)}}$ is increased by one as shown in line 3, 4 of Algorithm~\ref{update2}. 
 Phase2 completes as soon as $\sum_{u\in[2]}K_{u}^{\text{(rem)}} \leq 0$. 

 By our protocol, the packets are transmitted in two forms: uncoded and coded.
 Each layer first transmits the uncoded packets and these packets can be decoded by the users if they are received. We just need to confirm that all the remaining packets can be recovered successfully from the coded packets by each user. Note that the overheard uncoded packets are also stored at the users to serve as side information to help recover the coded packets in the future. 
By random linear network coding of $K$ packets,  each coded packet is associated with an encoding vector over a finite field of size $q$. 
 The probability of successfully decoding $K$ packets from $K$ received  coded packets is~\cite{Exact-decoding-probability}
 \begin{align}
 \Pr = \prod_{i=1}^{K}\left(1-\frac{1}{q^i}\right).
 \label{decodable-probability}
\end{align} 
Assuming $q$ 
is large enough, any received packet is linearly independent from previously received (sums of) packets with high probability.
Following the same idea, for our protocol, we can show that all coded packets received by the users are linearly independent with high probability. 
Then, we only need to count the number of coded packets received by each user in order to make sure they can recover all the destined packets.

Based on our two-phase protocol, the achievable region  is written in the following theorem.

\begin{theorem}[Achievable region of two-phase protocol]
\label{thm:Prototype}
The following region is achievable for the LPE-BC with COF and $\Ksf=2$ users:
\begin{subequations}
\begin{align}
\mathcal{C}^\text{in} :=
& \cup_{ t\geq0, k_{u,q}\geq 0, q\in[\Qsf], u\in[2]} 
\Big\{ (R_1,R_2) : t^\text{\rm(unc)} + t^\text{(NC)} \leq t,
\\
& t^\text{\rm(unc)} := \max_{q\in[\Qsf]} \left( t^\text{\rm(unc)}_{q} \right), \ \text{(duration of Phase1)},
\label{eq:schemePrototype-T1}
\\
& t^\text{(NC)} := \max_{u\in[2]}\left( t^\text{(NC)}_{u} \right), \ \text{(duration of Phase2)},
\label{eq:schemePrototype-T2}
\\
& t^\text{\rm(unc)}_{q} := \frac{k_{1,q} + k_{2,q}}{\Pr[  \max(N_{1},N_{2}) \geq q]}, 
\forall q\in[\Qsf], 
\label{eq:schemePrototype-T1q}
\\
& t^\text{(NC)}_{u} :=  \frac{k^\text{\rm(rem)}_{u}}{\mathbb{E}[N_{u}]},
\forall u\in[2], 
\label{eq:schemePrototype-T2u}
\\
& k^\text{\rm(rem)}_{u} := \Bigg[ \sum_{q \in [\Qsf]} k_{u,q}^{\text{\rm(rem)}} - (t^\text{\rm(unc)} - t^\text{\rm(unc)}_{q})\Pr[N_{u}\geq q]\Bigg] ^+, u \in[2], 
\label{eq:schemePrototype-KuNotDefined}
\\
& k^\text{\rm(rem)}_{u,q} := k_{u,q}\left( 1 - \frac{\Pr[N_{u}\geq q]}{\Pr[  \max(N_{1},N_{2}) \geq q]} \right),   
\begin{array}{l}
\forall q\in[\Qsf], \\
\forall u\in[2], \\
\end{array}
\label{eq:schemePrototype-Kuqremaining}
\\
 &R_u := \frac{\sum_{q\in[\Qsf]}k_{u,q}}{t}, 
\forall u\in[2], \ \text{(rate)}  \Big\}.
\end{align}
\label{eq:schemePrototype}
\end{subequations}
\end{theorem}
\vspace{-0.62cm}

\paragraph*{Analysis of Two-phase Protocol}
We focus on the situation where no common message needs to be sent over the channel. 
Fix $k_{u,q}, \,  u\in[2], \ q\in[\Qsf]$  and let $k_{u,q} \gg 1$ so that we can invoke the Law of Large Numbers in the following analysis (loosely speaking, we ``replace'' random processes with their statistical averages). Then $k_u = \sum_{q\in[\Qsf]} k_{u,q}, \ u\in[2]$ is the number of packets destined to user $u$. The special cases that $\exists q\in[\Qsf]$ or $u\in[2]: k_{u,q} = 0$ are discussed in Appendix~\ref{sec:multiple-layers}.

{\it Phase1}.
The transmitter sends $k_{u,q}$ uncoded packets on layer $q\in[\Qsf]$ for user $u\in[2]$, one by one according to a predetermined order in line 5 of Algorithm~\ref{prototype}, known to all users, until at least one of the two users has received it. 
 $t^\text{\rm(unc)}_{q}, q\in[\Qsf]$ is the number of time slots until all $k_{1,q} + k_{2,q}$ uncoded packets on layer $q$ are received by at least one user. 
Recall that the channel is characterized by the random variable  $N_{u,t}, u\in[2]$ and $N_{u,t} \geq q$ indicates that, at time slot $t$, the packet transmitted on layer $q$ has been received by user $u$. 
If $ \max(N_{1,t},N_{2,t}) \geq q$, a packet on layer $q$ has been received by at least one of the users at time slot $t$.  Therefore, it takes on average $\frac{1}{\Pr[ \max(N_{1},N_{2}) \geq q]}$ time slots to deliver one uncoded packet to some user on layer $q\in[\Qsf]$, and $t^\text{\rm(unc)}_{q}$ is hence as given in~\eqref{eq:schemePrototype-T1q}. All the received uncoded packets can be decoded by both users successfully.
Meanwhile, $k^\text{\rm(rem)}_{u,q}$ in~\eqref{eq:schemePrototype-Kuqremaining}, is the number of packets not successfully received by user $u$ on layer $q$ and is also the number of packets overheard by the other user $\bar{u}\neq u$ on layer $q$. 
By time $t^\text{\rm(unc)}$ in~\eqref{eq:schemePrototype-T1}, all layers are done sending uncoded packets, and Phase1 ends. There are $k^\text{\rm(rem)}_{u}$ in~\eqref{eq:schemePrototype-KuNotDefined} 
packets that have not been received by user $u\in[2]$, which will be sent in a network coded manner on any layer. If user $u$ receives $k^\text{\rm(rem)}_{u}$ linearly independent combinations of the overheard packets, it is able to decode the remaining $k^\text{\rm(rem)}_{u}$  packets. 

{\it Sub-phases}.
By~\eqref{time-permutation}, we have 
\begin{align}
0\equiv t^\text{\rm(unc)}_{\pi(0)}\leq t^\text{\rm(unc)}_{\pi(1)} \leq t^\text{\rm(unc)}_{\pi(2)} \ldots \leq t^\text{\rm(unc)}_{\pi(\Qsf)} \equiv t^\text{\rm(unc)},
\end{align} 
where we recall that at time $t^\text{\rm(unc)}_{\pi(j)}$, the layers $\pi(1), \ldots, \pi(j)$ have finished their uncoded phase. 
Phase1 is hence composed of $\Qsf$ sub-phases, where the $j$-th sub-phase has duration $\Delta_j := [t^\text{\rm(unc)}_{\pi(j-1)} - t^\text{\rm(unc)}_{\pi(j)}), \ j\in[\Qsf]$. 

As stated in line 5 of Algorithm~\ref{prototype}, the order in which packets are sent on layer $q\in[\Qsf]$ during the uncoded phase (that is, time interval $[0,t^\text{\rm(unc)}_{q}]$) is randomized, that is, the probability of a user being picked to be served in a given time slot is proportional to the number of uncoded packets that the user needs to receive on that layer. 
During the uncoded phase of Phase1, $A_q$ is assumed to be i.i.d. over time and independent of everything else with
\begin{align}
&\Pr[A_q=\{u\}] = \frac{k_{u,q}}{k_{1,q}+k_{2,q}}, u\in[2].
\label{eq:pr Aq}
\end{align}
With~\eqref{eq:pr Aq} and~\eqref{eq:schemePrototype-T1q}, we write
\begin{align}
&\Pr[A_q=\{u\}, M \geq q]
 = \frac{k_{u,q}}{t^\text{\rm(unc)}_{q}},
\label{eq:pr Aq max}
\\
&\Pr[A_q=\{u\}, M \geq q, N_u<q] 
 = \frac{k_{u,q}}{t^\text{\rm(unc)}_{q}} \ \eta_{u,q},
\label{eq:pr Aq max Nu}
\\
&
\eta_{u,q} := \frac{k^\text{\rm(rem)}_{u,q}}{k_{u,q}} = 1-\frac{\Pr[N_{u}\geq q]}{\Pr[\max(N_{1},N_{2})\geq q]}\in[0,1],
\label{eq:DTOct04proposed_eta_uq}
\end{align}
where~\eqref{eq:pr Aq max} is the probability that a packet destined to user $u\in[2]$ is assigned on layer $q\in[\Qsf]$ and its uncoded packet is received by at least one of the users; 
similarly,~\eqref{eq:pr Aq max Nu} is the probability that a packet destined to user $u\in[2]$ is assigned on layer $q\in[\Qsf]$ and its uncoded packet is received by the other user only. The quantity in~\eqref{eq:DTOct04proposed_eta_uq} can be thought of as the fraction of overheard packets for user $u\in[2]$ on layer $q\in[\Qsf]$.

Let $k^\text{\rm(unc)}_{u,q}[j]$ be the number of uncoded packets that have not yet been transmitted to user $u\in[2]$ on layer $q\in[\Qsf]$ at the end of the $j$-th sub-phase; these packets must be still sent on layer $q\in[\Qsf]$.
Also, let $k^\text{(rtx)}_{u}[j]$ be the number of overheard packets left to be delivered to user $u\in[2]$ at the end of the $j$-th sub-phase; we can send these packets in a network coded way on any layer.
Initialize $k^\text{\rm(unc)}_{u,q}[0] = k_{u,q} \geq 0$ and $k^\text{(rtx)}_{u}[0] = 0$. 
We have the following recursive equation for $j\in[\Qsf]$
\begin{align}
   k^\text{\rm(unc)}_{u,q}[j]  
  & = \left[ k^\text{\rm(unc)}_{u,q}[j-1] - \Delta_j \Pr[A_q=\{u\}, \max(N_1,N_2) \geq q] \right]^+
\notag\\ &= k_{u,q} \left[ 1-\frac{t^\text{\rm(unc)}_{\pi(j)}}{t^\text{\rm(unc)}_{q}} \right]^+.
\label{eq:DTOct04proposed_k_unc}
\end{align}
The update equation for $k^\text{\rm(unc)}_{u,q}[j]$ in~\eqref{eq:DTOct04proposed_k_unc} says that the number of uncoded packets for user $u\in[2]$ on layer $q\in[\Qsf]$ decreases with ``time'' $j\in[\Qsf]$. In particular, at the end of the $j$-th sub-phase, $k^\text{\rm(unc)}_{u,q}[j-1]$ is reduced by the number of packets that can be received by either user during the time interval $\Delta_j$ whenever user $u\in[2]$ is scheduled for transmission on layer $q\in[\Qsf]$. The final expression in~\eqref{eq:DTOct04proposed_k_unc} simply says that by time $t^\text{\rm(unc)}_{\pi(j)}$ the fraction of uncoded packets left to be transmitted is proportional to $1-t^\text{\rm(unc)}_{\pi(j)} / t^\text{\rm(unc)}_{q}$ if $t_{\pi(j)} <t_q$ and zero, otherwise.
Similarly, we have for $j\in[\Qsf]$
\begin{align}
k^\text{(rtx)}_{u}[j] &
= \Big[ k^\text{(rtx)}_{u}[j-1] - \Delta_j \sum_{\ell=1}^{j-1} \Pr[N_u\geq \pi(\ell)] 
\notag\\& 
   +  \sum_{q\in[\Qsf]}\min
        \Big(  p_{u,q,j}, k^\text{\rm(unc)}_{u,q}[j-1] \Big) 
        \Big]^+ 
\label{eq:DTOct04proposed_k_rtx}
    \\
& = \Big[  \sum_{ q\in[\Qsf]:t_q \geq t_{\pi(j)} } k_{u,q}^{\text{\rm(rem)}} \frac{t_{\pi(j)}^{\text{\rm(unc)}}}{t_q^{\text{\rm(unc)}}}   
\notag\\& 
+ \sum_{ q\in[\Qsf]:t_q < t_{\pi(j)} } 
    \left(k_{u,q}^{\text{\rm(rem)}} - (t_{\pi(j)}^{\text{\rm(unc)}} - t_q^{\text{\rm(unc)}})\Pr[N_u \geq q] \right) \Big]^+,
\label{eq:DTOct04proposed_k_rtx_final}
\\
p_{u,q,j} &:= \Delta_j\Pr[A_q=\{u\}, \max(N_1,N_2) \geq q, N_u<q].
\end{align}

The update equation for $k^\text{(rtx)}_{u}[j]$ in~\eqref{eq:DTOct04proposed_k_rtx} says that the number of coded packets for user $u\in[2]$ can either increase or decrease over ``time'' $j\in[\Qsf]$, depending on the difference of the number of coded packets delivered to user $u$ and the number of uncoded packets received at the other user only within the $j$-th sub-phase. 
Specifically, at the end of the $j$-th sub-phase, $k^\text{(rtx)}_{u}[j-1]$ is decreased by the number of packets that can be received by user $u\in[2]$ during the time interval $\Delta_j$ on the layers that have already completed their uncoded phase (which is proportional to $\sum_{\ell=1}^{j-1} \Pr[N_u\geq \pi(\ell)]$), or increased by the number of overheard packets during the time interval $\Delta_j$ across any of the layers. The ``min'' in~\eqref{eq:DTOct04proposed_k_rtx} simply says that the number of overheard packets for user $u\in[2]$ on layer $q\in[\Qsf]$ cannot exceed the number of uncoded packets left for transmission at the end of the $(j-1)$-th sub-phase, $k^\text{\rm(unc)}_{u,q}[j-1]$.
The much simplified expression in~\eqref{eq:DTOct04proposed_k_rtx_final}  is derived in Appendix~\ref{sec:multiple-layers}. 

At the end of the $\Qsf$-th sub-phase, all the uncoded packets on all layers are done, i.e., we have $k^\text{\rm(unc)}_{u,q}[\Qsf]=0$ for all $q$, but possibly some $k^\text{(rtx)}_{u}[\Qsf]>0$. Therefore, we still have $k^\text{\rm(rem)}_{u}= $ $k^\text{(rtx)}_{u}[\Qsf]$ in~\eqref{eq:schemePrototype} coded packets to deliver to user $u\in[2]$ during Phase2. 
The expression in~\eqref{eq:schemePrototype-KuNotDefined} can be obtained from~\eqref{eq:DTOct04proposed_k_rtx_final} with $j = \Qsf$ as follows 
\begin{align}
k^\text{(rtx)}_{u}[\Qsf]&
= \Big[  k_{u,\pi(\Qsf)}^{\text{\rm(rem)}} \frac{t_{\pi(\Qsf)}^{\text{\rm(unc)}}}{t_{\pi(\Qsf)} ^{\text{\rm(unc)}}}   
\notag\\& 
+ \sum_{q\in[\Qsf]:t_q < t_{\pi( \Qsf)}} 
    \left(k_{u,q}^{\text{\rm(rem)}} - (t_{\pi(\Qsf)}^{\text{\rm(unc)}} - t_q^{\text{\rm(unc)}})\Pr[N_u \geq q] \right) \Big]^+ \notag
\\& =   \Big[ \sum_{q \in [\Qsf]} k_{u,q}^{\text{\rm(rem)}} - (t^\text{\rm(unc)} - t^\text{\rm(unc)}_{q})\Pr[N_{u}\geq q]\Big] ^+,
\label{eq:DTOct04proposed_k_rtx_Q}
\end{align}
as claimed. 
We also give an alternative proof of the achievable region for the single-layer case in Appendix~\ref{sec:proof of time concentration}, as an alternative to the Markov chain based analysis  in~\cite{Sagduyu-unusual}. 

{\it Phase2}.
Once all layers are done sending their uncoded packets at time $t^\text{\rm(unc)}$ in~\eqref{eq:schemePrototype-T1}, on each layer we send different linearly independent  linear combinations of the overheard packets, as defined in Section~\ref{capacity:inner-bounds}. 
If the coded packets are transmitted so, user  $u\in[2]$ will eventually receive $k^\text{\rm(rem)}_{u}$ packets, which it uses to recover its desired messages by solving a  linear system that has a full-rank matrix with probability 1. 
In each time slot, user  $u$ receives on average $\sum_{q\in[\Qsf]} q \Pr[N_u = q] = \mathbb{E}[N_{u}]$ packets.
Therefore, $t^\text{(NC)}_{u}$, the average time needed to receive the remaining $k^\text{\rm(rem)}_{u}$ packets  in~\eqref{eq:schemePrototype-KuNotDefined}  is given in~\eqref{eq:schemePrototype-T2u}.



 The following lemma shows the correctness of the two-phase protocol in Theorem~\ref{thm:Prototype}.
\begin{lemma} At any time $t\in[T_{\pi(1)}^{\text{(unc)}}, T^{\text{(unc)}} + T^{\text{(NC)}} ]$, for any $\varepsilon >0$,
 \label{lemma-mechanism3}
 \begin{align}
\Pr[&\text{``the encoding vectors of the received $\sum_{q\in[\Qsf]}K_{u,q}^{\text{(NC)}}(t) = \sum_{q\in[\Qsf]}k_{u,q}^{\text{(NC)}}(t)$,} \notag
\\ &\forall u\in[2] \text{ packets are linearly independent''} ] > 1-\varepsilon.
  \end{align}
  \end{lemma}

Lemma~\ref{lemma-mechanism3} guarantees that after $T^{\text{(unc)}} + T^{\text{(NC)}}$ time slots with high probability user $u\in[2]$  can decode the desired $k_u$ packets.     
The proof of this Lemma is relegated to Appendix~\ref{sec: proof of lemma1}.

\begin{subequations}

Note that the outer bound in Theorem~\ref{thm:COFnewOuter} is for any number of users, while the inner bound in Theorem~\ref{thm:Prototype} is for $\Ksf=2$ users only.
Extension of the scheme that attains the inner bounds to more than $\Ksf = 2$ users requires being able to track which subset of non-intended users has received a certain packet; this is the same stumbling block as in the single-layer case in~\cite{on-the-capacity-of-1-to-k} for $\Ksf\geq 4$.
\end{subequations}

We end this section with a trivial inner bound result which is the simple extension of the algorithm in~\cite{Sagduyu-unusual}, where the erasure channel model studied in~\cite{Sagduyu-unusual} is the special case of $\Qsf=1$ in our LPE-BC model.  This will be used as a comparison point for our proposed Theorem~\ref{thm:Prototype}.
\begin{lemma}[Trivial scheme]
\label{thm:scheme1}
The following region is achievable for the LPE-BC with COF and $\Ksf=2$ users
\begin{subequations}
\begin{align}
& \cup_{R_{u,q}\geq0, q\in[\Qsf], u\in[2]} 
\Big\{ (R_1,R_2) : \max_{q\in[\Qsf]} \left( v_{q} \right) \leq 1,
\\
&v_{q} := \max\left(
 \frac{R_{1,q}}{\Pr[\max(N_1,N_2) \geq q]} + \frac{R_{2,q}}{\Pr[N_2 \geq q]},
\right.\notag\\&\quad\left. 
 \frac{R_{1,q}}{\Pr[N_1 \geq q]}  + \frac{R_{2,q}}{\Pr[\max(N_1,N_2) \geq q]}
\right), \ q\in[\Qsf], 
\label{trivial_achievable_region}
\\
&R_u := R_{u,1}+\ldots+R_{u,\Qsf}, \ u\in[2] \Big\}.
\end{align}
\label{eq:scheme1}
\end{subequations}
\end{lemma}
\vspace{-0.5cm}
The region in~\eqref{eq:scheme1} is achievable for the LPE-BC with COF by employing the two-phase algorithm in~\cite{Sagduyu-unusual} independently on each layer. To map the notation used in~\cite{Sagduyu-unusual} to ours, please note that $\epsilon_{u,q} = 1-\Pr[N_u \geq q], \ u\in[2], q\in[\Qsf]$  is the probability that layer $q$ is erased for user $u$, and $\epsilon_{12,q}= 1-\Pr[ \max(N_1,N_2) \geq q],  \ q\in[\Qsf]$ is the probability that layer $q$ is erased at both users.

Note that the extension of Lemma~\ref{thm:scheme1} to more than $\Ksf =2$ users requires knowing the capacity of the single-layer model for $\Ksf$ users, which is open at present in general. The scheme in~\cite{on-the-capacity-of-1-to-k} is tight (i.e., it achieves the outer bound in Theorem~\ref{thm:COFnewOuter}) for $\Qsf=1$ and $\Ksf\leq 3$ users, and also for $\Qsf=1$ and $\Ksf\geq 4$ in some symmetric settings. Paper~\cite{on-the-capacity-of-1-to-k} claims that the scheme matches (up to numerical precision) the outer bound for all simulated case of $\Ksf\leq 6$ users; if the scheme were indeed optimal for any number of users, then Lemma~\ref{thm:scheme1} could give a scheme for any number of layers and users, and would prove the tightness of Theorem~\ref{thm:COFnewOuter} for $\Qsf=1$.

\section{Stability Region} 
\label{subsec:stochastic}

In this section, we assume that packets arrive randomly to the system rather than assuming backlogged packet queues as was done above. The general assumptions remain the same as was presented in Section~\ref{capacity:inner-bounds}.
We first describe how Theorem~\ref{thm:Prototype} can be adapted to such an environment, then propose an achievable stability region of this modified protocol. 
\paragraph*{Protocol Description} 
The protocol works in epochs. During each epoch, a certain (random) number of packets have to be successfully delivered to the users by employing the coding scheme for the backlogged case in Theorem~\ref{thm:Prototype}. The beginning of a new epoch is a \emph{renewal event} for the system. Initially, epoch~$1$ starts at time $T[1] =0$. 
Epoch~$m\in\mathbb{N}$ starts at time  $L[m]$ and ends at time  $L[m+1]$.  Denote $T[m] := L[m+1]-L[m]$ as the number of time slots in epoch $m$, where each packet is transmitted in one slot. Epoch~$m+1$ starts at time $L[m+1]$ (right after the end of epoch~$m$), employing the same procedure as in epoch~$m$. Let $A_{u,t}$ be the number of exogenous arrival  packets for user $u$ at slot $t$. At the beginning of epoch~$m$, $K_{u}[m] := \sum_{t\in T[m-1]} A_{u,t}$ new exogenous packets need to be transmitted to user $u\in[2]$. If $K_{1}[m] = K_{2}[m] = 0$,  the $m$-th epoch ends at the time $T[m-1]$, which is the time that the $(m-1)$-th epoch ends.
In general, epoch~$m$ ends when all $K_{u}[m]$ packets have been decoded successfully by user $u$, thus $T[m], m\in \mathbb{N}$ are random.

The transmitter maintains $\Qsf + 2$ queues of infinite size, denoted by $Q_{01}, Q_{02}, \cdots, Q_{0\Qsf}, Q_{1}, Q_2$. 
New arrival packets assigned on layer $q$ are queued in $Q_{0q}, q\in[\Qsf]$, but are not transmitted until the epoch ends. 
Queue $Q_{u}$ is used to store the packets destined to user $u$ but only received by user $\bar{u}, u\neq \bar{u}$.
For each user $u$, the average arrival rate $\lambda_u = \mathbb{E}[A_{u,t}]$ is expressed as $\lambda_u = \sum_{q \in [\Qsf]}\lambda_{u,q} =  \sum_{q \in [\Qsf]}\mathbb{E}[A_{u,q;t}] $  where $A_{u,q;t}$ is the number of exogenous packets assigned for user $u$ on layer $q$ at slot $t$, for some $\lambda_{u,q}\geq0$ and $q\in[\Qsf]$. 
At the beginning of epoch~$m$, 
no packet is assigned to queue $Q_{u}$ (i.e., there are no overheard packets at the start of an epoch as the previous epoch ends after all packets are delivered), and each of the $K_{u}[m]$ packets is assigned independently at random with probability $\lambda_{u,q}/\lambda_u$ to queue $Q_{0q}$.  
Let $K_{u,q}[m]$ be the (random) number of packets that are assigned to queue $Q_{0q},$ and destined to user~$u$. 

The protocol works as follows. All packets, whether they are destined to user~1 or~2, are transmitted on a first-come-first-served policy from $Q_{0q}$.
The users send the COF to the transmitter after each transmission based on their receipt. 
The queue management policy according to the COF from both users, 
is given as follows. For $K_1[m] + K_2[m] \neq 0$, 
\begin{itemize}
\item If an uncoded packet transmitted from $Q_{0q}$ is not received by either of the two receivers, it remains in queue $Q_{0q}$;
\item If an uncoded packet, destined to user $u$ and transmitted from $Q_{0q}$, is successfully received by user $u$, it leaves $Q_{0q}$ (regardless whether the other user $\bar{u}$ receives it or not);
\item If an uncoded packet, destined to user $u$ and transmitted from $Q_{0q}$, is erased at user $u$ and received by the other user $\bar{u}$, it is moved from queue $Q_{0q}$ to queue $Q_u$;
\item If a coded packet, which is a linear combination of packets $p_1$ from $Q_1$ and packets $p_2$ from $Q_2$, is received by user $u$, then packets $p_u$ leaves $Q_u$;
\item If an uncoded packet transmitted from $Q_u$ is received by user $u$, it leaves $Q_u$ (regardless whether the other user $\bar{u}$ receives it or not). 
\end{itemize}
At any time, the transmission policy is the following:
\begin{itemize}

\item All queues are empty: this epoch ends. 

\item All $Q_{0q}$'s are non-empty: 
a packet from $Q_{0q}$ is transmitted  on layer $q$.  
This corresponds to the  initial uncoded transmission from queues $Q_{\{1\},q}$ and $Q_{\{2\},q}$ of Phase1 in Theorem~\ref{thm:Prototype}.

\item Some $Q_{0q}$'s are empty, and all $Q_{u} $'s are non-empty: if $\exists q,p\in[\Qsf] : p\neq q, Q_{0q} = \emptyset, Q_{0p} \neq \emptyset$ and $Q_{u} \neq \emptyset, \forall u\in[2]$, then we transmit a network coded packet 
on layer $q$, 
and an uncoded packet from $Q_{0p}$ on layer $p$.  
This corresponds to the  coded transmission from queues $\cup_{q\in[\Qsf]}  Q_{\{1,2\},q}$  in sub-phases of Phase1 in Theorem~\ref{thm:Prototype}.

\item Some $Q_{0q}$'s are empty, and some $Q_{u} $'s are empty: if $\exists q,p\in[\Qsf] : p\neq q, Q_{0q} = \emptyset, Q_{0p} \neq \emptyset$ and $\exists u \neq \bar{u} : Q_{u} \neq \emptyset, Q_{\bar{u}} = \emptyset$, then we transmit an uncoded packet from $Q_u$ on layer $q$, and an uncoded packet from $Q_{0p}$ on layer $p$.
This also corresponds to the  coded transmission from queues $\cup_{q\in[\Qsf]}  Q_{\{1,2\},q}$  in sub-phases of Phase1 in Theorem~\ref{thm:Prototype}. This happens when  the number of  packets intended to user $u$ only received by user $\bar{u}$ exceeds the  number of  packets intended to user $\bar{u}$ only received by user $u$, $u \neq \bar{u}$.

\item All $Q_{0q}$'s are empty, and all $Q_{u} $'s are non-empty: if $Q_{0q} = \emptyset, \forall q \in[\Qsf]$ and $Q_{u} \neq \emptyset,\forall u\in[2]$, then we transmit different coded 
packets from all the non-empty $Q_{u}$'s on all layers 
This corresponds to the  coded transmission from queues $\cup_{q\in[\Qsf]}  Q_{\{1,2\},q}$  of  Phase2 in Theorem~\ref{thm:Prototype}. 

\item All $Q_{0q}$'s are empty, and some $Q_{u} $'s are non-empty: if $Q_{0q} = \emptyset, \forall q \in[\Qsf]$ and $\exists u \neq \bar{u} : Q_{u} \neq \emptyset, Q_{\bar{u}} = \emptyset$, then we transmit different uncoded packets from the non-empty $Q_{u}$'s on all layers. 
This corresponds to the  coded transmission from queues $\cup_{q\in[\Qsf]}  Q_{\{1,2\},q}$  of  Phase2 in Theorem~\ref{thm:Prototype}. This happens when  the number of  packets intended to user $u$ only received by user $\bar{u}$ exceeds the  number of  packets intended to user $\bar{u}$ only received by user $u$, $u \neq \bar{u}$ after Phase1.
\end{itemize}

This protocol stated above is a natural adaptation of  Theorem~\ref{thm:Prototype}.
It is easy to  check that when all $\Qsf+2$ queues are empty, all the received packets can be decoded successfully. 
An achievable stability region is as follows.
\begin{theorem}[Achievable Stability Region (novel result)] 
For the LPE-BC with COF and $\Ksf = 2$ users, the following region is an inner bound to the stability region

\label{thm:achievable-stability-region}
\begin{align}
\mathcal{S}^{\text{\rm in}} &:= \big\{ (\lambda_1, \lambda_2)\in\mathbb{R}^{2}_{+}: 
\lambda_u := \sum_{q \in [\Qsf]}\lambda_{u,q}, 
\label{eq:achievable-stability}
\\&
\frac{\lambda_{1,q}+\lambda_{2,q}}{\Pr[ \max(N_1,N_2) \geq q]} < 1, 
\forall q\in[\Qsf], \notag
\\&
 \sum_{q\in[\Qsf]}\lambda_{u,q}  + \lambda_{\bar{u},q} \frac{\Pr[N_u \geq q]}{\Pr[ \max(N_1,N_2) \geq q]} < \mathbb{E}[N_u],  
\notag\\&
(u,\bar{u})\in[2]^2 : u \neq \bar{u}, \text{ for some $\lambda_{u,q}\geq0, u\in[2], q\in[\Qsf]$}
\big\}. \notag
\end{align}
\end{theorem}

The proof can be found in Appendix~\ref{sec:proof of thm:achievable-stability-region}.

Note that extension of $\mathcal{S}^{\text{\rm in}}$ to more than two users incurs the same problem as discussed for the inner bounds earlier.  
The region in~\eqref{eq:achievable-stability} recovers the result in~\cite{Sagduyu-unusual} when $\Qsf =1$.

\section{Optimality conditions} 
\label{subsec:optimality}
In this section, we focus on the two-layer and two-user case, compare the regions of the capacity outer bound obtained in Section~\ref{sec:outer-bound} and the stability inner bound presented in Section~\ref{subsec:stochastic}, and further state the sufficient conditions under which these two regions coincide. 
Let $\mathcal{C}^{\text{\rm in}}$ be the inner bound of Theorem~\ref{thm:Prototype}.  We have $\mathcal{C}^{\text{\rm in}} \subseteq \mathcal{C}\subseteq  \mathcal{C}^{\text{\rm out}}$ and $\mathcal{S}^{\text{\rm in}} \subseteq \mathcal{S}$ from Theorem~\ref{thm:achievable-stability-region}, where $\mathcal{S}$ is the stability region and $\mathcal{C}$ the capacity region. 
We also know~\cite{dynamic-power-allocation} that $\mathcal{S} \subseteq \mathcal{C}$. 
We first show that $\mathcal{C}^{\text{\rm in}}$ can be written in the same form as $\mathcal{S}^{\text{\rm in}}$, by replacing message rates with average arrival rates.
%
Next, we find conditions under which $\mathcal{S}^{\text{\rm in}} = \mathcal{C}^{\text{\rm out}}$,
 for $\Qsf = 2$ layers and $\Ksf = 2$ users, thus showing that under such conditions one has $\mathcal{S}=\mathcal{C}$. 
This result confirms the similarity between the capacity and stability regions already observed in~\cite{on-the-equivalence-of-shannon-capacity-and-stable-capacity, throughput-capacity-stability-regions, Sagduyu-unusual}. 

\begin{lemma}
 $\mathcal{S}^{\text{in}}$ in~\eqref{eq:achievable-stability} coincides with $\mathcal{C}^{\text{\rm in}}$ in~\eqref{eq:schemePrototype}.  
\end{lemma}
\begin{IEEEproof}
By~\eqref{eq:schemePrototype},  after some simple algebra, we obtain
\begin{align}
&t \geq t^{\text{(unc)}} + \max_{u\in[2]}(t^\text{(NC)}_u) \notag
\\
& \geq \max_{q \in [\Qsf]}\! \Bigg(\frac{k_{1,q} + k_{2,q}}{\Pr[ \max(N_1,N_2) \geq q]}\! \Bigg)\! + \!\max_{u\in[2]}\!\Bigg(\!\frac{1}{\mathbb{E}[N_u]}\Big[ \sum_{q \in [\Qsf]}k_{u,q} \notag
\\ 
& \left(\!1\!-\!\frac{\Pr[N_{u}\geq q]}{\Pr[\max(N_1,N_2) \geq q]}\!\right)\! \!- \! \!\Big(\!\max_{q \in [\Qsf]} (\!\frac{k_{1,q} + k_{2,q}}{\Pr[ \max(N_1,N_2) \geq q]}\! )\notag
\\
& - \frac{k_{1,q} + k_{2,q}}{\Pr[ \max(N_1,N_2) \geq q]}\Big) \Pr[N_{u}\geq q]\Big] ^+ \Bigg). 
 \label{total-time}
\end{align}

Assume layer $w \in[\Qsf]$ is the ``slowest" (i.e., $t^\text{(unc)} = t^\text{(unc)}_{w}$). Continuing from~\eqref{total-time}, we can write
 \vspace{-0.16cm}
 \begin{align}
& t \geq \! \frac{k_{1,w} + k_{2,w}}{\Pr[ \max(N_1,N_2) \geq w]}\! + \! \max_{u\in[2]}\!\Bigg(\!\frac{1}{\mathbb{E}[N_u]} \Big[ \sum_{q\in[\Qsf]}k_{u,q} \!
\\
& + \!\sum_{q\in [\Qsf]}\!   \notag k_{\bar{u},q} \frac{\Pr[N_u\geq q]}{\Pr[ \max(N_1,N_2) \geq w] }-  \mathbb{E}[N_u]\frac{k_{1,w} + k_{2,w}}{\Pr[ \max(N_1,N_2) \geq w]} \Big]^+\! \Bigg),
\notag\\
&(u,\bar{u})\in[2]^2 : u \neq \bar{u}
\end{align}

Let $R_{u,q} = \frac{k_{u,q}}{t}$ ,  the achievable region in~\eqref{eq:schemePrototype}  becomes
\begin{align}
&\Bigg \{ (R_1,R_2) \geq 0 : 
\max_{q\in[\Qsf]} \left( \frac{R_{1,q} + R_{2,q}}{\Pr[ \max(N_1,N_2) \geq q]}  \right) \leq 1,  
\notag 
\\
&\max_{u\in[2]} \left( \frac{R_{u}}{\mathbb{E}[N_u]} + \sum_{q \in [\Qsf]}\frac{\Pr[N_u\geq q]}{\Pr[ \max(N_1,N_2) \geq q]} \frac{R_{\bar{u},q}}{\mathbb{E}[N_u]} \right) \leq 1 \notag
\\
 &\hspace{0.5cm}  (u,\bar{u})\in[2]^2 : u \neq \bar{u}, \text{ for some } R_{u,q}\geq0 \Bigg \}, \notag
\\
&R_u := R_{u,1}+\ldots+R_{u,\Qsf}, \ u\in[2]. 
\end{align}
\label{eq:conclusion-innerbound}
As a result, 
the achievable region $\mathcal{C}^{\text{\rm in}}$ in~\eqref{eq:schemePrototype} is the same as $\mathcal{S}^{\text{in}}$ presented in~\eqref{eq:achievable-stability}. 
\end{IEEEproof}

Before  proceeding to ddetermine the sufficient optimality conditions,  
for convenience, we define
\begin{subequations}
\begin{align}
  & \xi_1 := \max_{q\in[2]}\left(
  \frac{\Pr[\max(N_1,N_2) \geq q]}{\Pr[N_1\geq q]}   
  \right), 
\\& \xi_2 := \min_{q\in[2]}\left(
    \frac{\Pr[ \max(N_1,N_2) \geq q]}{\Pr[N_1\geq q]}     
    \right), 
\\& \xi_3 := \max_{q\in[2]}\left(
    \frac{\Pr[N_2\geq q]}{\Pr[ \max(N_1,N_2) \geq q]}     
\right), 
\\& \xi_4 := \min_{q\in[2]}\left(
    \frac{\Pr[N_2\geq q]}{\Pr[\max(N_1,N_2) \geq q]}     
\right),
\end{align}
where clearly $\xi_1 \geq \xi_2 \geq 1 \geq \xi_3 \geq \xi_4 \geq0$.
\label{eq:def A B C D}
\end{subequations}
Rewrite the outer bound in~\eqref{eq:converse thm:COFnewOuter} as 
\begin{subequations}
\begin{align}
&R_{1} + \frac{R_{2}}{\xi_1} \leq  \mathbb{E}[N_1], \label{outer-bound.A}
\\
&\xi_2 R_{1} + R_{2} \leq \mathbb{E}[\max(N_1,N_2)], \label{outer-bound.B}
\\
&R_{1} + \frac{R_{2}}{\xi_3}  \leq  \mathbb{E}[\max(N_1,N_2) ], \label{outer-bound.C}
\\
&\xi_4 R_{1}  + R_{2}  \leq  \mathbb{E}[N_2]. \label{outer-bound.D}
\end{align}
\label{eq:outerbound}
\end{subequations}
Now we give the sufficient optimality conditions as follows.  
\begin{theorem}
\label{thm:optimality-conditions}
The stability region inner bound in Theorem~\ref{thm:achievable-stability-region} coincides with capacity region outer bound in Theorem~\ref{thm:COFnewOuter} for the LPE-BC with COF for the case of $\Ksf =2$ users and $\Qsf = 2$ layers when the following two conditions are verified:

(C1)  the joint channel statistic is as in Table~\ref{tab:newcondition} with $x_2\geq 2x_4$ and 

\begin{table}[H]
\caption{\small Joint PMF $\Pr[(N_1,N_2)=(i,j)]$ with $x_1 + 2x_2 + x_3 + x_4 = 1$.}
\label{tab:newcondition}
\begin{align*}
\begin{array}{| c | c |  c | c | c |}
\hline
&j = 0 &j = 1  &j = 2  \\
\hline
i = 0& x_1 & x_2 & 0  \\
i = 1& x_2 & x_3 & 0  \\
i = 2& 0 & 0 & x_4\\
\hline
\end{array} 
\end{align*}
\vspace*{-.4cm}
\end{table} 
(C2) either bound~\eqref{outer-bound.B} or bound~\eqref{outer-bound.C} is redundant, and 

(C3) either $\frac{\Pr[N_1\geq 2]}{\Pr[N_1\geq 1]} \geq \frac{\Pr[\max(N_1,N_2) \geq 2]}{\Pr[ \max(N_1,N_2) \geq 1]} \geq \frac{\Pr[N_2\geq 2]}{\Pr[N_2\geq 1]}$ or  $\frac{\Pr[N_2\geq 2]}{\Pr[N_2\geq 1]} \geq \frac{\Pr[ \max(N_1,N_2) \geq 2]}{\Pr[ \max(N_1,N_2) \geq 1]} \geq \frac{\Pr[N_1\geq 2]}{\Pr[N_1\geq 1]}$ hold.
\end{theorem}
The proof can be found in Appendix~\ref{sec:optimal-condition}.

{\it Intuition}:  
The condition (C1) in Theorem~\ref{thm:optimality-conditions} is inspired by~\cite[Theorem 3]{Geng-Quan}, which is different from the LPEBC model we use here, since the layers (subchannels) are independent in~\cite{Geng-Quan} but correlated in the LPEBC model.   
This class of channel in Table~\ref{tab:newcondition} may be interpreted as follows. User 1 and 2 have the same ability to receive packets on layer 1 and 2 separately,  that is $\Pr[N_1 \geq 1] = \Pr[N_2\geq 1]$ and $\Pr[N_1 \geq 2] = \Pr[N_2\geq 2]$; user 1 and 2 either receive or erase a packet on layer 2  at the same time, that is $\Pr[N_1\geq 2] = \Pr[N_2 \geq 2] = \Pr[\max(N_1,N_2) \geq 2]$.

Conditions (C2) and (C3) in Theorem~\ref{thm:optimality-conditions} may be interpreted as follows.
User $u\in[2]$ is more likely to receive a packet from layer~1 than user $\bar{u} \neq u$, 
while at the same time user $\bar{u}$ is more likely to receive a packet from layer~2 than user $u$.
It is fairly straightforward to see that when $\frac{\Pr[N_1\geq 2]}{\Pr[N_1\geq 1]} =\frac{\Pr[N_2\geq 2]}{\Pr[N_2\geq 1]} = \frac{\Pr[\max(N_1,N_2) \geq 2]}{\Pr[\max(N_1,N_2) \geq 1]}$, both~\eqref{outer-bound.B} and~\eqref{outer-bound.C}  are redundant and the outer bound 
becomes identical to the inner bound; under this condition we obtain the capacity region $\mathcal{C} = \{ (R_1, R_2)\in\mathbb{R}^2_+: \max\left(\frac{R_1}{\mathbb{E}[N_1]} + \frac{R_2}{\mathbb{E}[\max(N_1,N_2) ]}, \frac{R_2}{\mathbb{E}[\max(N_1,N_2) ]} + \frac{R_2}{\mathbb{E}[N_2]} \right) \leq 1\}$ that has the same form as the capacity region derived in~\cite{Sagduyu-unusual} for the single layer BEC-BC with COF; in other words, in this special case, the two-layer LPE-BC behaves as the one-layer BEC-BC where
$1-\varepsilon_{1} = \Pr[N_1\geq 1] + \Pr[N_1\geq2] = \mathbb{E}[N_1]$, 
$1-\varepsilon_{2} = \mathbb{E}[N_2]$, 
$1-\varepsilon_{12}= \mathbb{E}[\max(N_1,N_2)]$ 
correspond to the notation in~\cite{Sagduyu-unusual}.  

\section{Numerical Evaluations}
\label{sec:numerical}


\subsection{Example~1 (the channel of user 1 and 2 are independent)} 
\afterpage{
\begin{table}
\caption{\small Joint PMF $\Pr[(N_1,N_2)=(i,j)]$.}
\label{tab:example1}
\begin{align*}
\begin{array}{| c | c |  c | c | c |}
\hline
&j = 0 &j = 1  &j = 2 & \Pr[N_{1} = i] \\
\hline
i = 0& 0.125 & 0 & 0.125 & 0.250 \\
i = 1& 0.250 & 0 & 0.250 & 0.500 \\
i = 2& 0.125 & 0 & 0.125 & 0.250 \\
\hline
\Pr[N_{2} = j] & 0.500   &  0   &  0.500&  \\
\hline
\end{array} 
\end{align*}
\vspace*{-.4cm}
\end{table} 

\begin{figure}
  \centering
  \includegraphics[width=0.7\columnwidth]{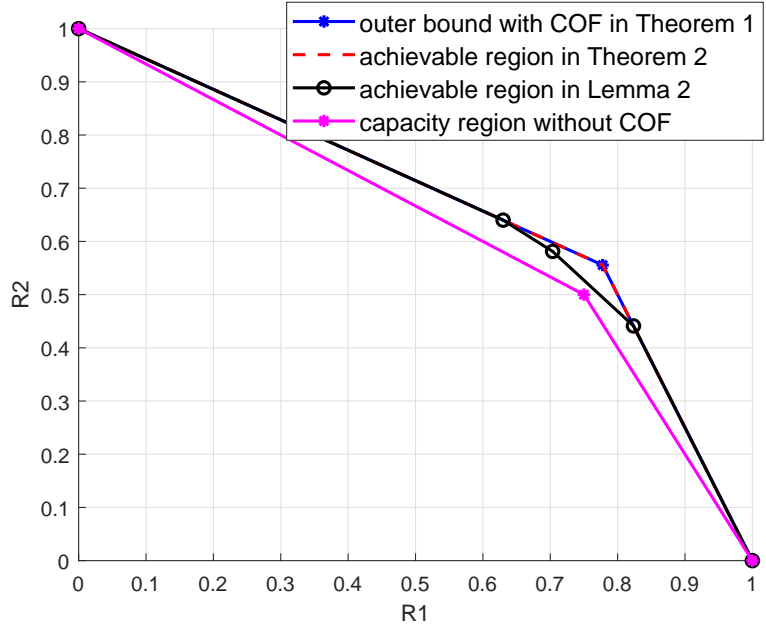}
  \caption{\small Outer and inner bounds for the channel in Table~\ref{tab:example1}.}
  \label{fig:example1}
\end{figure}
}
This example considers the case of $\Ksf=2$ users and $\Qsf=2$ layers, with $N_1$ independent of $N_2 $ 
and with the joint channel statistics as in Table~\ref{tab:example1}. 
Fig.~\ref{fig:example1} illustrates the capacity region in~\eqref{thm:Tse-Yates ITA-2011}, 
the outer bound region of Theorem~\ref{thm:COFnewOuter}, and 
the inner bound regions of Lemma~\ref{thm:scheme1},
and Theorem~\ref{thm:Prototype}. 
Without CSIT, the capacity region in~\eqref{thm:Tse-Yates ITA-2011} has three corner points $(R_1,R_2)\in\{ 
(0,1),
(\frac{3}{4},\frac{1}{2}),
(1,0)\},
$
where $1=\mathbb{E}[N_1]=\mathbb{E}[N_2].$
The corner point $(\frac{3}{4},\frac{1}{2})$ is achieved by assigning layer~1 to user~1 and layer~2 to user~2~\cite{FadingBC}.
With COF, it can be shown analytically that the outer bound in Theorem~\ref{thm:COFnewOuter} has three corner points $(R_1,R_2)\in\{ 
(0,1),
(\frac{7}{9},\frac{5}{9}),
(1,0)\},
$
and that Lemma~\ref{thm:scheme1} does not achieve the corner point $(\frac{7}{9},\frac{5}{9})$, while 
Theorem~\ref{thm:Prototype}
does (with $R_1 = R_{1,1}$ and $R_2 = R_{2,2}$ in region $\mathcal{C}^{\text{\rm in}}$). 
This is an example where our inner and outer bounds match. 
Notice that COF enlarges the capacity region for this example.

\subsection{Example~2 (the channel of user 1 and 2 are correlated)}

\afterpage{
\begin{table}
\caption{\small Joint PMF $\Pr[(N_1,N_2)=(i,j)]$.}
\label{tab:threecorners}
\begin{align*}
\begin{array}{| c | c |  c | c | c |}
\hline
&j = 0 &j = 1  &j = 2 & \Pr[N_{1} = i] \\
\hline
i = 0& 0.0497 & 0.2443 & 0.0321 & 0.3261 \\
i = 1& 0.1483 & 0.2251 & 0.1222 & 0.4956 \\
i = 2& 0.0435 & 0.0728 & 0.0620 & 0.1783 \\
\hline
\Pr[N_{2} = j] & 0.2415    &  0.5422   &   0.2163 &  \\
\hline
\end{array} 
\end{align*}
\vspace*{-.4cm}
\end{table} 

\begin{figure}
  \centering
  \includegraphics[width=0.7\columnwidth]{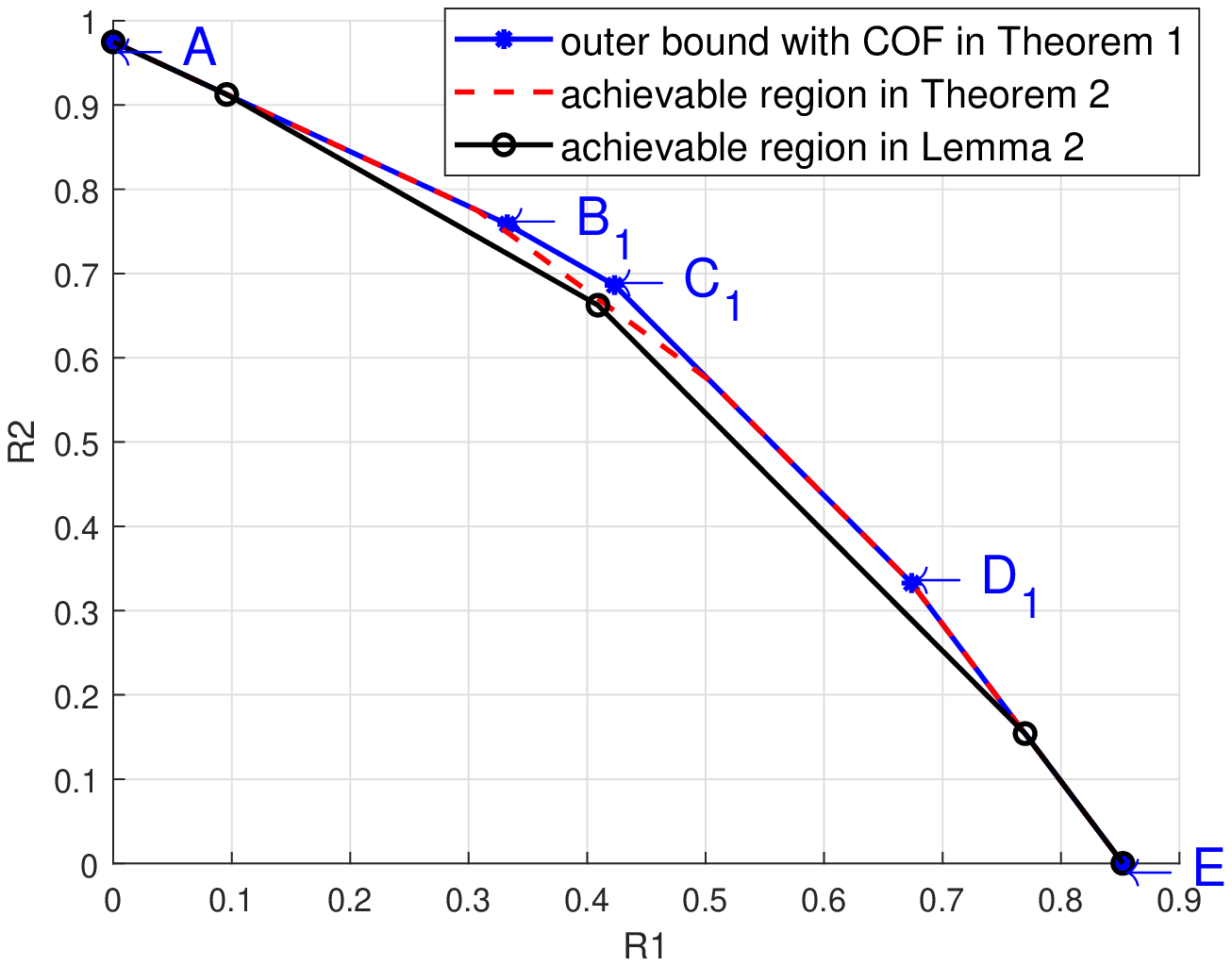}
  \caption{\small Outer and inner bounds for the channel in Table~\ref{tab:threecorners}.}
  \label{fig:example2}
\end{figure}
}

This example illustrates the case of $\Ksf=2$ users and $\Qsf=2$ layers, with $N_1$ correlated with $N_2 $. 
The inner and outer bound regions for the channel described in Table~\ref{tab:threecorners} are evaluated in Fig.~\ref{fig:example2}, in which both users have a more reliable look at layer~1 than at layer~2.

The outer bound in Theorem~\ref{thm:COFnewOuter} is the convex-hull of the following rate points: 
$A  =(0,0.9748)$,
$B_1=(0.3326,0.7585)$, 
$C_1=(0.4231,0.6862)$,
$D_1=(0.6739,0.3326)$, 
$E  =(0.8522,0)$.
Corner points $A$ and $E$ are always trivially achievable, so we will not list them in the following.
The achievable region in Lemma~\ref{thm:scheme1} has non-trivial corner points:
$B_2=(0.0957,0.9125)$,
$C_2=(0.4091,0.6624)$,
$D_2=(0.7697,0.1540)$.
%
%
%
The achievable region in Theorem~\ref{thm:Prototype} has non-trivial corner points:
$B_3=(0.3069,0.7752)$,
$C_3=(0.5035,0.5729)$,
$D_3=(0.6739,0.3326)$.
%
Note that Theorem~\ref{thm:Prototype} achieves one of the corner points ($D_1$) of the outer bound in Theorem~\ref{thm:COFnewOuter}.

An interesting observation from the numerical optimization for this example is that at the corner points either $k_{1,q}=0$ or $k_{2,q}=0$ in the various achievable regions across layers (i.e., a layer is assigned to one user only, as it was the case in Example~1), with the only exception of the C-points; for the C-points, the `more reliable'  layer~1 is shared by both users. We also remark from Fig.~\ref{fig:example2} that the inner and outer bounds are the furthest apart around the C-points. Why this is the case is subject of current investigation.
 In general, inner and out bounds coincide when one of the two rates is not too large, i.e., around the trivially achievable corner points  $A$ and $E$  which are the equivalent rates of point-to-point channels.

\subsection{Example~3 (the channel statistics satisfy the conditions in Theorem~\ref{thm:achievable-stability-region})}



This example demonstrates that the achievable stability region in Theorem~\ref{thm:achievable-stability-region} coincides with the outer bound of capacity region in Theorem~\ref{thm:COFnewOuter}, i.e., the conditions in Theorem~\ref{thm:optimality-conditions} are satisfied.

Consider the channel in Table~\ref{tab:twocorners}, in which both users have a more reliable look at layer~2 than at layer~1; here the channel states are correlated. 
The outer bound in Theorem~\ref{thm:COFnewOuter} is the convex-hull of the following rate pairs: 
$P_1=(0,1.234)$,
$P_2=(0.302,1.035)$,
$P_3=(0.366,0.912)$, 
$P_4=(0.836,$ $0)$. 

If all four bounds in~\eqref{eq:outerbound} were active,  the outer bound would be a convex hull of at most 6 corner points (including the point (0,0), two corner points on the $R_1$ and $R_2$ axes, and 3 other non-trivial corner points). Here, we only have two non-trivial corner points, points $P_2, P_3$. We know the bound in~\eqref{outer-bound.A} and the one in~\eqref{outer-bound.D} are always active. Hence, either the bound in~\eqref{outer-bound.B} or the one in~\eqref{outer-bound.C} is redundant. 
Here is a case where either~\eqref{outer-bound.B}  or~\eqref{outer-bound.C} is redundant. 
For this channel, 
$\frac{\Pr[\max(N_1,N_2) \geq 1]}{\Pr[N_1\geq 1]} > \frac{\Pr[\max(N_1,N_2) \geq 2]}{\Pr[N_1\geq 2]} $,
$\frac{\Pr[N_2\geq 1]}{\Pr[ \max(N_1,N_2) \geq 1]} > \frac{\Pr[N_2\geq 2]}{\Pr[ \max(N_1,N_2) \geq 2]}$
and
$\xi_1 = 1.940$, 
$\xi_2 = 1.926$,
$\xi_3 = 0.844$,
$\xi_4 = 0.658$ in~\eqref{eq:def A B C D}. 
This is an example where the erasures are correlated and for which we obtain the optimal capacity region which coincides with the stability region.
\afterpage{
\begin{table}
\caption{\small Joint PMF $\Pr[(N_1,N_2)=(i,j)]$.}
\label{tab:twocorners}
\vspace*{-.2cm}
\begin{align*}
\begin{array}{| c | c |  c | c | c |}
\hline
&j = 0 &j = 1  &j = 2 & \Pr[N_{1} = i] \\
\hline
i = 0& 0.088 & 0.178 & 0.264 & 0.530 \\
i = 1& 0.011 & 0.018 & 0.075 & 0.104 \\
i = 2& 0.131 & 0.110 & 0.125 & 0.366 \\
\hline
\Pr[N_{2} = j] & 0.230    &  0.306   &   0.464 &  \\
\hline
\end{array} 
\end{align*}
\vspace*{-.2cm}
\end{table}

\begin{figure}
\vspace{-0.4cm}
  \centering
  \includegraphics[width=0.7\columnwidth]{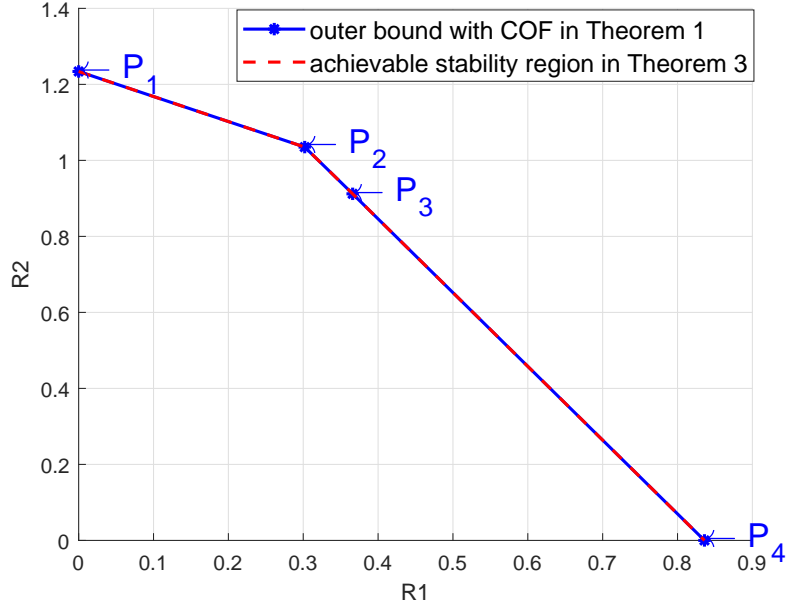}
  \caption{\small Capacity and stability region for the channel in Table~\ref{tab:twocorners}.}
  \label{fig:example}
\vspace*{-.4cm}
\end{figure}
}

\section{Conclusions}
\label{sec:concl}
This paper derived achievable regions for the LPE-BC with COF both in information-theoretic terms (i.e., capacity region) and queueing-theoretic terms (i.e., stability region).
The studied LPE-BC extends the classical (single-layer) binary erasure BC and can be connected to the Gaussian fading BC at high SNR. Our capacity inner bound and achievable stability region make use of network coded retransmissions when the sender, through COF, realizes that a packet has been received only by unintended users.  What this work shows is the necessity of network coding across users (a key element also for the single-layer binary erasure BC with COF) and across layers. Conditions under which the obtained stability region inner bound coincides with the capacity region outer bound are given, thus establishing optimality. 
Future work includes determining a set of conditions under which the proposed scheme is optimal, extending the analysis to more than two users, and ultimately deriving constant gap approximations to the capacity of the fading AWGN-BC without CSIT but with COF.


\begin{appendices}

\section{Proof of Lemma~\ref{lemma-mechanism3}}
\label{sec: proof of lemma1}
We focus on proving a statement that is slightly stronger than Lemma~\ref{lemma-mechanism3}:  At any time $t\in[T_{\pi(1)}^{\text{(unc)}}, T^{\text{(unc)}}+ T^{\text{(NC)}}]$, 
 \begin{align}
\Pr[\mathcal{K}(t)] > 1-\varepsilon, \  \forall \varepsilon \in (0, 1). 
\label{lemma:strong}
  \end{align}
  where $\mathcal{K}(t)$ denotes the event that ``the encoding vectors of received $\sum_{q\in[\Qsf]}\sum_{u\in[2]}K_{u,q}^{\text{(NC)}}(t) = \sum_{q\in[\Qsf]}\sum_{u\in[2]}$ $ k_{u,q}^{\text{(NC)}}(t)$ packets are linearly independent''.
We prove the above statement by induction on time $t$. At the end of time $t = T_{\pi(1)}^{\text{(unc)}}$, since no coded packet is transmitted previously and we initialize $K_{1,q}^{\text{(NC)}}(0) = 0 $ and $K_{2,q}^{\text{(NC)}}(0)=0, \forall q\in[\Qsf]$, we have 
$ \Pr[\mathcal{K}(T_{\pi(1)}^{\text{(unc)}}) ] =1.$
The condition in~\eqref{lemma:strong} is thus satisfied.
  Consider now time $t \in (T_{\pi(1)}^{\text{(unc)}}, T^{\text{(unc)}}+ T^{\text{(NC)}} ).$ By induction assumption, 
 \begin{align}
\Pr[\mathcal{K}(t) ] > 1-\varepsilon,  \  \forall \varepsilon \in (0, 1). 
 \label{assumption}
    \end{align}
    
    By our Theorem~\ref{thm:Prototype}, for any time $t$, we always have 
 \begin{align} 
 \sum_{q\in[\Qsf]}\left| Q_{\{1,2\},q}(t) \right| = \sum_{u\in[2]}  \left( K_{u}^{\text{(rem)}}(t) + \sum_{q\in[\Qsf]} K_{u,q}^{\text{(NC)}}(t) \right).
 \label{relationship}
     \end{align}
\begin{subequations}     
At time $t$, denote  the realization of the RVs as follows
\begin{align}
k_{\{1,2\}}(t) &= \sum_{q\in[\Qsf]}\left| Q_{\{1,2\},q}(t) \right|, \label{lemma-term1}
\\
k^{\text{(rem)}}(t) & =  \sum_{u\in[2]} K_{u}^{\text{(rem)}}(t), \label{lemma-term2}
\\ 
k^{\text{(NC)}}(t)& = \sum_{u\in[2]} \sum_{q\in[\Qsf]}K_{u,q}^{\text{(NC)}}(t), \label{lemma-term3}
\end{align} 
\label{three-terms}
\end{subequations}   
which  are always non-negative.
Rewrite~\eqref{relationship} for slot $t+1$ as 
 \begin{align} 
k_{\{1,2\}}(t+1)  = k^{\text{(rem)}}(t+1) + k^{\text{(NC)}}(t+1).
 \label{relation-t+1}
 \end{align} 
Based on the values of the three terms in~\eqref{three-terms},  at the end of slots $t+1$ and  $t$, we partition them into 
\begin{align}
k_{\{1,2\}}(t+1) 
& \begin{cases}
= k_{\{1,2\}}(t), 
 \\
> k_{\{1,2\}}(t); 
\end{cases}
\notag \\
k^{\text{(rem)}}(t+1) 
& \begin{cases}
 >  k^{\text{(rem)}}(t) ,
 \\
 =  k^{\text{(rem)}}(t),
 \\
 < k^{\text{(rem)}}(t);  
 \end{cases}
\notag \\
 k^{\text{(NC)}}(t+1)
& \begin{cases}
   = k^{\text{(NC)}}(t),
\\
 > k^{\text{(NC)}}(t). 
 \notag
\end{cases}
\end{align}
 Thus, we have $2\times3\times2 = 12$ combinations in total.
However, some of the combinations  are impossible and some of the combinations can be merged together. We list them as follows. 
\begin{subequations}
 \paragraph*{Case 1}  
  $k^{\text{(NC)}}(t+1) = k^{\text{(NC)}}(t)$. 
  This implies that no new coded packets are received by the users, 
 and the received $ k^{\text{(NC)}}(t+1)$ packets are linearly independent by the induction hypothesis in~\eqref{assumption}. Therefore, 
  \begin{align*}
\Pr\left[\mathcal{K}(t+1) \big| k^{\text{(NC)}}(t+1) = k^{\text{(NC)}}(t),  \mathcal{K}(t) \right] = 1.
     \end{align*}

  \paragraph*{Case 2}  $k^{\text{(NC)}}(t+1) > k^{\text{(NC)}}(t)$. This implies that there are $k^{\text{(NC)}}(t+1) - k^{\text{(NC)}}(t)>0$ coded packets received by the two users in slot $t+1$.
    We have two sub-cases depending on the values of $| Q_{\{1,2\}}(t+1)|$ and  $| Q_{\{1,2\}}(t)|$.
      \paragraph*{Case 2.1}     $ k_{\{1,2\}}(t+1)=  k_{\{1,2\}}(t)$. This implies that no more overheard packets are added to the queues $Q_{\{1,2\},q}, \forall q\in[\Qsf]$ in slot $t+1$. This may happen in some sub-phases of Phase1 when the uncoded packets are successfully received by the destined user or the uncoded packets are erased at both users; this may also happen in Phase2.
        We have three sub-cases depending on the realizations of  $K^{\text{(rem)}}(t+1)$ and $K^{\text{(rem)}}(t)$. 
   	  \paragraph*{Case 2.1.1}  $k^{\text{(rem)}}(t+1) > k^{\text{(rem)}}(t)$. This implies that the number of packets added to $Q_{\{1,2\},q}$ is larger than the number of packets received by the users in slot $t+1$.  
	 Since  $ k_{\{1,2\}}(t+1)=  k_{\{1,2\}}(t)$  and according to~\eqref{relation-t+1}, here we have
	$  k^{\text{(rem)}}(t+1) + k^{\text{(NC)}}(t+1) = k^{\text{(rem)}}(t) + k^{\text{(NC)}}(t).$ 
	This is violated by the conditions of $k^{\text{(NC)}}(t+1) > k^{\text{(NC)}}(t)$ and $k_{\{1,2\}}(t+1) > k_{\{1,2\}}(t)$, and hence          leads to a contradiction and hence
	\begin{align*}
         \Pr\left[k^{\text{(NC)}}(t+1) > k^{\text{(NC)}}(t), k_{\{1,2\}}(t+1)=  k_{\{1,2\}}(t), k^{\text{(rem)}}(t+1) > k^{\text{(rem)}}(t)\right] =0.
          \end{align*}
  \paragraph*{Case 2.1.2}   $k^{\text{(rem)}}(t+1) = k^{\text{(rem)}}(t)$. This implies that  the number of packets added to $Q_{\{1,2\},q}$ is equal to the number of packets received by the users in slot $t+1$. 
 This case is again impossible, which follows similarly to Case 2.1.1. Thus,
  \begin{align*}
         \Pr\left[k^{\text{(NC)}}(t+1) > k^{\text{(NC)}}(t), k_{\{1,2\}}(t+1)=  k_{\{1,2\}}(t), k^{\text{(rem)}}(t+1) = k^{\text{(rem)}}(t)\right] =0.
          \end{align*}

  \paragraph*{Case 2.1.3}   $0 \leq k^{\text{(rem)}}(t+1) < k^{\text{(rem)}}(t)$. This implies that  the number of packets added to $Q_{\{1,2\},q}$ is less than the number of packets received by the users in slot $t+1$. 
By the induction hypothesis, we have  
 \begin{align} 
& \Pr\left[\mathcal{K}(t+1) \Big| k^{\text{(NC)}}(t+1) > k^{\text{(NC)}}(t), k_{\{1,2\}}(t+1)=  k_{\{1,2\}}(t), k^{\text{(rem)}}(t+1) <k^{\text{(rem)}}(t), \mathcal{K}(t)  \right]  \notag
\\& = \prod_{i=k^{\text{(NC)}}(t)}^{ k^{\text{(NC)}}(t+1) -1} \left( 1- \frac{|\mathcal{X}|^i}{|\mathcal{X}|^{k_{\{1,2\}}(t+1)}} \right) \notag
\\& =  \prod_{i=1}^{ k^{\text{(NC)}}(t+1) -  k^{\text{(NC)}}(t) } \left( 1- \frac{1}{|\mathcal{X}|^{i+k^{\text{(rem)}}(t+1)} } \right)
\label{change-variable-case1}
\\&\geq \left( 1- \frac{1}{|\mathcal{X}| } \right)^{\Delta k^{\text{(NC)}}(t+1)}
\label{lower-bound-case1}
     \end{align}
  where~\eqref{change-variable-case1} follows from~\eqref{relation-t+1}; 
  where in~\eqref{lower-bound-case1}, 
   $  \Delta k^{\text{(NC)}}(t+1) = k^{\text{(NC)}}(t+1) - k^{\text{(NC)}}(t) >0$,
    which is the number of coded packets received by both users in slot $t+1$. Since we assume $\mathcal{X}$ is a finite field, $|\mathcal{X}|$ must be a prime power.       
   \paragraph*{Case 2.2} $k_{\{1,2\}}(t+1) > k_{\{1,2\}}(t)$. This implies that there are new overheard packets  added to $Q_{\{1,2\},q}, \forall q\in[\Qsf]$ in slot $t+1$. 
Similar to Case 2.1, we have three sub-cases depending on the realizations of  $K^{\text{(rem)}}(t+1)$ and $K^{\text{(rem)}}(t)$. 
 \paragraph*{Case 2.2.1}  $k^{\text{(rem)}}(t+1) > k^{\text{(rem)}}(t) \geq 0$.  We have  
 \begin{align} 
& \Pr\left[\mathcal{K}(t+1) \Big| k^{\text{(NC)}}(t+1) > k^{\text{(NC)}}(t), k_{\{1,2\}}(t+1) > k_{\{1,2\}}(t), k^{\text{(rem)}}(t+1) > k^{\text{(rem)}}(t), \mathcal{K}(t)  \right]  \notag
\\& =  \prod_{i=1}^{ k^{\text{(NC)}}(t+1) - k^{\text{(NC)}}(t)  } \left( 1- \frac{1}{|\mathcal{X}|^{k^{\text{(rem)}}(t+1) +i} } \right) \notag
\\& \geq  \left( 1- \frac{1}{|\mathcal{X}|^2 } \right)^{\Delta k^{\text{(NC)}}(t+1)}  \notag
     \end{align} 
      \paragraph*{Case 2.2.2}  $k^{\text{(rem)}}(t+1) = k^{\text{(rem)}}(t) \geq 0$.  We have 
       \begin{align} 
& \Pr\left[\mathcal{K}(t+1) \Big| k^{\text{(NC)}}(t+1) > k^{\text{(NC)}}(t), k_{\{1,2\}}(t+1) > k_{\{1,2\}}(t), k^{\text{(rem)}}(t+1) = k^{\text{(rem)}}(t), \mathcal{K}(t)  \right]  \notag
\\& =  \prod_{i=1}^{ k^{\text{(NC)}}(t+1) - k^{\text{(NC)}}(t)  } \left( 1- \frac{1}{|\mathcal{X}|^{k^{\text{(rem)}}(t+1) +i} } \right) \notag
\\& \geq  \left( 1- \frac{1}{|\mathcal{X}| } \right)^{\Delta k^{\text{(NC)}}(t+1)} \notag
     \end{align} 
 \paragraph*{Case 2.2.3}  $0 \leq k^{\text{(rem)}}(t+1) < k^{\text{(rem)}}(t) $.  We have 
       \begin{align} 
& \Pr\left[\mathcal{K}(t+1) \Big| k^{\text{(NC)}}(t+1) > k^{\text{(NC)}}(t), k_{\{1,2\}}(t+1) > k_{\{1,2\}}(t), k^{\text{(rem)}}(t+1) <k^{\text{(rem)}}(t), \mathcal{K}(t)  \right]  \notag
\\& =  \prod_{i=1}^{ k^{\text{(NC)}}(t+1) - k^{\text{(NC)}}(t)  } \left( 1- \frac{1}{|\mathcal{X}|^{k^{\text{(rem)}}(t+1) +i} } \right) \notag
\\& \geq  \left( 1- \frac{1}{|\mathcal{X}| } \right)^{\Delta k^{\text{(NC)}}(t+1)} \notag
     \end{align} 

    \paragraph*{Combining all cases} Based on the discussion of Cases 1 to 2.2.3,  for any $k^{\text{(NC)}}(t+1), k^{\text{(NC)}}(t), k_{\{1,2\}}(t+1), k_{\{1,2\}}(t), k^{\text{(rem)}}(t+1), k^{\text{(rem)}}(t)$, we have the following inequality 
      \begin{align} 
     & \Pr\left[\mathcal{K}(t+1) \Big| k^{\text{(NC)}}(t+1), k^{\text{(NC)}}(t), k_{\{1,2\}}(t+1), k_{\{1,2\}}(t), k^{\text{(rem)}}(t+1), k^{\text{(rem)}}(t), \mathcal{K}(t)\right]  \label{lemma-probability}
      \\& \geq  \left( 1- \frac{1}{|\mathcal{X}|} \right)^{\Delta k^{\text{(NC)}}(t+1)}. \notag
          \end{align}   
By considering all the conditional probabilities, we thus have
        \begin{align}  
      \Pr[\mathcal{K}(t+1)] &\geq \prod_{i=T_{\pi(1)}^{\text{(unc)}}}^{t}  \bigcup_{ k^{\text{(NC)}}(t+1), k^{\text{(NC)}}(t), k_{\{1,2\}}(t+1), k_{\{1,2\}}(t), k^{\text{(rem)}}(t+1), k^{\text{(rem)}}(t) }  \notag
      \\& \Big(  ~\eqref{lemma-probability} \times \Pr\left[k^{\text{(NC)}}(t+1), k^{\text{(NC)}}(t), k_{\{1,2\}}(t+1), k_{\{1,2\}}(t), k^{\text{(rem)}}(t+1), k^{\text{(rem)}}(t) \right]  \Big) \notag
      \\& \geq  \left( 1- \frac{1}{|\mathcal{X}| } \right)^{\sum_{i = T_{\pi(1)}^{\text{(unc)}}}^{t}\Delta k^{\text{(NC)}}(i+1)} \notag
      \\& = \left( 1- \frac{1}{|\mathcal{X}| } \right)^{ k^{\text{(NC)}}(t+1)} \notag
      \\& \geq  \left( 1- \frac{1}{|\mathcal{X}| } \right)^{k_1+k_2}.
      \label{lemma-combine}      
               \end{align}   
In conclusion, for any fixed $k_1$ and $k_2$, we can choose a sufficiently large input alphabet $\mathcal{X}$ such that~\eqref{lemma-combine} approaches one. \eqref{lemma:strong} thus holds for all $t\in[T_{\pi(1)}^{\text{(unc)}}, T^{\text{(unc)}}+ T^{\text{(NC)}}]$.
\end{subequations}

\section{A new proof for the single-layer case in~\cite{Sagduyu-unusual}}
\label{sec:proof of time concentration}

\subsection{Aim}
The authors of~\cite{Sagduyu-unusual} proved the optimality of Theorem~\ref{thm:Prototype} (i.e., achieving the region in Theorem~\ref{thm:COFnewOuter}) for the case of a single layer and two receivers by using an analysis based on Markov chains. Our attempt to extend this approach to the multi-layer of multi-receiver case failed due to the complexity of the state space of the Markov chain. Here we give an alternative proof of the result in~\cite{Sagduyu-unusual}, which we shall then extend to the case of any number of layers and two receivers. The extension to multiple receivers follows by the same reasoning.

We seek a concentration result of the following form. Let $k_{u} := \lfloor n R_u \rfloor$, for fixed $R_u > 0$, be the total number of packets that must be delivered to user $u\in[K]$. Let $T$ be the random variable that represents the time needed to deliver all these packets according to Theorem~\ref{thm:Prototype}. We aim to show convergence in probability of the random variable (RV) $T$ to its mean, 
namely
\begin{align}
\lim_{n \to \infty} \Pr\left[|T-\mathbb{E}[T]| > \varepsilon\right] = 0,  \ \forall \varepsilon>0. 
\label{eq:time-goal}
\end{align}
Since the total time $T$ is the summation of the time needed for Phase1 (indicated as $T_{\rm p1}$ next, corresponds to $T^{(\text{unc})}$ used earlier) and Phase2 (indicated as $T_{\rm p2}$ next, corresponds to $T^{(\text{NC})}$ used earlier), the idea is to show that the RVs $T_{\rm p1}$ and $T_{\rm p2}$ are independent, and that each concentrates to its mean with probability 1 
for large enough $n$. Actually, we aim to show that a convergence is exponentially fast in $n$. 
In Appendix~\ref{App-A-phase1}, we show that $T_{\rm p1}$ concentrates to its expectation by using the Chernoff bound for geometric RVs.
In order to show the concentration of $T_{\rm p2}$ to its mean, we also need to determine the behavior of two other RVs: the numbers of packets $B_1$ and $B_2$ to be received at user~1 and~2, respectively, in Phase2. 
In Appendix~\ref{App-A-Bu}, we aim to show that both  $B_1$ and $B_2$ converge to their respective means exponentially fast in $n$. We shall show that the RVs $(B_1, B_2)$ are mutually independent, where independence follows because $B_1$ and $B_2$ are defined over disjoint time intervals during Phase1 and because the channel is memoryless. 
In Appendix~\ref{App-A-T2},
in order to show the RV $T_{\rm p2}$ has a sharp concentration to its expectation, we  create an event  which contains all undesirable events related to this RV. We shall refer to such events as ``outages.'' 
Based on the Chernoff and/or Hoeffding bounds, we show that the probability of outage vanishes exponentially fast in some parameters. This is inspired by the asymptotic equipartition property or method of types in information theory, namely, for an event $\Ec$ and an outage event $\Oc$, we seek bounds of the form
\begin{align*}
\Pr[\Ec]
  &= \Pr[\Ec|\Oc]\Pr[\Oc]+\Pr[\Ec|\Oc^c]\Pr[\Oc^c]
\\&\leq \Pr[\Oc]+\Pr[\Ec|\Oc^c],
\end{align*}
where the event $\Ec$ is either the right or the left tail of a RV of interest. Given the complement of the outage event, we can bound each of  the range of the RVs in the event $\mathcal{E}$. 
By appropriately combining all these pieces, we will obtain our desired result in~\eqref{eq:time-goal}. 
To simplify notations, in the rest of this paper, we use $M$ to denote $\max(N_{1},N_{2})$ and $M_t$ to represent the value of $M$ at time $t$. 

\subsection{Duration of Phase1}
\label{App-A-phase1}
In Phase1 the transmitter keeps transmitting $k_1$ and $k_2$ packets destined to receivers 1 and 2, respectively,  until at least one of the receivers receives it;
the transmission time needed to complete Phase1 is the RV 
\begin{align}
T_{\rm p1} :=  \sum_{i=1}^{k_{1}+k_{2}} G_{0,i}, 
\label{eq:COF Q=1 T1}
\end{align}
where the $G_{0,i}$ are i.i.d. geometric RVs with parameter $\left(1-\epsilon_{12}\right)$ where $\epsilon_{12} := \Pr[ M =0]$ for $i\in[k_{1}+k_{2}]$ and  
the mean of $T_{\rm p1}$ is
$\mathbb{E}[T_{\rm p1}] = \frac{k_1+k_2}{1-\epsilon_{12}}.$

\paragraph*{Concentration Result for $T_{\rm p1}$}
\begin{subequations}
By~\cite[Theorem 10.32]{GeometricChernoff}, for every $\varepsilon_1 >0$, 
\begin{align}
&\Pr \left[ T_{\rm p1}-\mathbb{E}[T_{\rm p1}] > \varepsilon_1 \right] 
\leq   \exp \Big( -\frac{\varepsilon_1^{2}}{2\mathbb{E}[T_{\rm p1}]^2} \frac{k_1+k_2-1}{1+ \varepsilon_1/\mathbb{E}[T_{\rm p1}]} \Big), \label{single-Tp1-right}
\\&
\Pr \left[ T_{\rm p1}-\mathbb{E}[T_{\rm p1}] < -\varepsilon_1 \right] 
\leq   \exp \Big( -\frac{2(k_1+k_2)^2\varepsilon_1^{2}}{\mathbb{E}[T_{\rm p1}]^3}\Big). \label{single-Tp1-left}
\end{align} 
\label{single-Tp1}
The bounds in~\eqref{single-Tp1}  imply that $T_{\rm p1}$ converges to its expectation $\mathbb{E}[T_{\rm p1}]$ with probability 1 as $k_1+k_2\to \infty$. 

\end{subequations}
\subsection{Number of packets not delivered by the end of Phase1}
\label{App-A-Bu}
At the end of Phase1, there are $B_1$ packets destined to user~1 that were received at user~2 but not at user~1, where
\begin{align*}
   &B_{1}\sim\text{Binomial}\left(k_{1}, p_1\right), 
\\ &\mathbb{E}[B_1] = k_1p_1, 
\\ &p_1 =  \frac{\Pr[N_1=0,N_2 =1]}{1- \epsilon_{12}}, 
\\ &\Pr[N_1=0,N_2 =1] = \epsilon_1 - \epsilon_{12}.
\end{align*}
Similarly, there are $B_2$ packets destined to user~2 that were received at user~1 but not at user~2, where
\begin{align*}
   &B_{2}\sim\text{Binomial}\left(k_{2},p_2 \right),  
\\ &\mathbb{E}[B_2] = k_2p_2, 
\\ &p_2 =\frac{\Pr[N_{1}=1,N_{2}=0]}{1-\epsilon_{12}},
\\ &\Pr[N_{1}=1,N_{2}=0] = \epsilon_{2}-\epsilon_{12}.
\end{align*}

Now, since in Phase1 packets are sent uncoded to the users on different slots, 
we have immediately that $(B_1,B_2)$ are independent.
Next we show that $(B_1,B_2)$ is independent of $T_{\rm p1}$.
This follows because, for any $k\in\mathbb{N}$ 
\begin{align*}
  &\Pr[\text{`packet received at user~2 but not at user~1'}|G_{0}=k] 
\\&= \frac{\Pr[M_1=0,\ldots,M_{k-1}=0,N_{1,k}=0,N_{2,k}=1]}{\Pr[M_1=0,\ldots,M_{k-1}=0,M_k\not=0]}
\\&= \frac{\Pr[N_{1,k}=0,N_{2,k}=1]}{\Pr[M_k\not=0]} 
\\&= \frac{\epsilon_{1}-\epsilon_{12}}{1-\epsilon_{12}}, 
\\  &\epsilon_{1} :=\Pr[N_{1}=0].
\end{align*}
Similarly, $B_2$ is independent of $T_{\rm p1}$ because
\begin{align*}
  &\Pr[\text{`packet received at user~1 but not at user~2'}|G_{0}=k] 
\\&= \frac{\epsilon_{2}-\epsilon_{12}}{1-\epsilon_{12}}, 
\\&\epsilon_{2} :=\Pr[N_{2}=0].
\end{align*}
This shows that the RVs $B_1,B_2,T_{\rm p1}$ are mutually independent.  

\paragraph*{Concentration Result for $B_1$ and $B_2$} We next show that both $B_1$ and $B_2$ concentrate to their means. 
In order to do so, we use bounding ideas routinely used in the method-of-type type of proofs in information theory. 
Define the outage event for $\Ksf=2$, 
$ \mathcal{A} := \left\{ \exists u\in[\Ksf] : \left|B_{u} -\mathbb{E}[B_{u}] \right| >\varepsilon \right\}. $ 
By the Chernoff bound, for every $\varepsilon >0$, we have
\begin{align*}
&\Pr \left[ |B_{u}-\mathbb{E}[B_{u}]|  > \varepsilon \right] \leq
   2 \exp(-\frac{\varepsilon^{2}}{3\mathbb{E}[B_{u}]}), u\in[2].
\end{align*}
By the union of events bound, for $\varepsilon = o\left( (\max_{u\in[2]}\mathbb{E}[B_u])^{\frac{1}{2}+\epsilon} \right)$, we obtain that
\begin{align}
 \Pr[\mathcal{A}] \leq 4 \exp\left(-\frac{(\max_{u\in[2]}\mathbb{E}[B_u])^{2\epsilon}}{3} \right),
\label{bound-on-Ak1k2}
\end{align}
which implies that $B_u$ concentrates to its expectation $\mathbb{E}[B_u] = k_u p_u$ when $k_u$ is large enough, $u\in[2]$.
\subsection{Duration of Phase2}
\label{App-A-T2}
In Phase2 the transmitter keeps sending linear combinations of the overheard $B_1+B_2$ packets until 
each user successfully decodes its intended packets. Since Phase2 occurs over a disjoint time interval compared to Phase1, we immediately have that $T_{\rm p2}$ is independent of $T_{\rm p1}$, where by the working of our protocol we have
\begin{align}
 T_{\rm p2} &:=
 \min\left(t : \sum_{j=1}^{t} N_{u,j} \geq B_{u}, \forall u\in[\Ksf]\right), 
\label{eq:T_p2}
\end{align}
where the `time index' $j$ for the RVs $ N_{1,j}$ and $N_{2,j}$ over Phase2 has be reset to one in order not to clutter the notation. Also in order to cover the case  $B_1=B_2=0$, we define $\sum_{j=1}^{0} \ldots = 0$. 

The tricky part is to find the left and right tail probabilities of $T_{\rm p2}$, which are needed to show the concentration of $T_{\rm p2}$ to its mean. 
To show this, we define  the following outage-like events, here $\Ksf=2$, $\mathbb{E}[N_u] = 1- \epsilon_u$, 
\begin{align}
  \mathcal{B}_\ell &:= \left\{ \exists u\in[\Ksf] : \sum_{j=1}^{\ell} N_{u,j} < B_{u} \right\},
  \label{Tp2:step1}
\\
\mathcal{C}_\ell &:= \left\{ \exists u\in[\Ksf] : \left|\sum_{j=1}^{\ell} N_{u,j} -\ell \mathbb{E}[N_{u}] \right| > \varepsilon   \right\},
\end{align}
where $\mathcal{B}_\ell$ is the complement event of $T_{\rm p2}$ defined in~\eqref{eq:T_p2} and where $\mathcal{C}_\ell$ is the complement of the event that the number of packets received by each user concentrates to its mean.

Next we want to upper bound the probabilities of these outage-like events.
By the Chernoff bound and the union bound, we have
\begin{align}
&\Pr[\mathcal{C}_\ell] \leq  4\exp(-\frac{\varepsilon^2}{3 \ell} \min_{u\in[2]}(1/\mathbb{E}[N_u]) ). 
\label{bound-of-bl}
\end{align}  
\begin{subequations}
Bounding $ \Pr[\mathcal{B}_\ell] $ is done as follows
  \begin{align}
 \Pr[\mathcal{B}_\ell] &
 \leq \Pr\left[ \mathcal{C}_\ell \right] + \Pr\left[  \mathcal{B}_\ell | \mathcal{C}_\ell^c \right] 
\label{less-than-one}
\\&\leq \Pr\left[ \mathcal{C}_\ell \right] + \Pr\left[ \exists u\in[\Ksf]: B_{u} > \ell \mathbb{E}[N_{u}]-\varepsilon  \right] 
\label{note2}
\\&\leq \Pr\left[ \mathcal{C}_\ell \right] + \Pr\left[ \mathcal{A} \right]
\\& 
+\Pr\left[ \exists u\in[\Ksf]: B_{u} > \ell \mathbb{E}[N_{u}]-\varepsilon \Big| \mathcal{A}^c \right]
\label{note3}
\\&\leq \Pr\left[ \mathcal{C}_\ell \right] + \Pr\left[ \mathcal{A}\right]
\label{condition-on-Abefore}
\\&+ \Pr\left[ \exists u\in[\Ksf]: \mathbb{E}[B_{u}]+\varepsilon > \ell \mathbb{E}[N_{u}]-\varepsilon  \right]
\label{condition-on-A}
\end{align}
where~\eqref{less-than-one} follows the axioms of probability; 
where~\eqref{note2} holds since conditioning the event $\mathcal{C}_\ell^c$ allows us to lower bound the term $\sum_{j=1}^{\ell}N_{u,j}, \forall u\in[\Ksf]$;
in addition,  $\mathcal{B}_\ell$ is independent of $\mathcal{C}_\ell^c$ since the channel is memoryless;
where~\eqref{note3} follows  similar steps as in~\eqref{less-than-one}; 
where~\eqref{condition-on-A} holds since conditioning on the event $\mathcal{A}^c$ allows us to upper bound  $B_u, \forall u\in[\Ksf]$.

Next, the terms in~\eqref{condition-on-Abefore} can be bounded with~\eqref{bound-on-Ak1k2} and~\eqref{bound-of-bl};
while the term in~\eqref{condition-on-A} is zero if 
\begin{align}
\ell \geq \mathbb{E}[T_{\rm p2}] + \max_{u\in[\Ksf]} \left(  \frac{2\varepsilon}{\mathbb{E}[N_u]} \right), 
\label{eq:Condition ell LESS derivation}
\end{align}
\label{eqset: ell LARGE} 
\end{subequations}
since
\begin{align}
\mathbb{E}[T_{\rm p2}]= \max_{u\in[\Ksf]}  \left(  \frac{\mathbb{E}[B_u]}{\mathbb{E}[N_{u}]} \right) = \max_{u\in[\Ksf]}\left(\frac{k_{u}}{1-\epsilon_u} \frac{\epsilon_u-\epsilon_{12}}{1-\epsilon_{12}}\right).
 \label{expect:Tp2}
\end{align}
\begin{subequations}
By similar steps to those in~\eqref{eqset: ell LARGE}, we obtain 
\begin{align}
\Pr\left[ \mathcal{B}_{\ell-1}^c \right]
&\leq \Pr\left[ \mathcal{C}_{\ell-1} \right] + \Pr\left[ \mathcal{A} \right]
\\&+ \Pr\left[ \mathbb{E}[B_{u}]-\varepsilon\leq (\ell-1) \mathbb{E}[N_{u}]+\varepsilon, \forall u\in[\Ksf] \right],
\label{eq: Condition ell MORE derivation}
\end{align}
and the term in~\eqref{eq: Condition ell MORE derivation} is zero if 
$ \ell < \mathbb{E}[T_{\rm p2}] - \max_{u\in[\Ksf]} \left(  \frac{2\varepsilon}{\mathbb{E}[N_u]} \right) +1. $

Now we have all the components to compute the left and right tail of $T_{\rm p2}$.
\label{eqset: ell SMALL}
\end{subequations}
\paragraph*{Right Tail Bound for $ T_{\rm p2}$}
Recall that $( T_{\rm p1},B_{1},B_{2})$, which are functions of the channels gains in Phase1, are independent of the channel gains in phase2 because the channel is memoryless. According to~\eqref{eq:T_p2}, we have
$ \Pr[T_{\rm p2} > \ell] = \Pr\left[ \mathcal{B}_\ell \right].$ 
By~\eqref{eqset: ell LARGE}, we have obtained
$\Pr[T_{\rm p2} > \ell] \leq  o(\varepsilon) $ if  
$ \ell   \geq \mathbb{E}[T_{\rm p2}]+ \max_{u\in[\Ksf]} \left(  \frac{2\varepsilon}{\mathbb{E}[N_u]} \right). $

\paragraph*{Left Tail Bound for $T_{\rm p2}$}
Since $T_{\rm p2} < \ell$ implies that $\sum_{j=1}^{\ell -1} N_{u,j} \geq B_{u}, \forall u\in[\Ksf]$, 
we bound the left tail of $T_{\rm p2}$ as
$\Pr[T_{\rm p2} < \ell] = \Pr\left[ \mathcal{B}_{\ell-1}^c \right].$ 
By~\eqref{eqset: ell SMALL}, we have obtained
$ \Pr[T_{\rm p2} < \ell] \leq  o(\varepsilon) $ if  
$\ell   < \mathbb{E}[T_{\rm p2}]- \max_{u\in[\Ksf]} \left(  \frac{2\varepsilon}{\mathbb{E}[N_u]} \right) +1. $

\paragraph*{Concentration Result for $T_{\rm p2}$} Combining the left and right tail bound of $T_{\rm p2}$, we have
\begin{align}
  &\Pr\left[\Big|T_{\rm p2}- \mathbb{E}[T_{\rm p2}] \Big| > \varepsilon_{2} \right]  \notag
\\&=\Pr\left[ T_{\rm p2} > \mathbb{E}[T_{\rm p2}] +\varepsilon_{2}   \right] \notag
   +\Pr\left[ T_{\rm p2} < \mathbb{E}[T_{\rm p2}] -\varepsilon_{2} \right] \notag
\\& \leq 2o(\varepsilon) 
\label{Tp2-steps-end}
\end{align}
 if 
 \begin{align} 
   \begin{cases}
\mathbb{E}[T_{\rm p2}] +\varepsilon_{2}  \geq \mathbb{E}[T_{\rm p2}]+ \max_{u\in[\Ksf]} \left(  \frac{2\varepsilon}{\mathbb{E}[N_u]} \right) 
\\ 
\mathbb{E}[T_{\rm p2}] - \varepsilon_{2}  <  \mathbb{E}[T_{\rm p2}]- \max_{u\in[\Ksf]} \left(  \frac{2\varepsilon}{\mathbb{E}[N_u]} \right) +1. 
\end{cases} \label{concentration:Tp2}
\end{align}
Thus, by choosing $ \varepsilon_{2} = \max_{u\in[\Ksf]} \left(  \frac{2\varepsilon}{\mathbb{E}[N_u]} \right)$ in~\eqref{concentration:Tp2}, 
we have $T_{\rm p2}$ concentrates to its expected value given by $\mathbb{E}[T_{\rm p2}]$ in~\eqref{expect:Tp2}.  

\subsection{Total duration}
Finally, we are interested in $T: = T_{\rm p1} + T_{\rm p2}$, for which we have
\begin{align}
  \mathbb{E}[T]
  &= \mathbb{E}[T_{\rm p1}] + \mathbb{E}[T_{\rm p2}] \notag
\\&=   \frac{k_1+k_2}{1-\epsilon_{12}} +  \max_{u\in[\Ksf]}\left(\frac{k_{u}}{1-\epsilon_u} \frac{\epsilon_u-\epsilon_{12}}{1-\epsilon_{12}}\right) \notag
\\& = \max\left(\frac{k_1}{1-\epsilon_1} + \frac{k_2}{1-\epsilon_{12}}, \frac{k_1}{1-\epsilon_{12}} + \frac{k_2}{1-\epsilon_{2}}  \right).
\end{align}
Now, since $T_{\rm p1}$ and $T_{\rm p2}$ concentrate to their mean, also $T$ does -- recall that $T_{\rm p1}$ and $T_{\rm p2}$ are independent so $T$'s distribution is the convolution of the distributions of $T_{\rm p1}$ and $T_{\rm p2}$.

Thus, we have derived the result for the infinite backlog case as in~\cite{Sagduyu-unusual} without the use of Markov chains. We next generalize this idea to the case of multiple layers.

\section{Proof of Theorem~\ref{thm:Prototype} }
\label{sec:multiple-layers}

\subsection{Aim}
We aim to generalize the proof of the single-layer case in Appendix~\ref{sec:proof of time concentration}, to the case of any number of layers $\Qsf$, and $\Ksf = 2$ users. 
We still use the two-phase protocol, but now Phase1 is composed of $\Qsf$ sub-phases, since the time needed for each layer to complete the transmission of its uncoded packets may be different. 
In Appendix~\ref{App-B-Tp1}, the idea is to show that for Phase1, the time $T_{q}^{\text{(unc)}}$ defined in~\eqref{def:Tq}, of sub-phase $q\in[\Qsf],$ concentrates to its expectation. The total time of Phase1 is $T^{\text{(unc)}}$ defined in~\eqref{def:Tp1}, as it is the time needed for the ``slowest'' layer to finish its uncoded phase. 
In general, the expectation of the maximum of some RVs is not equal to the maximum of the expected value of the RVs. 
Herein, we attain that $t^{\text{(unc)}} = \max_{q\in[\Qsf]} t_{q}^{\text{(unc)}}$ by showing that each RV $T_{q}^{\text{(unc)}}$ has a sharp concentration to its expectation by using the outage-event technique  introduced in Appendix~\ref{sec:proof of time concentration}.
At the end of each sub-phase, the number of ``overheard packets'' for each user is indicated by $K^\text{(rtx)}_{u}[j], u \in [\Ksf], j\in[\Qsf]$ defined in~\eqref{def:Bu-j}. 
In Appendix~\ref{sub:specialcases}, we aim to show that the RVs $(K^\text{(rtx)}_{1}[\Qsf], K^\text{(rtx)}_{2}[\Qsf])$ also concentrate to their expectations. 
 We discuss the expectations of these RVs in two separate cases and obtain the general simplified expression in~\eqref{eq:DTOct04proposed_k_rtx_final}.
The key technique to show the concentration of these RVs still follows the method of types in information theory.
Then, proving a concentration for Phase2 (where only network coded packets are transmitted on each layer) for a given $K^\text{(rtx)}_{u}[\Qsf], u \in [\Ksf],$ in Appendix~\ref{App-B-Tp2}, will follow along the same line as proof  in Appendix~\ref{sec:proof of time concentration}. This will conclude the proof that the total time $T:=T^{\text{(unc)}} + T^{\text{(NC)}}$ has a sharp concentration at its expectation. 

\subsection{Duration of Phase1}
\label{App-B-Tp1}
In Phase1 the transmitter keeps sending each of the $k_{1,q}+k_{2,q}$ packets on layer $q, q\in[\Qsf],$ until one of the receivers receives it. We are given $\{k_{u,q} \geq 0, q\in[\Qsf], u\in[\Ksf]\}, \ \Qsf\geq 1, \ \Ksf=2$.
In the following, we shall first analyze the case $k_{1,q}+k_{2,q}>0$ for all $q\in[\Qsf]$.
 Appendix~\ref{sub:specialcases} describes how the analysis should be modified if $k_{1,q}+k_{2,q}=0$ for some $q\in[\Qsf]$.
Recall that the packets are transmitted randomly on each layer, and that the RV $A_{q,t}$ indicates which user is being served in the uncoded phase at time $t$ on layer $q$, with probability 
\begin{align}
\Pr[A_{q} = u] = \frac{k_{u,q}}{k_{1,q}+k_{2,q}}, \ 
 u\in[2], q\in[\Qsf],
\label{eq:def Pr[Aqt = u]} 
\end{align}
which is well defined since we assume here that $k_{1,q}+k_{2,q}>0$ for all $q\in[\Qsf]$.

To simplify notation, we define
  $ X_{u,q,t} := 1_{\{A_{q,t} = u, \ M_{t} \geq q\}}, \ t\in\mathbb{N}. $
The event  $\{ X_{u,q,t} = 1 \}, u\in [2],$ indicates that a packet destined to user $u$ has been transmitted on layer $q$ and at least one of the receivers has received that packet at time slot $t$. Since $A_{q,t}$ and $M_t$ are independent and each is i.i.d. over time, we have 
$ \mathbb{E}[X_{u,q}] = \Pr[A_{q} =u, M \geq q] = \frac{k_{u,q}}{k_{1,q}+k_{2,q}} \Pr[M \geq q].$
  The time needed to send all $k_{1,q}+k_{2,q}>0$ uncoded packets on layer $q\in[\Qsf]$ is the RV
\begin{align}
   T_{q}^{\text{(unc)}} &: =  \min\left\{ j : \sum_{t=1}^j  X_{u,q,j}  \geq k_{u,q}, \forall u\in[2] \right\}.
\label{def:Tq}
\end{align}
The time needed to complete Phase1 is the RV
\begin{align}   
   T^{\text{(unc)}} : =\max_{q\in[\Qsf]}\left\{ T_{q}^{\text{(unc)}} \right\}. 
\label{def:Tp1}
\end{align}

We partition the layers into
\begin{align*}
   V_{0} &= \{q\in[\Qsf]: k_{1,q}=k_{2,q}=0\},
\\ V_{1} &= \{q\in[\Qsf]: k_{1,q}>k_{2,q}=0\},
\\ V_{2} &= \{q\in[\Qsf]: k_{1,q}=0<k_{2,q}\},
\\ V_{12}&= \{q\in[\Qsf]: k_{1,q}>0, k_{2,q}>0\}.
\end{align*}
$|V|$ gives the number of layers belongs to set $V$. In the following, we shall do the analysis for $|V_{12}|=\Qsf$ (so as to avoid doing a very similar analysis three times)\footnote{%
In the general case, the condition in~\eqref{eq:Tp1-ccdf-large-ell} reads 
\begin{align*}
& \max\left\{
\max_{q\in V_{1}} \frac{k_{1,q}+\varepsilon}{\Pr[M \geq q]} ,
\max_{q\in V_{2}} \frac{k_{2,q}+\varepsilon}{\Pr[M \geq q]} , \right.
\\& \left.
\max_{u\in[2],q\in V_{12}} \frac{k_{u,q}+\varepsilon}{\mathbb{E}[ X_{u,q}]} \right\}
= 
\max_{u\in[2]} \left\{ \frac{k_{u,q}+\varepsilon}{\mathbb{E}[X_{u,q}]} \right\}
\leq \ell_q,
\end{align*}
and similarly for the condition in~\eqref{eq:Tp1-ccdf-small-ell}.
Also  
\begin{align*}
 \Pr\left[ \Oc^{-}_{\ell_q} \cup \Oc^{+}_{\ell_q} \right]
  &\leq \sum_{u\in[2]} \Pr\left[q\in V_{12}: \left| \sum_{t=1}^{\ell_q} X_{u,q,t} - \ell_q \mathbb{E}[X_{u,q}] \right|
> \varepsilon \right]
\\&+  \Pr\left[q\in V_{1}\cup V_{2}: \left| \sum_{t=1}^{\ell} 1_{\{M_{t} \geq q\}} - \ell_q \Pr[M \geq q] \right|
> \varepsilon  \right]
\\&\leq 4\exp(-\frac{\varepsilon^2}{3 \ell_q} \min_{u\in[2] }(1/\mathbb{E}[X_{u,q}]) ) + 4\exp(-\frac{\varepsilon^2}{3 \ell_q \Pr[M\geq q]}).
\end{align*}
}.

\begin{subequations}
To bound the left and right tails of $T_{q}^{\text{(unc)}}$, for all $q\in[\Qsf]$, $\varepsilon >0$, we define the outage-like events
\begin{align}
\Oc^{-}_{\ell_q} &:= \{ \exists u\in[2]: \sum_{t=1}^{\ell_q} X_{u,q,t} <   \ell_q \mathbb{E}[X_{u,q}] -\varepsilon \}, 
\\
\Oc^{+}_{\ell_q} &:= \{ \exists u\in[2]: \sum_{t=1}^{\ell_q} X_{u,q,t} >   \ell_q \mathbb{E}[X_{u,q}] +\varepsilon \}.
\end{align}
By the Chernoff bound and the union of events bound, for every $\varepsilon > 0$, we have
\begin{align} 
\Pr\left[ \Oc^{-}_{ \ell_{q}} \right] &\leq \sum_{u\in[2]}  \Pr\left[ 
 \sum_{t=1}^{\ell_q} X_{u,q,t} - \ell_q \mathbb{E}[X_{u,q}] 
<- \varepsilon 
\right] \notag
\\&  \leq 2 \exp(-\frac{\varepsilon^2}{3 \ell_q} \min_{u\in[2] }(1/\mathbb{E}[X_{u,q}]) ),
\label{eq:expE1-def}
\\
\Pr\left[ \Oc^{+}_{ \ell_{q}} \right] &\leq \sum_{u\in[2]}  \Pr\left[ 
 \sum_{t=1}^{\ell_q} X_{u,q,t} - \ell_q \mathbb{E}[X_{u,q}] 
> \varepsilon
\right] \notag
\\&  \leq 2 \exp(-\frac{\varepsilon^2}{2 \ell_q} \min_{u\in[2] }(1/\mathbb{E}[X_{u,q}]) ).
\end{align}
\end{subequations}
Following similar steps as in the bounding of the left and right tails of $T_{\rm p2}$ in the single-layer case, we next give tail probabilities of each $T_q^{\text{(unc)}}$, $q\in[\Qsf]$.

\begin{subequations}
\paragraph*{Right Tail Bound for $ T_{q}^{\text{(unc)}}$} 
According to~\eqref{def:Tq}, we have for all $ q\in[\Qsf]$
\begin{align}
  &\Pr\left[ T_{q}^{\text{(unc)}} > \ell_q \right] \notag
\\&=\Pr\left[ \exists u\in[2] :  \sum_{t=1}^{\ell_q} X_{u,q,t} < k_{u,q} \right] \notag
\\&\leq \Pr\left[ \Oc^{-}_{\ell_q} \right]  
  + \Pr\left[ \exists u\in[2]:  \sum_{t=1}^{\ell_q} X_{u,q,t} < k_{u,q} \Big| (\Oc^{-}_{\ell_q})^c \right]  \label{ieq:T1qAxiom}
\\&\leq \Pr\left[ \Oc^{-}_{\ell_{q}} \right] \notag
\\&  + \Pr\left[ \exists u\in[2] :  \ell_q  \mathbb{E}[X_{u,q}]-\varepsilon < k_{u,q} \right]
\label{eq:Tp1-ccdf-large-ell}
\end{align}
where~\eqref{ieq:T1qAxiom} follows the axioms of probability; where~\eqref{eq:Tp1-ccdf-large-ell} holds since conditioning on the event $(\Oc^{-}_{  \ell_{q}} )^c$ allows us to  lower bound the term $ \sum_{t=1}^{\ell_q} X_{u,q,t}, u\in[2], q\in[\Qsf]$; in addition, the term in~\eqref{eq:Tp1-ccdf-large-ell} becomes zero if  
$\ell_q \geq t_q^{\text{(unc)}}+ \frac{\varepsilon}{ \mathbb{E}[X_{u,q}]}  ,\  \forall u\in[2]$
where 
$t_q^{\text{(unc)}} 
:= \frac{k_{u,q}}{ \mathbb{E}[X_{u,q}]}
 =  \frac{k_{1,q}+k_{2,q}}{\Pr[M \geq q]}.$
\label{eq:Tp1-cdf-large-ell}
\end{subequations}

\paragraph*{Left Tail Bound for $T_{q}^{\text{(unc)}}$}
Similarly, for every $q\in[\Qsf]$, we have 
\begin{align}
  &\Pr\left[ T_q^{\text{(unc)}} < \ell_q \right] \notag
\\&=\Pr\left[  \sum_{t=1}^{\ell_q-1} X_{u,q,t} \geq k_{u,q} , \forall u\in[2] \right] \notag
\\&\leq \Pr\left[ \Oc^{+}_{ \ell_{q} -1}  \right]
+ \Pr\Bigg[  \sum_{t=1}^{\ell_q-1} X_{u,q,t}  \geq k_{u,q}, \forall u\in[2]  \Big| (\Oc^{+}_{ \ell_{q}}  )^c \Bigg] \notag
\\&\leq \Pr\left[ \Oc^{+}_{ \ell_{q}-1}  \right] 
+ \Pr\Bigg[  (\ell_q-1) \mathbb{E}[X_{u,q}] + \varepsilon \geq k_{u,q} ,\forall u\in[2]  \Bigg] \label{eq:Tp1-ccdf-small-ell}
\\&= \Pr\left[ \Oc^{+}_{  \ell_{q}-1}  \right] + 0, \notag
\end{align}
 if 
$  \exists u\in[2]: \ell_q < t_q^{\text{(unc)}}- \frac{\varepsilon}{ \mathbb{E}[X_{u,q}] } +1. $
\paragraph*{Concentration Result for $T_{q}^{\text{(unc)}}$} 
Now we want to show that each $T_q^{\text{(unc)}}, q\in[\Qsf]$ has a sharp concentration to its mean $ t_q^{\text{(unc)}}$. Since 
$\Pr\left[  |T_q^{\text{(unc)}} - t_q^{\text{(unc)}}| \leq  \varepsilon_3, \forall q\in[\Qsf]  \right] = 1-\Pr\left[ \exists q\in[\Qsf]:  |T_q^{\text{(unc)}} - t_q^{\text{(unc)}}| >  \varepsilon_3  \right],$ 
based on the results of~\eqref{eq:Tp1-cdf-large-ell} and~\eqref{eq:Tp1-ccdf-small-ell}, we have
\begin{subequations}
\begin{align}
&  \Pr\left[ \exists q\in[\Qsf]:  |T_q^{\text{(unc)}} - t_q^{\text{(unc)}}| >  \varepsilon_3  \right] \notag
\\&= \Pr\left[ \exists q\in[\Qsf]:T_q^{\text{(unc)}}  >  t_q^{\text{(unc)}} +\varepsilon_3 \right] \notag
 \\&  + \Pr\left[ \exists q\in[\Qsf]: T_q^{\text{(unc)}}  <  t_q^{\text{(unc)}} -\varepsilon_3 \right] \notag
\\&\leq \sum_{q\in[\Qsf]}(\Pr\left[ \Oc^{-}_{ \ell_{q}} \right] + \Pr\left[\Oc^{+}_{\ell_{q}-1}  \right]) , 
\label{eq:chepena-1}
\end{align}
where~\eqref{eq:chepena-1} holds if 
\begin{align}
\begin{cases}
 t_q^{\text{(unc)}}+\varepsilon_3  \geq t_q^{\text{(unc)}}+ \frac{\varepsilon}{\mathbb{E}[X_{u,q}]}  \\ %
 t_q^{\text{(unc)}}-\varepsilon_3   < t_q^{\text{(unc)}}- \frac{\varepsilon}{\mathbb{E}[X_{u,q}]} +1, \\
\end{cases}
\forall q\in[\Qsf].
\end{align}
\label{sub-T-unc}
\end{subequations}
Thus, by choosing $\varepsilon_3 =\max_{(u,q)\in[2]\times[\Qsf]}(\frac{\varepsilon}{\mathbb{E}[X_{u,q}]})$, the bound in~\eqref{eq:chepena-1} implies that  for every $q\in[\Qsf]$, $T_q^{\text{(unc)}}$ concentrates to its expected value given by $t_q^{\text{(unc)}}$. According to the definition of $T^{\text{(unc)}}$ in~\eqref{def:Tp1}, this also implies that  $T^{\text{(unc)}}$ concentrates to its expected value given by $t^{\text{(unc)}}: = \max_{q\in[\Qsf]} \{ t_q^{\text{(unc)}}\}$. More precisely,  for every $\varepsilon_3 >0$,
\begin{align*}
  \lim_{\min_{u\in[2],q\in[\Qsf] : k_{u,p}>0}\{k_{u,p}\}\to\infty} \Pr\left[ \Big|T^{\text{(unc)}} - t^{\text{(unc)}}\Big| >  \varepsilon_3  \right] = 0.
\end{align*}

\subsection{Number of packets not delivered by the end of a sub-phase}
\label{sub:specialcases}
Next, we want to show that $K^\text{(rtx)}_{u}[\Qsf]$, the number of packets only received by the other user $\bar{u}, \bar{u}\neq u, u\in[2]$ concentrates to its expectation.
Recall that at time $T^\text{\rm(unc)}_{\pi(j)}$, the layers $\pi(1), \ldots, \pi(j)$ have finished their uncoded phase; the interval $[T^\text{\rm(unc)}_{\pi(j-1)},T^\text{\rm(unc)}_{\pi(j)}), \ j\in[\Qsf],$ is where of the $j$-th sub-phase of Phase1 takes place. 
We indicate the average duration of the $j$-th sub-phase of Phase1 as 
$  \Delta_j := \mathbb{E}[T_{\pi(j)}^{\text{(unc)}} -T_{\pi(j-1)}^{\text{(unc)}} ], \ j\in[\Qsf].$
Based on our protocol and considering the notation used in Theorem~\ref{thm:Prototype} , we define for $u\in[2], q\in[\Qsf], j\in\mathbb{N}$
\begin{align}   
 K_{u,q}^\text{(unc)}[j]&
                          = \left[k_{u,q} - \sum_{t=1}^{T_{\pi(j)}^{\text{(unc)}}} X_{u,q,t}\right]^+   
                         \label{def:K-unc}
\end{align}
where 
$K_{u,q}^\text{(unc)}[j]$ is the number of packets destined to user $u$ on layer $q$ that have not been received by any user by the $j$-th sub-phase. Let $ K_{u,q}^\text{(unc)}[0] =0$. By the definition of the RVs ($T_{\pi(j)}^{\text{(unc)}}, j\in[\Qsf]$) in~\eqref{def:Tq}, we have when $q= j$, layer $\pi(q)$ finishes its uncoded packets and $K_{u,\pi(q)}^\text{(unc)}[j] = 0$. Thus, $K_{u,\pi(q)}^\text{(unc)}[j] \geq 0$, when $q\geq j$.

We define $K^\text{(rtx)}_{u}[j]$ as the number of packets destined to user~$u\in[2]$ but not yet successfully received by user~$u$ at time $T^\text{\rm(unc)}_{\pi(j)}$.
Let us initialize $K^\text{(rtx)}_{u}[0] = 0, u\in[2]$. 
By the description of our protocol, $K^\text{(rtx)}_{u}[j]$ is the number of overheard packets from the previous sub-phase $K^\text{(rtx)}_{u}[j-1]$, plus the number of overheard packets that were sent uncoded during the time window $[T^\text{\rm(unc)}_{\pi(j-1)},T^\text{\rm(unc)}_{\pi(j)})$, minus the number of packets delivered in a network-coded manner $[T^\text{\rm(unc)}_{\pi(j-1)},T^\text{\rm(unc)}_{\pi(j)})$. 
Let $I_{u,q,t}$ be an indicator that user $u\in[2]$ is scheduled to transmit an uncoded packet on layer $q\in[\Qsf]$ at time $t\in\mathbb{N}$ and that this uncoded packet is received by the other user only; we have 
\begin{align*}
\Pr[I_{u,q,t} = 1] = \frac{k_{u,q}}{k_{1,q}+k_{2,q}}\Pr[M\geq q, N_u<q].
\end{align*}
By our protocol
\begin{align}
&K^\text{(rtx)}_{u}[j] := \Bigg[ 
K^\text{(rtx)}_{u}[j-1] 
- 
\sum_{q=1}^{j-1} \sum_{t=T_{\pi(j-1)}^{\text{(unc)}} +1}^{T_{\pi(j)}^{\text{(unc)}} }   
 1_{\{N_{u,t} \geq \pi(q)\}} \notag
\\&+ 
\sum_{q=j}^{\Qsf} \min\Big( \sum_{t=T_{\pi(j-1)}^{\text{(unc)}} +1}^{T_{\pi(j)}^{\text{(unc)}} } \!\!\!\!\!\!\!\!\!
   I_{u,\pi(q),t},    K_{u,\pi(q)}^\text{(unc)}[j-1] \Big)  
\Bigg]^+, \ j\in[\Qsf].
\label{def:Bu-j}
\end{align}
\begin{subequations}
It can be easily checked that $B_u[j] = B_u[j-1]$ if two layers complete their uncoded phase at the same time (i.e., if $T_{\pi(j)}^{\text{(unc)}}  = T_{\pi(j-1)}^{\text{(unc)}}$); the same holds if more that two layers finish their uncoded phase at the same time. This is because we define $\sum_{t=T+1}^{T} \ldots = 0$.  For $q\geq j$, we have 
\begin{align}
 &K_{u,\pi(q)}^\text{(unc)}[j-1] - \sum_{t=T_{\pi(j-1)}^{\text{(unc)}} +1}^{T_{\pi(j)}^{\text{(unc)}} }
  I_{u,\pi(q),t}  \notag
   \\&  = k_{u,\pi(q)} - \sum_{t=1}^{T_{\pi(j-1)}^{\text{(unc)}} } X_{u,\pi(q),t} 
 - \sum_{t=T_{\pi(j-1)}^{\text{(unc)}} +1}^{T_{\pi(j)}^{\text{(unc)}} }
   I_{u,\pi(q),t}  
   \label{compare:note2}
   \\& \geq k_{u,\pi(q)} - \sum_{t=1}^{T_{\pi(j-1)}^{\text{(unc)}} } X_{u,\pi(q),t} 
- \sum_{t=T_{\pi(j-1)}^{\text{(unc)}} +1}^{T_{\pi(j)}^{\text{(unc)}} }
   X_{u,\pi(q),t} 
     \label{compare:note3}
   \\& = k_{u,\pi(q)} - \sum_{t=1}^{T_{\pi(j)}^{\text{(unc)}} } X_{u,\pi(q),t} 
     \label{compare:note4}
  \\& = K_{u,\pi(q)}^\text{(unc)}[j] 
 \geq 0
      \label{compare:note5}
   \end{align}
   where~\eqref{compare:note2} and~\eqref{compare:note5} follow~\eqref{def:K-unc} and $K_{u,\pi(q)}^\text{(unc)}[j-1] > 0, K_{u,\pi(q)}^\text{(unc)}[j] \geq 0$ when $q\geq j$; where~\eqref{compare:note3} holds because in the $j$-th sub-phase, on layer $\pi(q)$, the number of packets destined to user $u$ has been received by at least one of the users  is no less than the number of packets destined to user $u$ has been received by the unintended user only; where~\eqref{compare:note4} simply groups the packets in different sub-phases together. 
   \end{subequations}  
Therefore,~\eqref{def:Bu-j} can be simplified as
\begin{align}
K^\text{(rtx)}_{u}[j] 
 & = \Bigg[ K^\text{(rtx)}_{u}[j-1] + D_{u}[j] \Bigg]^+
 \label{express:B_u[j]}
\end{align} 
where 
\begin{align} 
&D_{u}[j]: =\sum_{q=j}^{\Qsf} \sum_{t=T_{\pi(j-1)}^{\text{(unc)}} +1}^{T_{\pi(j)}^{\text{(unc)}} }  
\!\!\!\!\!\!\!\!      I_{u,\pi(q),t} -    
\sum_{q=1}^{j-1} \sum_{t=T_{\pi(j-1)}^{\text{(unc)}} +1}^{T_{\pi(j)}^{\text{(unc)}} }  
\!\!\!\!\!\!\!     1_{\{N_{u,t} \geq \pi(q)\}},
     \label{express:D_u[j]}
     \\&\mathbb{E}[D_{u}[j]] =\Delta_j \left( \sum_{q=j}^{\Qsf} \Pr[ I_{u,\pi(q)} =1] - \sum_{q=1}^{j-1} \Pr[N_{u} \geq \pi(q)] \right).
     \label{express:meanD_u[j]}
     \end{align} 

Before analyzing the concentration of RVs ($K^\text{(rtx)}_{u}[j], u\in[2], j\in[\Qsf]$) in~\eqref{express:B_u[j]}, let us clarify some special cases.
\paragraph*{Special cases} $\exists q\in[\Qsf]: k_{1,q}+k_{2,q}=0$. 
The above random variables and processes are well defined if $\forall q\in[\Qsf]: k_{1,q}+k_{2,q}>0$ (in which case the probabilities in~\eqref{eq:def Pr[Aqt = u]} are well defined and the RVs in~\eqref{def:Tq} are strictly positive).
We describe here how the above has to be changed if $\exists q\in[\Qsf]: k_{1,q}+k_{2,q}=0$.
Let $V_{0} := \{q\in[\Qsf] : k_{1,q} + k_{2,q} = 0 \}$.
There are no packets to be sent on the layers indexed by $V_{0}$. 
We thus set $T_{q} = 0$ for all $q \in V_{0}$, and $T_{\pi(1)} = \ldots =T_{\pi(|V_{0}|)} =  K^\text{(rtx)}_{u}[1]= \ldots = K^\text{(rtx)}_{u}[|V_{0}|] = 0$. 
Note that the case $|V_{0}|=\Qsf$ means that there are no packets to transmit at all.

\begin{subequations}
Next, we focus on the expectation of $B_u[j]$ and define an outage-like event, for all $\varepsilon >0$,
\begin{align}
&\mathcal{E}_j := \Bigg\{
\text{ either  $\exists u\in[2]$ or $q\in[j-1]$:} \notag
\\&
 \left| \sum_{t=T_{\pi(j-1)}^{\text{(unc)}}+1}^{T_{\pi(j)}^{\text{(unc)}}} 1_{\{N_{u,t} \geq \pi(q)\}} 
- \Delta_j Pr[N_{u} \geq \pi(q)] \right| >   \varepsilon  , \label{term1}
\\&\text{or $\exists u\in[2]$ or $q\in[j:\Qsf]$:} \notag
\\&
\left| \sum_{t=T_{\pi(j-1)}^{\text{(unc)}}+1}^{T_{\pi(j)}^{\text{(unc)}}}  I_{u,\pi(q),t} -
\Delta_j \Pr[ I_{u,\pi(q),t} =1] \right| >  \varepsilon  
\Bigg\}, j\in[\Qsf]. \label{term2}
\end{align}
\label{def:epsilon_j}
\end{subequations}
\begin{subequations}
Now, we give the probability upper bound of the terms in~\eqref{term1} and~\eqref{term2} of $\mathcal{E}_j$. Let $\mathcal{O}_{\ell_{\pi(j-1)}} := \mathcal{O}_{\ell_{\pi(j-1)}}^- \cup \mathcal{O}_{\ell_{\pi(j-1)}}^+$ and $\mathcal{O}_{\ell_{\pi(j)}} := \mathcal{O}_{\ell_{\pi(j)}}^- \cup \mathcal{O}_{\ell_{\pi(j)}}^+$, we have
\begin{align}
&\Pr\Bigg[ \text{ $\exists u\in[2]$ or $q\in[j-1]$:} \notag
\\& \left| \sum_{t=T_{\pi(j-1)}^{\text{(unc)}}+1}^{T_{\pi(j)}^{\text{(unc)}}}  1_{\{ N_{u,t} \geq \pi(q)\}} -
 \Delta_j \Pr[ N_{u} \geq \pi(q)] \right| > \varepsilon \Bigg] \notag
  \\& 
  \leq \Pr[\mathcal{O}_{\ell_{\pi(j-1)}} \cup \mathcal{O}_{\ell_{\pi(j)}}] \notag
  \\&+ \Pr\Bigg[ \text{ $\exists u\in[2]$ or $q\in[j-1]$:} \notag
\left| \sum_{t=T_{\pi(j-1)}^{\text{(unc)}}+1}^{T_{\pi(j)}^{\text{(unc)}}}  1_{\{ N_{u,t} \geq \pi(q)\}} \right. \notag
\\&\left. -
 \Delta_j \Pr[ N_{u} \geq \pi(q)] \right| > \varepsilon  \Big| (\mathcal{O}_{\ell_{\pi(j-1)}} \cup \mathcal{O}_{\ell_{\pi(j)}})^c \Bigg] 
 \label{ieq:N_u-axiom}
  \\&
  \leq\Pr[\mathcal{O}_{\ell_{\pi(j-1)}} \cup \mathcal{O}_{\ell_{\pi(j)}}] \notag
  \\& +\sum_{(u,q) \in[2]\times [j-1]} \Pr\Bigg[ \sum_{t=\floor{t_{\pi(j-1)}^{\text{(unc)}}-\varepsilon_3 }+1}^{\floor{t_{\pi(j)}^{\text{(unc)}}+\varepsilon_3}}  1_{\{N_{u,t} \geq \pi(q)\}} \notag
  \\&
- \Delta_j^+ \Pr[N_{u} \geq \pi(q)] >  \varepsilon -(\Delta_j^+ - \Delta_j)\Pr[N_{u} \geq \pi(q)]  \Bigg]
\label{ieq:N_u-condition1}
\\& + \sum_{(u,q) \in[2]\times [j-1]} \Pr\Bigg[ \sum_{t=\floor{t_{\pi(j-1)}^{\text{(unc)}}+\varepsilon_3}+1}
^{\floor{t_{\pi(j)}^{\text{(unc)}}-\varepsilon_3}}  1_{\{N_{u,t} \geq \pi(q)\}} 
- \Delta_j^-  \notag
\\& \times \Pr[N_{u} \geq \pi(q)]< - \Big( \varepsilon + (\Delta_j^- - \Delta_j)\Pr[N_{u} \geq \pi(q)] \Big) \Bigg]
\label{ieq:N_u-condition2}
\\&
\leq \Pr[\mathcal{O}_{\ell_{\pi(j-1)}} \cup \mathcal{O}_{\ell_{\pi(j)}}]  \notag
\\& 
+\sum_{(u,q) \in[2]\times [j-1]} \left( \exp(- \frac{\lambda_{u,q,1}}{3\Delta_j^+}) +  \exp(- \frac{\lambda_{u,q,2}}{2\Delta_j^-}) \right)
\label{ieq:N_u-bound}
\end{align}
where 
\begin{align}
&\Delta_j^+: =\floor{t_{\pi(j)}^{\text{(unc)}}+ \varepsilon_3} - \floor{t_{\pi(j-1)}^{\text{(unc)}}-\varepsilon_3}, 
\\
&\Delta_j^-: =\floor{t_{\pi(j)}^{\text{(unc)}}- \varepsilon_3} - \floor{t_{\pi(j-1)}^{\text{(unc)}}+\varepsilon_3}, \label{delta-}
\\&
g_1(p): = \frac{ \left(\varepsilon - \Big( \Delta_{j}^+ - \Delta_j \Big) p\right)^2 }{p},  \label{g1}
\\&
g_2(p): =\frac{\left(\varepsilon + \Big( \Delta_{j}^- - \Delta_j \Big) p\right)^2}{p}, \label{g2}
\\&
\lambda_{u,q,1} :=  g_1(\Pr[N_u \geq \pi(q)]), \notag
\\&
\lambda_{u,q,2} := g_2(\Pr[N_u \geq \pi(q)]), \notag
\end{align}
 \end{subequations}
where~\eqref{ieq:N_u-axiom} follows the axioms of probability; 
\eqref{ieq:N_u-condition1} and~\eqref{ieq:N_u-condition2} hold since conditioning the event $(\mathcal{O}_{\ell_{\pi(j-1)}} \cup \mathcal{O}_{\ell_{\pi(j)}})^c$ allows us to lower and upper bound the term $T_{\pi(j-1)}^{\text{(unc)}}$ and $T_{\pi(j)}^{\text{(unc)}}$ (recall that $T_{\pi(j-1)}^{\text{(unc)}}$ and $T_{\pi(j)}^{\text{(unc)}}$ are integer-valued RVs); 
in addition, to construct the same form as the Chernoff bound, we subtract $(\Delta_j^+ - \Delta_j)\Pr[N_{u} \geq \pi(q)]$  and $(\Delta_j^- - \Delta_j)\Pr[N_{u} \geq \pi(q)]$ from both sides of the inequalities within the probability of~\eqref{ieq:N_u-condition1} and~\eqref{ieq:N_u-condition2}, respectively; 
by the Chernoff bound, we obtain~\eqref{ieq:N_u-bound};
we also select $\varepsilon_3$ to make $\Delta_j^- > 0$ in~\eqref{delta-}. Note that when the number of packets to be transmitted is large enough, we have shown that $T_{\pi(j)}^{\text{(unc)}}$ has a sharp concentration to its expectation, which implies  $\Delta_j^- \approx \Delta_j \approx \Delta_j^+$. In particular, in~\eqref{g1} and~\eqref{g2}, for constant $p\in(0,1)$ and $\varepsilon = o(\Delta_j^{\frac{1}{2} + \epsilon})$, continue with~\eqref{ieq:N_u-bound}, we have
\begin{align*}
&\Pr\Bigg[ \text{ $\exists u\in[2]$ or $q\in[j-1]$:} 
\\& \left| \sum_{t=T_{\pi(j-1)}^{\text{(unc)}}+1}^{T_{\pi(j)}^{\text{(unc)}}}  1_{\{ N_{u,t} \geq \pi(q)\}} -
 \Delta_j \Pr[ N_{u} \geq \pi(q)] \right| > \varepsilon \Bigg]
 \\& \leq  \Pr[\mathcal{O}_{\ell_{\pi(j-1)}} \cup \mathcal{O}_{\ell_{\pi(j)}}]  
\\& 
+\!\!\!\!\!\! \!\! \sum_{(u,q) \in[2]\times [j-1]} \left( \exp\left(- o(\frac{\Delta_j^{2\epsilon}}{3\Pr[N_u \geq \pi(q)]})\right) + \exp\left(- o(\frac{\Delta_j^{2\epsilon}}{2\Pr[N_u \geq \pi(q)]}) \right) \right).
 \end{align*} 
Similarly,
\begin{align*}
&\Pr\Bigg[ \text{ $\exists u\in[2]$ or $q\in[j:\Qsf]$:} 
\\&
\left| \sum_{t=T_{\pi(j-1)}^{\text{(unc)}}+1}^{T_{\pi(j)}^{\text{(unc)}}}   I_{u,\pi(q),t}  -
\Delta_j \Pr[ I_{u,\pi(q)} =1] \right| > \varepsilon \Bigg] 
\\&
\leq  \Pr\left[ \mathcal{O}_{\ell_{\pi(j-1)}} \cup \mathcal{O}_{\ell_{\pi(j)}} \right] 
\\&
+\!\!\!\!\!\! \!\! \sum_{(u,q) \in[2]\times [j:\Qsf]} \left( \exp\left(- o(\frac{\Delta_j^{2\epsilon}}{3\Pr[I_{u,\pi(q)} =1]})\right) + \exp\left(- o(\frac{\Delta_j^{2\epsilon}}{2\Pr[I_{u,\pi(q)} =1]}) \right) \right).
\end{align*}
By the union bound, we have 
\begin{align}
&\Pr[\mathcal{E}_j] \leq 2\Pr\left[ \mathcal{O}_{\ell_{\pi(j-1)}} \cup \mathcal{O}_{\ell_{\pi(j)}} \right]
\notag
\\&+\!\!\!\!\!\! \!\! \sum_{(u,q) \in[2]\times [j-1]} \left( \exp\left(- o(\frac{\Delta_j^{2\epsilon}}{3\Pr[N_u \geq \pi(q)]})\right) + \exp\left(- o(\frac{\Delta_j^{2\epsilon}}{2\Pr[N_u \geq \pi(q)]}) \right) \right) \notag
\\& +\!\!\!\!\!\! \!\! \sum_{(u,q) \in[2]\times [j:\Qsf]} \left( \exp\left(- o(\frac{\Delta_j^{2\epsilon}}{3\Pr[I_{u,\pi(q)} =1]})\right) + \exp\left(- o(\frac{\Delta_j^{2\epsilon}}{2\Pr[I_{u,\pi(q)} =1]}) \right) \right).
\end{align}
This implies that the outage event $\mathcal{E}_j, \forall j\in [\Qsf]$ has vanishing probability when $\Delta_j$, the expectation of the number of slots of sub-phase $j$ is large.  
Given $\mathcal{E}_j^c$, by~\eqref{express:D_u[j]}, we obtain
 \begin{align}
D_{u}[j] \in \left[\mathbb{E}[D_u[j]]-\Qsf \varepsilon , \mathbb{E}[D_u[j]]+\Qsf \varepsilon \right].
\label{range:D_u}
\end{align}
Following the same technique as in~\eqref{eq:Tp1-cdf-large-ell} and~\eqref{eq:Tp1-ccdf-small-ell}, we have
\begin{subequations}
\begin{align}
&\Pr[\exists u\in[2]: \left| D_u[j] - \mathbb{E}[D_u[j]] \right| > \varepsilon_4 ] \notag
\\& \leq \Pr[\mathcal{E}_j] + \Pr\left[\exists u\in[2]: \left| D_u[j] - \mathbb{E}[D_u[j]] \right| > \varepsilon_4 \Big| \mathcal{E}_j^c \right] 
\label{concentration:D_u-j-axiom}
\\&
\leq  \Pr[\mathcal{E}_j] +  \Pr\left[\exists u\in[2]: \left| \mathbb{E}[D_u[j]] + \Qsf\varepsilon - \mathbb{E}[D_u[j]] \right| > \varepsilon_4 \right],
\label{concentration:D_u-j-bound}
\end{align}
where~\eqref{concentration:D_u-j-axiom} follows the axioms of probability; 
where~\eqref{concentration:D_u-j-bound} follows~\eqref{range:D_u}; in addition, by choosing $\varepsilon_4 = \Qsf \varepsilon$, the second term in~\eqref{concentration:D_u-j-bound} becomes zero. Thus, we have shown that the RVs ($D_u[j], u\in[2], j\in[\Qsf]$) concentrate to their expectations, and by~\eqref{express:B_u[j]}, we obtain that 
\end{subequations}
 \begin{align}
\mathbb{E}[K^\text{(rtx)}_{u}[j]] = \left[ \mathbb{E}[K^\text{(rtx)}_{u}[j-1]] +  \mathbb{E}[D_u[j]] \right]^+ \geq 0. \label{eq:mean Bu}
\end{align}

In the following, we  simplify the expression of $\mathbb{E}[K^\text{(rtx)}_{u}[j]]$. 
Recall that RVs ($K^\text{(rtx)}_{u}[j], u\in[2], j\in[\Qsf])$ are nonnegative. We partition the expectations of RVs ($K^\text{(rtx)}_{u}[j], u\in[2], j\in[\Qsf]$) into two sets $W$ and $W^c$ where
$W = \left\{ \mathbb{E}[K^\text{(rtx)}_{u}[j]] >0, \forall u\in[2], j\in[\Qsf]\right\},$ and 
$W^c = \left\{ \exists u\in[2], j\in[\Qsf]: \mathbb{E}[K^\text{(rtx)}_{u}[j]]=0\right\}.$ 
Since we are working on the case that $|V_{12}|=\Qsf$, we have $\mathbb{E}[K^\text{(rtx)}_{u}[1]] =  t_{\pi(1)}^{\text{(unc)}}  \sum_{q=1}^{\Qsf}
     \Pr[I_{u,\pi(q)} =1] >0$, 
we can write  
\begin{align}
W^c& = \left\{ \exists u\in[2], j\in[2:\Qsf]: \mathbb{E}[K^\text{(rtx)}_{u}[j]]=0\right\}.
\end{align}
Recall that $\mathbb{E}[D_{u}[1]] =t_{\pi(1)}^{\text{(unc)}}  \sum_{q=j}^{\Qsf} \Pr[ I_{u,\pi(q)} =1]$ by~\eqref{express:meanD_u[j]}. 
Therefore, in set $W$, according to~\eqref{eq:mean Bu} and $\mathbb{E}[K^\text{(rtx)}_{u}[0]]=0$, we have for $j\in[\Qsf]$,
\begin{align}
&\mathbb{E}[K^\text{(rtx)}_{u}[j] ] 
=  \sum_{q\in[j]} \mathbb{E}[D_u[q] ].
\label{setW:E-B_u}
\end{align}
Next, we give the expression of $\mathbb{E}[K^\text{(rtx)}_{u}[\Qsf] ]$ in set $W^c$ by induction.
We want to show that
\begin{align}
&\mathbb{E}[K^\text{(rtx)}_{u}[j+1] ]  = 0, \notag
\\&  \text{if  } \exists u\in[2], \min( j\in[2:\Qsf-1]): \mathbb{E}[ K^\text{(rtx)}_{u}[j] ]=0 ,
\label{alternative:B_Q=0}
\end{align}
which implies
$\mathbb{E}[K^\text{(rtx)}_{u}[\Qsf] ]= 0, $ when  $  W^c$ is true. 
By~\eqref{eq:mean Bu},
 if $\exists u\in[2], \min( j\in[2:\Qsf-1]): \mathbb{E}[ K^\text{(rtx)}_{u}[j] ]=0$, we have 
 \begin{align}
 \mathbb{E}[K^\text{(rtx)}_{u}[i]] = 
 \begin{cases}
\sum_{q\in[i]} \mathbb{E}[D_u[q]]  >0, \qquad \qquad \  \forall i<j,
 \\
 \left[\mathbb{E}[ K^\text{(rtx)}_{u}[i-1] ] + \mathbb{E}[ D_u[i] ]\right]^+ = 0,\  i \geq j,
 \end{cases}
 \end{align}
  which implies
  \begin{align*}
 & \mathbb{E}[ D_u[j] ]
  \\&=  \Delta_{j}\! \left(\sum_{q=j}^{\Qsf}  \Pr\left[I_{u,\pi(q)} =1 \right] - \sum_{q=1}^{j-1}\Pr\left[N_{u} \geq \pi(q)\right] \!\right) 
 <0,
  \\
&  \mathbb{E}[ K^\text{(rtx)}_{u}[j+1] ] = \left[  \mathbb{E}[ D_u[j+1] ] \right]^+ = 0
 \end{align*}
 where 
 \begin{align*}
 & \mathbb{E}[ D_u[j+1] ] 
 \\&= \Delta_{j+1}\left(\sum_{q=j+1}^{\Qsf}  \Pr\left[I_{u,\pi(q)} =1\right] - \sum_{q=1}^{j}\Pr\left[N_{u} \geq \pi(q)\right] \right)
  \\
  & \leq  \mathbb{E}[ D_u[j] ] < 0.
 \end{align*}
 Thus, we have shown~\eqref{alternative:B_Q=0}, which implies that for set $W$ we have
  \begin{align}
   \mathbb{E}[ K^\text{(rtx)}_{u}[j] ]  = \left[ \sum_{q\in[j]} \mathbb{E}[D_u[q]] \right]^+, \ j\in[\Qsf].
   \label{setWc:E-B_u}
    \end{align}
By considering~\eqref{setW:E-B_u} and~\eqref{setWc:E-B_u}, we conclude that
$\mathbb{E}[K^\text{(rtx)}_{u}[j] ] = \left[ \sum_{q\in[j]} \mathbb{E}[D_u[q] ] \right]^+, \ j\in[\Qsf],$ 
which is the expression in~\eqref{eq:DTOct04proposed_k_rtx_final}.
Also, $\mathbb{E}[K^\text{(rtx)}_{u}[\Qsf] ]$ can be written in closed form as follows
\begin{align}
&\mathbb{E}[K^\text{(rtx)}_{u}[\Qsf] ] =  \left[ \sum_{q\in[\Qsf]} k_{u,\pi(q)} + \sum_{q\in[\Qsf]} k_{\bar{u},\pi(q)}\frac{\Pr[N_u\geq \pi(q)]}{\Pr[M\geq \pi(q)]} \right. \notag
 \\&\left. - \mathbb{E}[N_u] \frac{k_{1,\pi(\Qsf)} + k_{2,\pi(\Qsf)}}{\Pr[M\geq \pi(\Qsf)]} \right]^+,
 \label{expectation:B_u-Q}
\end{align}
which is the expression in~\eqref{eq:DTOct04proposed_k_rtx_Q}.
\begin{subequations}
Next, given $(\cup_{j\in[\Qsf]} \mathcal{E}_j)^c$ defined in~\eqref{def:epsilon_j}, we want to find the range of $K^\text{(rtx)}_{u}[\Qsf], u\in[2]$.
By~\eqref{express:B_u[j]} and~\eqref{range:D_u}, starting with $B_u[1]$, given $\mathcal{E}_1^c$, we have
 \begin{align}
K^\text{(rtx)}_{u}[1] = D_u[1] \in \left[ \left[\mathbb{E}[D_u[1]]-\Qsf \varepsilon \right]^+ , \mathbb{E}[D_u[1]]+\Qsf \varepsilon \right],
\label{eq:B_u[1]} 
\end{align}
where $\mathbb{E}[D_u[1]] = \Delta_1  \sum_{q=1}^{\Qsf} \Pr[ I_{u,\pi(q)} =1] >0$. For $K^\text{(rtx)}_{u}[2]$, given $(\mathcal{E}_1 \cup \mathcal{E}_2)^2$ and continuing with~\eqref{eq:B_u[1]},  
\begin{align}
K^\text{(rtx)}_{u}[2] &= \left[ K^\text{(rtx)}_{u}[1] + D_u[2] \right]^+ \notag
\\&
\in \left[ \left[ \left[\mathbb{E}[D_u[1]]-\Qsf \varepsilon \right]^+ + \mathbb{E}[D_u[2]] - \Qsf\varepsilon \right]^+ , \right.
\notag
\\& \left.
\qquad \left[ \mathbb{E}[D_u[1]]+\Qsf \varepsilon + \mathbb{E}[D_u[2]] + \Qsf \varepsilon \right]^+ \right]
\label{range-B2-o1}
\\&
\subseteq \left[ \left[ \mathbb{E}[D_u[1]] + \mathbb{E}[D_u[2]] -2\Qsf \varepsilon \right]^+ , \right.
\notag
\\& \left.
\qquad \left[ \mathbb{E}[D_u[1]]+ \mathbb{E}[D_u[2]] +2 \Qsf \varepsilon  \right]^+ \right]
\label{range-B2-o2}
\end{align}
where 
$\mathbb{E}[D_{u}[2]] =\Delta_2 \left( \sum_{q=2}^{\Qsf} \Pr[ I_{u,\pi(q)} =1] -  \Pr[N_{u} \geq \pi(1)] \right),$ 
 which may be positive or negative depending on the channel statistics. Note that~\eqref{range-B2-o1} is a subset of~\eqref{range-B2-o2} because $[a+b]^+ \leq \left[ [a]^+ +b \right]^+,$ for  $a, b \in \mathbb{R}$. 
Similarly, after $\Qsf$ iterations, we can obtain that
 \begin{align}
  &K^\text{(rtx)}_{u}[\Qsf] \notag \\&
  \in  
  \left[  \left[ \sum_{q\in[\Qsf]}\mathbb{E}[D_u[q]] - \Qsf^2 \varepsilon  \right]^+ , 
   \left[ \sum_{q\in[\Qsf]}\mathbb{E}[D_u[q]] + \Qsf^2 \varepsilon  \right]^+ \right]^+.
   \label{range:B_u}
\end{align}
\end{subequations}

\begin{subequations}
Now that we have all the components to show that the RVs ($K^\text{(rtx)}_{u}[\Qsf], u\in[2]$) concentrate to their mean, i.e., 
\begin{align}
&\Pr[\exists u\in[2]: \left| K^\text{(rtx)}_{u}[\Qsf] - \mathbb{E}[K^\text{(rtx)}_{u}[\Qsf]] \right| > \varepsilon_5 ] \notag
\\& \leq \Pr[\cup_{j\in[\Qsf]} \mathcal{E}_j ] \notag
\\& + \Pr\left[\exists u\in[2]: \left| K^\text{(rtx)}_{u}[\Qsf] - \mathbb{E}[K^\text{(rtx)}_{u}[\Qsf]] \right| > \varepsilon_5 \Big| (\cup_{j\in[\Qsf]} \mathcal{E}_j)^c \right] 
\label{concentration:B_u-Q-axiom}
\\&
\leq  \Pr[\cup_{j\in[\Qsf]} \mathcal{E}_j ] \notag
\\&+ \! \Pr\!\left[\exists u\in[2]: \!\! \left[ \sum_{q\in[\Qsf]}\mathbb{E}[D_u[q]] + \Qsf^2 \varepsilon  \right]^+ \!\!\!- \mathbb{E}[K^\text{(rtx)}_{u}[\Qsf]]  > \varepsilon_5 \right]
\label{concentration:B_u-1-bound}
\\&+\!  \Pr\!\left[\exists u\in[2]:\!\! \left[ \sum_{q\in[\Qsf]}\mathbb{E}[D_u[q]] - \Qsf^2 \varepsilon  \right]^+\!\!\!\!- \mathbb{E}[K^\text{(rtx)}_{u}[\Qsf]]  <-  \varepsilon_5 \right]
\label{concentration:B_u-2-bound}
\\&
\leq \Pr[\cup_{j\in[\Qsf]} \mathcal{E}_j ] + 0, 
\text{ if } \varepsilon_5 = \Qsf^2 \varepsilon,
\label{concentration:B_u-final}
\end{align}
where~\eqref{concentration:B_u-Q-axiom} follows by the axioms of probability;~\eqref{concentration:B_u-1-bound} and~\eqref{concentration:B_u-2-bound} hold since conditioning on $(\cup_{j\in[\Qsf]} \mathcal{E}_j)^c $ gives the range of $B_u[\Qsf]$ characterized in~\eqref{range:B_u}. By choosing $\varepsilon_5 = \Qsf^2 \varepsilon$, the terms in~\eqref{concentration:B_u-1-bound} and~\eqref{concentration:B_u-2-bound} become zero. The reason is explained below.
\end{subequations}

\begin{subequations}
To simplify the notations, let $x = \sum_{q\in[\Qsf]}\mathbb{E}[D_u[q]] \in \mathbb{R}$. We want to show that 
\begin{align}
0 \leq [x+ \varepsilon_5 ]^+ -[x]^+ \leq \varepsilon_5 , \label{prove:case1}
\\
-\varepsilon_5 \leq [x-\varepsilon_5 ]^+ -[x]^+ \leq 0. \label{prove:case2}
\end{align}
\label{lemma}
It can be proved by considering the following cases.
\begin{enumerate}
\item $x\in(-\infty, -\varepsilon_5]$: 

$[x+\varepsilon_5 ]^+ -[x]^+ = 0-0 = 0$,

$[x-\varepsilon_5 ]^+ -[x]^+ = 0-0 = 0$.
\item $x\in(-\varepsilon_5, 0]$: 

$[x+\varepsilon_5]^+ -[x]^+ = x+ \varepsilon_5 -0 = x+ \varepsilon_5 \in (0, \varepsilon_5 ]$,

$[x-\varepsilon_5 ]^+ -[x]^+ = 0-0 = 0$.
\item $x\in(0, \varepsilon_5 ]$: 

$[x+\varepsilon_5 ]^+ -[x]^+ = x+\varepsilon_5 -x= \varepsilon_5 $,

$[x-\varepsilon_5 ]^+ -[x]^+ = 0-x =-x\in[-\varepsilon_5 ,0)$.
\item $x\in(\varepsilon_5 , +\infty)$: 

$[x+\varepsilon_5 ]^+ -[x]^+ = x+\varepsilon_5 -x= \varepsilon_5 $,

 $[x-\varepsilon_5 ]^+ -[x]^+ = x-\varepsilon_5-x = -\varepsilon_5 $.

\end{enumerate}
Thus, we have shown~\eqref{lemma} which implies 
$\left[ \sum_{q\in[\Qsf]}\mathbb{E}[D_u[q]] + \Qsf^2 \varepsilon  \right]^+ \!\!\!- \mathbb{E}[K^\text{(rtx)}_{u}[\Qsf]]  \in [0, \Qsf^2 \varepsilon ], u\in[2],$ and 
$\left[ \sum_{q\in[\Qsf]}\mathbb{E}[D_u[q]] - \Qsf^2 \varepsilon  \right]^+\!\!\!\!- \mathbb{E}[K^\text{(rtx)}_{u}[\Qsf]]  \in[-\Qsf^2 \varepsilon, 0]. $
Thus, we choose $\varepsilon_5 = \Qsf^2 \varepsilon$ and the bound in~\eqref{concentration:B_u-final} indicates that $K^\text{(rtx)}_{u}[\Qsf], u\in[2]$ has a sharp concentration to its mean $\mathbb{E}[K^\text{(rtx)}_{u}[\Qsf]]$ characterized by~\eqref{expectation:B_u-Q}.
\end{subequations}


\subsection{Duration of Phase2}
\label{App-B-Tp2}
In Phase2 the expression of the time needed for this phase is the same as the single-layer case, namely
\begin{align}
 T^{\text{(NC)}} &:=
 \min \left(\ell :  \sum_{j=1}^{\ell} N_{u,j} \geq K^\text{(rtx)}_{u}[\Qsf], \forall u\in[2]\right). 
\end{align}
Then, by similar steps as in~\eqref{Tp2:step1}--\eqref{Tp2-steps-end}, we obtain the right and left tail bounds for $T^{\text{(NC)}}$.
\begin{subequations} 
\paragraph*{Right Tail Bound for $T^{\text{(NC)}} $}
Recall that $(T^{\text{(unc)}},K^\text{(rtx)}_{1}[\Qsf], K^\text{(rtx)}_{2}[\Qsf])$, which are functions of the channels gains in Phase1, are independent of the channel gains in Phase2 because the channel is memoryless.
For all $\ell \geq 0$, we bound the right tail of $T^{\text{(NC)}}$ as
\begin{align}
&\Pr\left[ T^{\text{(NC)}} > \ell \right] \notag
\\&=  \Pr\left[\exists u\in[2] : \sum_{t=1}^{\ell}N_{u,t} <  K^\text{(rtx)}_{u}[\Qsf] \right] \notag
\\& \leq \Pr\left[ \exists u\in[2] : \sum_{t=1}^{\ell}N_{u,t} < \ell \mathbb{E}[N_{u}] -\varepsilon \right]  \notag
\\&+    \Pr\left[\exists u\in[2] : K^\text{(rtx)}_{u}[\Qsf] > \mathbb{E}[K^\text{(rtx)}_{u}[\Qsf]] + \varepsilon_5 \right] \notag
\\&+    \Pr\left[\exists u\in[2] : \ell \mathbb{E}[N_u] -\varepsilon <  \mathbb{E}[K^\text{(rtx)}_{u}[\Qsf]] + \varepsilon_5  \right]
\label{Tp2:bound1}
\end{align}
where the term in~\eqref{Tp2:bound1} is zero if
$ \ell  \geq  t^{\text{(NC)}} + (\varepsilon_5 + \varepsilon)\max_{u\in[2]}\left\{1/\mathbb{E}[N_u] \right\},$ 
 where 
 \begin{align}
  t^{\text{(NC)}} = \max_{u\in[2]}\left\{\frac{\mathbb{E}[K^\text{(rtx)}_{u}[\Qsf]]}{\mathbb{E}[N_{u}]}\right\}.
\label{expression:tNC}
\end{align}
\label{Tp2:rightTail}
\end{subequations}

\paragraph*{Left Tail Bound for $T^{\text{(NC)}} $}
Similarly, we bound the left tail of $T^{\text{(NC)}}$ as
\begin{align}
&\Pr\left[ T^{\text{(NC)}} < \ell \right] \notag
\\
&= \Pr\left[ \forall u\in[2] :  \sum_{t=1}^{\ell}N_{u,t} \geq  K^\text{(rtx)}_{u}[\Qsf] \right] \notag
\\&
\leq \Pr\left[ \exists u\in[2] : \sum_{t=1}^{\ell}N_{u,t} > \ell \mathbb{E}[N_{u}] +\varepsilon \right]  \notag
\\&+    \Pr\left[ \exists u\in[2] : K^\text{(rtx)}_{u}[\Qsf] <  \mathbb{E}[K^\text{(rtx)}_{u}[\Qsf]] - \varepsilon_5 \right] \notag
\\&+ \Pr\left[ \forall u\in[2] :  \ell \mathbb{E}[N_{u}] +\varepsilon >   \mathbb{E}[K^\text{(rtx)}_{u}[\Qsf]] - \varepsilon_5 \right]
\label{Tp2:bound2}
\end{align}
where the term in~\eqref{Tp2:bound2} is zero if
$ \ell  \leq  t^{\text{(NC)}} - (\varepsilon_5 + \varepsilon)\max_{u\in[2]}\left\{1/\mathbb{E}[N_u] \right\}.$
\paragraph*{Concentration Result for $T^{\text{(NC)}} $} Combining the results in~\eqref{Tp2:rightTail} and~\eqref{Tp2:bound2}, we have
\begin{align}
  &\Pr\left[\Big|T^{\text{(NC)}}- t^{\text{(NC)}} \Big| > \varepsilon_6  \right] \notag
\\&=\Pr\left[ T^{\text{(NC)}} >  t^{\text{(NC)}} +\varepsilon_6 \right]  \notag
   +\Pr\left[ T^{\text{(NC)}} <  t^{\text{(NC)}} -\varepsilon_6 \right] 
\\&\leq \Pr\left[ \exists u\in[2] : \left| \sum_{t=1}^{\ell}N_{u,t} - \ell \mathbb{E}[N_{u}] \right| > \varepsilon  \right]  \notag
\\&+ \Pr\left[\exists u\in[2] : |K^\text{(rtx)}_{u}[\Qsf] - \mathbb{E}[K^\text{(rtx)}_{u}[\Qsf]]| > \varepsilon_5  \right] \notag
+0
\\& \leq  \Pr[\mathcal{C}_\ell] +  \Pr\left[(\cup_{j\in[\Qsf]} \mathcal{E}_j)\right]
\label{Tp2:concentration}
\end{align}
if 
\begin{align*}
\begin{cases}
 t^{\text{(NC)}} +\varepsilon_6 \geq t^{\text{(NC)}} + (\varepsilon_5 + \varepsilon)\max_{u\in[2]}\left\{1/\mathbb{E}[N_u] \right\}  \\ %
t^{\text{(NC)}} -\varepsilon_6  \leq t^{\text{(NC)}} - (\varepsilon_5 + \varepsilon)\max_{u\in[2]}\left\{1/\mathbb{E}[N_u] \right\} \\
\end{cases}.
\end{align*}
By choosing $\varepsilon_6 = (\Qsf^2+1)\varepsilon\max_{u\in[2]}\left\{1/\mathbb{E}[N_u] \right\}$, the bound in~\eqref{Tp2:concentration} implies that $T^{\text{(NC)}}$ concentrates to its expected value given by $t^{\text{(NC)}} $ characterized in~\eqref{expression:tNC}.
Therefore, for all $\varepsilon_6 >0$,
\begin{align*}
  \lim_{\min_{u\in[2],q\in[\Qsf] : k_{u,p}>0}\{k_{u,p}\}\to\infty}
  \!\!\!\!\! \Pr\left[\Big|T^{\text{(NC)}}- t^{\text{(NC)}} \Big| > \varepsilon_6  \right] = 0.
\end{align*}

\subsection{Total Duration}
Finally, we are interested in $T: = T^{\text{(unc)}} + T^{\text{(NC)}}$, for which we have
\begin{subequations}
\begin{align}
&\Pr\left[
\left|T^{\text{(unc)}}+T^{\text{(NC)}}- (t^{\text{(unc)}} + t^{\text{(NC)}}) \right| > \varepsilon_7
\right] \notag
\\& = 1- \Pr\left[
\left|T^{\text{(unc)}}+T^{\text{(NC)}}-(t^{\text{(unc)}} + t^{\text{(NC)}})\right| \leq \varepsilon_7 \right]
\label{comp-rule}
\\& \leq 1 - \Pr\left[
\left|T^{\text{(unc)}}-t^{\text{(unc)}}\right| \leq \frac{\varepsilon_7 }{2} , \left|T^{\text{(NC)}}-t^{\text{(NC)}}\right| \leq \frac{\varepsilon_7 }{2}
\right]
\label{triangle}
\\& \leq  
\Pr\left[\left|T^{\text{(unc)}}-t^{\text{(unc)}}\right| > \frac{\varepsilon_7 }{2}  \right] + 
\Pr\left[\left|T^{\text{(NC)}}-t^{\text{(NC)}}\right| >\frac{\varepsilon_7 }{2} \right] 
\label{comp-union}
\\&\leq  \sum_{q\in[\Qsf]}(\Pr\left[ \Oc^{-}_{ \ell_{q}} \right] + \Pr\left[\Oc^{+}_{\ell_{q}}  \right]) + \Pr[\mathcal{C}_\ell] +  \Pr\left[(\cup_{j\in[\Qsf]} \mathcal{E}_j)\right]
\label{result}
\end{align}
if 
\begin{align}
\varepsilon_7 &=\max\{ 2\varepsilon_3 , 2\varepsilon_6 \} \notag
\\&  =\max\left\{ \frac{4\varepsilon}{1-\varepsilon} ,  2(\Qsf^2+1)\varepsilon \max_{u\in[2]}\{1/\mathbb{E}[N_u ] \} \right\},
\end{align}
\end{subequations}
where~\eqref{comp-rule} follows from the complement rule;~\eqref{triangle} follows from the triangle inequality;~\eqref{comp-union} follows from the  complement rule and the union bound;~\eqref{result} follows from the results in~\eqref{sub-T-unc} and~\eqref{Tp2:concentration}.
The bound in~\eqref{result} implies that the total time $T := T^{\text{(unc)}}+T^{\text{(NC)}}$ has a sharp concentration at the expected value $\mathbb{E}[T]:=  t^{\text{(unc)}} + t^{\text{(NC)}}$, where
\begin{align}
&t^{\text{(unc)}} + t^{\text{(NC)}} \notag
\\
& =  \frac{k_{1,\pi(\Qsf)}+ k_{2,\pi(\Qsf)}}{\Pr[M \geq \pi(\Qsf)]} + 
 \max_{u\in[2]} \Bigg( \frac{1}{\mathbb{E}[N_u]} \times  \notag
 \\
& \Big[ \sum_{q\in[\Qsf]}k_{u,q}  + \sum_{q\in [\Qsf]} k_{\bar{u},q} \frac{\Pr[N_u\geq q]}{\Pr[M\geq q]} - \notag
\\
&  \mathbb{E}[N_u]\frac{k_{1,\pi(\Qsf)} + k_{2,\pi(\Qsf)}}{\Pr[M\geq q]} \Big]^+  \Bigg),
(u,\bar{u})\in[2]^2 : u \neq \bar{u}.
\label{expectation:totalT}
\end{align}

We define the (long-term average zero-error) rates as 
$ R_{u,q} = \frac{ k_{u,q}}{ \mathbb{E}[T^{\text{(unc)}} + T^{\text{(NC)}}] },  \ u\in[2], \ q\in[\Qsf],
$ and 
$R_u = \frac{\sum_{q\in[\Qsf]} k_{u,q}}{ \mathbb{E}[T^{\text{(unc)}} + T^{\text{(NC)}}] } = \sum_{q\in[\Qsf]} R_{u,q},$ 
which are well defined since  $\mathbb{E}[T^{\text{(unc)}} + T^{\text{(NC)}}]=0$ if and only if all $k_{u,q}$'s are zero which is not interesting.
Thus, the achievable region is given by
\begin{align*}
\mathcal{C}^{\text{\rm in}} &:= \big\{ (R_1, R_2)\in\mathbb{R}^{2}_{+}: 
R_u := \sum_{q \in [\Qsf]}R_{u,q}, u\in[2],
\\&
  \frac{R_{1,\pi(\Qsf)}+ R_{2,\pi(\Qsf)}}{\Pr[M \geq \pi(\Qsf)]} + 
 \max_{u\in[2]} \Bigg( \frac{1}{\mathbb{E}[N_u]} \times  \\
& \Big[ \sum_{q\in[\Qsf]}R_{u,q}  + \sum_{q\in [\Qsf]} R_{\bar{u},q} \frac{\Pr[N_u\geq q]}{\Pr[M\geq q]} - 
\\
&  \mathbb{E}[N_u]\frac{R_{1,\pi(\Qsf)} + R_{2,\pi(\Qsf)}}{\Pr[M\geq q]} \Big]^+  \Bigg) \leq 1,
\\&
(u,\bar{u})\in[2]^2 : u \neq \bar{u}, \text{ for some $R_{u,q}\geq0, u\in[2], q\in[\Qsf]$}
\big\}. 
\end{align*}


\section{Proof of Theorem~\ref{thm:achievable-stability-region}}
\label{sec:proof of thm:achievable-stability-region}

\subsection{Aim}
In this section, we follow similar steps as in~\cite{Sagduyu-unusual} with Lyapunov drift analysis, to show that if the arrival rates are within the region $\mathcal{S}^{\text{\rm in}}$ characterized in~\eqref{eq:achievable-stability}, then the Markov chain $\{ K_{u,q}[m], u\in[2], q\in[\Qsf] \}_{m\in\mathbb{N}}$ is ergodic. Ergodicity implies that there exists a stationary distribution, which implies that the stochastic process $\left\{|Q_{01;t}|, |Q_{02;t}|,\ldots, |Q_{0\Qsf;t}|, |Q_{1;t}|, |Q_{2;t}|\right\}_{t\in\mathbb{N}}$, characterizing the number of packets in the queues, is stable. The difference is that~\cite{Sagduyu-unusual} is based on the long-term rewards of the Markov chains, while our work is  based on the concentration result shown in Appendix~\ref{sec:multiple-layers}.
%
\subsection{Proofs}
Let us define a Lyapunov function
\begin{align}
&v({\bf k}) := \max\Big\{ \frac{k_{1,q}+k_{2,q}}{\Pr[M\geq q]}, \forall q\in[\Qsf], 
\notag \\
&\sum_{q\in[\Qsf]}\frac{k_{1,q}}{\mathbb{E}[N_1]} + \sum_{q\in[\Qsf]}\frac{k_{2,q}}{\mathbb{E}[N_1]}\frac{\Pr[N_1\geq q]}{\Pr[M\geq q]}, 
\notag \\
&\sum_{q\in[\Qsf]}\frac{k_{2,q}}{\mathbb{E}[N_2]} + \sum_{q\in[\Qsf]}\frac{k_{1,q}}{\mathbb{E}[N_2]}\frac{\Pr[N_2\geq q]}{\Pr[M\geq q]}
\Big\}, 
\end{align}
where  ${\bf k}$ is the vector containing all the $k_{u,q}$'s,
where $k_{u,q}$ is the number of packets for user $u$ in queue $Q_{0q}$ at the beginning of epoch $m$. Let $|{\bf k}|: = \sqrt{(\sum_{q\in[\Qsf]}k_{1,q})^2 + (\sum_{q\in[\Qsf]}k_{2,q})^2}$,  ${\bf K}[m] := (K_{u,q}[m], \forall u, \forall q), m\in\mathbb{N}$. 
From~\cite[Theorem 6]{Sagduyu-unusual}, to  show the ergodicity of the Markov chain  we need that
\begin{align}
  \mathbb{E}\left[v({\bf K}[m+1]) \big| {\bf K}[m]={\bf k}\right] < \infty  \label{ieq:inside-region}
\end{align}
holds for ${\bf k}$ inside a bounded region for all $m\in \mathbb{N}$, and  for some $\epsilon>0$
\begin{align}
    \mathbb{E}\left[v({\bf K}[m+1]) \big| {\bf K}[m]={\bf k}\right] \leq (1-\epsilon)v({\bf k}) \label{ergodic-condition}
\end{align} 
 holds for ${\bf k}$ outside a bounded region for all $m \in \mathbb{N}$.


Next, we characterize $v({\bf K}[m+1])$. Denote by $\hat G_{u,q}({\bf k})$ the number of packets destined to receiver $u$ on layer $q$ that arrived to the system during epoch~$m$, given that there are ${\bf k}$ packets at the beginning of epoch~$m$. Let ${\bf \hat G(k}) := (\hat G_{u,q}({\bf k}), \forall u, \forall q)$. We use $T[m]$ to represent the total time needed for epoch $m$.
Considering the definitions and the operation of the scheme, we obtain  
$ K_{u,q}[m+1] = \hat G_{u,q}({\bf k}) = \sum_{l\in T[m]} |A_{u,q;l} |$ for $u\in[2], q\in[\Qsf]$ 
and 
$ v( {\bf K}[m+1]) = v({\bf\hat G(k})).$

Now that we showed $v( {\bf K}[m+1]) = v({\bf\hat G(k}))$, our next step is to characterize $\frac{\mathbb{E}[v({\bf \hat G}({\bf k}))]}{v({\bf k})}$. However, we should first define some important limits.  

We showed in~\eqref{expectation:totalT} in Appendix~\ref{sec:multiple-layers} that $T$, the time needed to successfully complete the transmission of ${\bf k}$ packets, has a sharp concentration to its mean value
$\mathbb{E}\left[T\right] = v({\bf k})$  
when $|{\bf k}|\to \infty$.
Since we assume that the number of packets to be transmitted is large enough at each epoch, also in~\eqref{expectation:totalT}, we showed that $T[m]$ has a sharp concentration at $v({\bf k})$, i.e., for every $\varepsilon >0$, we have
$\lim_{m \to {\infty}} \Pr[\cup_{i=m}^{\infty}\{|T[m] - v({\bf k})| > \varepsilon\} ] = 0.$ 
Based on~\cite[Proposition 1.1]{book:probabilityAGraduate}, we have $T[m] \xrightarrow{\text{a.s.}}  v({\bf k})$, as $m \!\to {\infty}$. Thus,   
\vspace{-0.3cm}
\begin{align}
\lim_{ |{\bf k}|\rightarrow \infty}\frac{T[m]}{v({\bf k})} = 1 
, \qquad
\lim_{ |{\bf k}|\rightarrow \infty}\frac{\mathbb{E}[T[m]]}{v({\bf k})} = 1. 
\label{eq:time}
\end{align}

As $ |{\bf k}|\to \infty, T[m] \to \infty$, and using the strong law of large numbers, we have 
$\lim_{ |{\bf k}|\rightarrow \infty}\frac{\sum_{l\in T[m]} |A_{u,q;l} |}{T[m]} = \lambda_{u,q}.$ 
 Now we return to our original goal of characterizing  $\frac{\mathbb{E}[v({\bf \hat G}({\bf k}))]}{v({\bf k})}$. By Wald's equation~\cite[Theorem 12]{wolff_1989}, 
\begin{align}
  \mathbb{E}[\hat G_{u,q}({\bf k})] = \mathbb{E} [\sum_{l\in T[m]}|A_{u,q;l} |] = \lambda_{u,q} \mathbb{E}[T[m]] \label{eq:wald}  
\end{align}
 and considering~\eqref{eq:time}, we have
\begin{align}
\lim_{ |{\bf k}|\rightarrow \infty}\!\!\frac{\hat G_{u,q}({\bf k})}{v({\bf k})} &= 
\lim_{ |{\bf k}|\rightarrow \infty}\!\!\frac{\hat G_{u,q}({\bf k}])}{ T[m]}\frac{T[m]} {v({\bf k})}= \lambda_{u,q}, \label{concentration}
\\
 \lim_{ |{\bf k}|\rightarrow \infty}\frac{\mathbb{E}[\hat G_{u,q}({\bf k})]}{v({\bf k})} &= 
\lim_{ |{\bf k}|\rightarrow \infty}\frac{ \lambda_{u,q} \mathbb{E}[T[m]]}{v({\bf k})} = \lambda_{u,q}.  \label{integrable}
\end{align}
According to~\cite[Corollary 4.1.]{book:probabilityAGraduate},~\eqref{integrable} implies that the sequence $\left\{ \frac{\hat G_{u,q}({\bf k})}{v({\bf k})}, u\in[2], q\in[\Qsf] \right\}$ is uniformly integrable.
Moreover, $\frac{v({\bf\hat G(k}))}{v({\bf k})} $ is uniformly integrable since the sum and the maximum of uniformly integrable functions are also uniformly integrable.
Let $ {\bm \lambda} := (\lambda_{u,q}, \forall u\in[2], q\in[\Qsf])$. 
By~\eqref{concentration}, we can write 
\begin{align*}
&\lim_{ |{\bf k}|\rightarrow \infty} \frac{v({\bf\hat G(k}))}{v({\bf k})} = 
\lim_{ |{\bf k}|\rightarrow \infty}\max\Big\{ 
\\&\frac{\hat G_{1,q}({\bf k})+ \hat G_{2,q}({\bf k})}{\Pr[M\geq q] v({\bf k})},\! \forall q\in[\Qsf], 
\sum_{q\in[\Qsf]}\!\frac{\hat G_{1,q}({\bf k})}{\mathbb{E}[N_1] v({\bf k})} +
\\&
 \sum_{q\in[\Qsf]}\!\frac{\hat G_{2,q}({\bf k})}{\mathbb{E}[N_1] v({\bf k})}\frac{\Pr[N_1\geq q]}{\Pr[M\geq q]}, \!
\sum_{q\in[\Qsf]}\!\frac{\hat G_{2,q}({\bf k})}{\mathbb{E}[N_2] v({\bf k})} 
\\&+ \sum_{q\in[\Qsf]}\frac{\hat G_{1,q}({\bf k})}{\mathbb{E}[N_2] v({\bf k})}\frac{\Pr[N_2\geq q]}{\Pr[M\geq q]}
\Big\} = v({\bm \lambda})
\end{align*}
and since $\frac{v({\bf\hat G(k}))}{v({\bf k})}$ is uniformly integrable, we have 
 $ \lim_{ |{\bf k}|\rightarrow \infty}\frac{\mathbb{E}[v({\bf \hat G}({\bf k}))]}{v({\bf k})} = v({\bm \lambda}).$ 

Next, for some $\epsilon >0$, pick $k(\epsilon)$ large enough such that $|{\bf k}|>k(\epsilon)$; pick ${\bm \lambda}$ in $\mathcal{S}^{\text{\rm in}}$, so that 
$v({\bm \lambda})\leq 1-\epsilon$. As a result,
\begin{align} \label{eq:EvvInq}
&\frac{\mathbb{E}[v({\bf \hat G}({\bf k}))]}{v({\bf k})} \leq  v({\bm \lambda})+ \frac{\epsilon}{2} 
\leq 1-\epsilon +  \frac{\epsilon}{2}=1- \frac{\epsilon}{2}.
\end{align}
The inequality in~\eqref{eq:EvvInq} shows that the condition in~\eqref{ergodic-condition} is satisfied. Let us now focus on the condition in~\eqref{ieq:inside-region}, and characterize $\mathbb{E}[v({\bf \hat G}({\bf k}))]$. It is quite straightforward to show that $\mathbb{E}[T[m]] < \infty$ when $|{\bf k}|\leq k(\epsilon)$. By~\eqref{eq:wald}, we have $\mathbb{E}[\hat G_{u,q}({\bf k})]<\infty$. This indicates that the sequence $\left\{ \hat G_{u,q}({\bf k}), u\in[2], q\in[\Qsf] \right\}$ is uniformly integrable by~\cite[Corollary 4.1]{book:probabilityAGraduate}, and $v({\bf\hat G(k}))$ is uniformly integrable since the sum and the maximum of uniformly integrable functions are also uniformly integrable. Thus, $\mathbb{E}[v({\bf \hat G}({\bf k}))]<\infty$, and this concludes that the condition in~\eqref{ieq:inside-region} holds.

Now that we showed that both conditions in~\eqref{ieq:inside-region} and \eqref{ergodic-condition} are satisfied, we can conclude that the Markov Chain $\{ K_{u,q}[m],  u\in[2], q\in[\Qsf] \}_{m\in\mathbb{N}}$ is geometrically ergodic by following~\cite[Theorem 6]{Sagduyu-unusual}.  Also, $\{|Q_{01;t}|,$ $|Q_{02;t}|,$ $\ldots, |Q_{0\Qsf;t}|, |Q_{1;t}|, |Q_{2;t}|\}_{t\in\mathbb{N}}$  is  regenerative concerning the renewal process characterizing the time needed for successive returns of the process $\{ (K_{u,q}[m], u\in[2], q\in[\Qsf]) \}_{m\in\mathbb{N}}$ to the all-zero state. The renewal process is nonlattice and the regenerative process is right-continuous and has left-hand limits. 
This implies that there exists a distribution function $F(\bf x)$ satisfying the conditions in definition such that $\left\{|Q_{01;t}|, |Q_{02;t}|,\ldots, |Q_{0\Qsf;t}|, |Q_{1;t}|, |Q_{2;t}|\right\}_{t\in\mathbb{N}}$ converges in distribution to it by~\cite[Theorem 20]{wolff_1989}.  Finally, we conclude that if the arrival rates are in the interior of the region $\mathcal{S}^{\text{\rm in}} $, then the stochastic process  $\big\{|Q_{01;t}|,$ $ |Q_{02;t}|,\ldots, |Q_{0\Qsf;t}|, |Q_{1;t}|, |Q_{2;t}|\big\}_{t\in\mathbb{N}}$ representing the length of queues  is stable.
This concludes the proof.

\section{Optimality conditions of Theorem~\ref{thm:optimality-conditions}}
\label{sec:optimal-condition}
\subsection{Aim}

In this section, we demonstrate sufficient conditions 
for which  the achievable region in Theorem~\ref{thm:achievable-stability-region} coincides with the outer bound in Theorem~\ref{thm:COFnewOuter} for $\Ksf =2 $ users and $\Qsf = 2$ layers. 
They are given as
(C1) either bound~\eqref{outer-bound.B} and bound~\eqref{outer-bound.C} are the same 
and bound~\eqref{outer-bound.A} and bound~\eqref{outer-bound.D} are symmetric, and
(C2) either bound~\eqref{outer-bound.B} or bound~\eqref{outer-bound.C} is redundant, 
(C3) either $\frac{\Pr[N_1\geq 2]}{\Pr[N_1\geq 1]} \geq \frac{\Pr[M\geq 2]}{\Pr[M\geq 1]} \geq \frac{\Pr[N_2\geq 2]}{\Pr[N_2\geq 1]}$ or  $\frac{\Pr[N_2\geq 2]}{\Pr[N_2\geq 1]} \geq \frac{\Pr[M\geq 2]}{\Pr[M\geq 1]} \geq \frac{\Pr[N_1\geq 2]}{\Pr[N_1\geq 1]}$.

In general, if all the inequalities in~\eqref{eq:outerbound} are active, the region of the outer bound has three corner points. The idea behind these conditions is that we are simplifying the outer bound in~\eqref{eq:outerbound}  by reducing the number of corner points. Specifically, (C1) has one corner point, (C2) and (C3) give two corner points. 
The key technique to prove (C1) is that we assign the number of packets to be transmitted on each layer specifically depends on the  property of the channel model in Table~\ref{tab:newcondition} (i.e., user 1 and 2 have the same ability to receive packets on layer 1 and 2 separately and user 1 and 2 either receive or erase a packet on layer 2  at the same time), and to prove (C2) and (C3) is the Fourier Motzkin Elimination (FME) procedure.

\subsection{Proof of (C1) }

Based on the channel statistics in Table~\ref{tab:newcondition}, we have the outer bound 
\begin{subequations}
\begin{align}
&R_{1} + \frac{x_2 + x_3+x_4}{2x_2+x_3+x_4}R_{2} \leq x_2+x_3+2x_4 , \label{special-bound.A}
\\
& R_{1} + R_{2} \leq 2x_2+x_3+2x_4, \label{special-bound.B}
\\
&  \frac{x_2 + x_3+x_4}{2x_2+x_3+x_4}  R_{1} + R_{2}  \leq   x_2+x_3+2x_4 . \label{special-bound.D}
\end{align}
The intersection point $P$ of~\eqref{special-bound.A} and~\eqref{special-bound.D} is
\label{eq:outerbound-special}
\end{subequations}
$(\frac{(2x_2+x_3+x_4)(x_2+x_3+2x_4)}{3x_2+2x_3+2x_4}, \frac{(2x_2+x_3+x_4)(x_2+x_3+2x_4)}{3x_2+2x_3+2x_4})$. After some simple linear algebra steps, we can obtain that the corner points of $(x_2+x_3+2x_4 , 0)$, $P$ and $(0, x_2+x_3+2x_4 )$ all satisfy~\eqref{special-bound.B}, which implies that~\eqref{special-bound.B} is redundant. The corner points on the axes are trivial, and we only need to check if $P$ is in the achievable stability region of Theorem~\ref{thm:achievable-stability-region}. 
In fact, the corner point $P$ is achievable when there are $k$ packets for user 1 and 2 respectively. All the uncoded packets are transmitted on layer 1; the overheard packets are transmitted on layer 2. Using the same notations as we introduced in Section~\ref{sec:schemes}, we assign $k_{1,1} = k_{2,1} = k,  k_{1,2} = k_{2,2} = 0$. The number of slots to finish the uncoded packets are
$ t_{\pi(1)}^{\text{(unc)}} = \frac{k+k}{\Pr[M\geq 1]},
 \  \ t_{\pi(2)}^{\text{(unc)}} = 0. $
Also, $t^{\text{(unc)}}  = t_{\pi(1)}^{\text{(unc)}}.$ The numbers of overheard packets to be transmitted in a network coded manner in Phase2 for two users are
$k_1^{\text{(rem)}} = \left[ k(1-\frac{\Pr[N_1\geq 1]}{\Pr[M\geq 1]}) - 2k\frac{\Pr[N_1\geq 2]}{\Pr[M\geq 1]} \right]^+$  and 
$k_2^{\text{(rem)}} = \left[ k(1-\frac{\Pr[N_2\geq 1]}{\Pr[M\geq 1]}) - 2k\frac{\Pr[N_2\geq 2]}{\Pr[M\geq 1]} \right]^+.$ 
Since $\Pr[N_1\geq 1] = \Pr[N_2 \geq 1]$ and $\Pr[N_1\geq 2] = \Pr[N_2 \geq 2]$, we have $k_1^{\text{(rem)}} = k_2^{\text{(rem)}}$, which is positive when $x_2\geq 2x_4$. The number of slots to finish Phase2 is
$t^{\text{(NC)}} = \frac{k_1^{\text{(rem)}}}{\mathbb{E}[N_1]}$.  
The total number of slots is 
$t = t^{\text{(unc)}} + t^{\text{(NC)}}.$ Therefore, 
$R_1 = R_2 = \frac{k}{t} = \frac{(2x_2+x_3+x_4)(x_2+x_3+2x_4)}{3x_2+2x_3+2x_4},$ 
which is the corner point $P$. Thus, with this packet assignment method, the single corner point $P$ besides the two on the axes in the outer bound is achievable, which implies that the optimality holds under condition (C1).

\subsection{Proof of (C2) and (C3)}
Since there are four bounds in~\eqref{eq:outerbound} for the capacity outer bound, besides the two axes, our idea is to simplify the outer bound of the capacity region by making some of the four bounds redundant.  Comparing the inequality set of the inner bound after FME with the outer bound, we obtain the conditions for these two regions coincide.
For the sake of notation simplicity, let
$ \Pr[N_1 \geq q] = a_q, \ 
    \Pr[N_2 \geq q] = b_q, \  \Pr[M \geq q]=c_q, \  \forall q \in [2].$  
Define~\eqref{outer-bound.A} as bound.A, ~\eqref{outer-bound.B} as bound.B, ~\eqref{outer-bound.C} as bound.C and~\eqref{outer-bound.D} as bound.D.
Rewrite the inner bound of  Theorem~\ref{thm:Prototype} as
\begin{subequations}
\begin{align}
  \frac{R_{1,1} + R_{2,1}}{c_1} < 1,
\qquad
  \frac{R_{1,2} + R_{2,2}}{c_2} < 1,
  \\
  \frac{R_{1}}{a_1 + a_2} + \frac{a_1}{c_1}\frac{R_{2,1}}{a_1+a_2} + \frac{a_2}{c_2}\frac{R_{2,2}}{a_1 + a_2} < 1,
  \\
  \frac{R_{2}}{b_1 + b_2}+ \frac{b_1}{c_1}\frac{R_{1,1}}{b_1 + b_2} + \frac{b_2}{c_2}\frac{R_{1,2}}{b_1 + b_2} < 1;
\end{align}  
rewrite~\eqref{eq:def A B C D} as
\label{eq:innerbound}
\end{subequations}
$ \xi_1 := \max\left(\frac{c_1}{a_1} , \frac{c_2}{a_2} \right) \geq 
    \xi_2 := \min\left(\frac{c_1}{a_1} , \frac{c_2}{a_2} \right) \geq 1 
    \geq \xi_3 := \max\left(\frac{b_1}{c_1} , \frac{b_2}{c_2} \right) \geq
    \xi_4 := \min\left(\frac{b_1}{c_1} , \frac{b_2}{c_2} \right) \geq 0.$
To easily express all the corner points, we also represent the axis of $R_1$ as bound.1, the axis of $R_2$ is as bound.2. 
Let $(R_1^{XY}, R_2^{XY})$ denote the intersection point between bound.X and bound.Y. 
It is easy to check that when 
$\frac{\Pr[N_1\geq 2]}{\Pr[N_1\geq 1]} =\frac{\Pr[N_2\geq 2]}{\Pr[N_2\geq 1]} = \frac{\Pr[M\geq 2]}{\Pr[M \geq 1]},$ 
 both bound.B 
 and bound.C 
 are redundant and the outer bound in~\eqref{eq:outerbound} becomes identical to the inner bound in~\eqref{eq:innerbound}. 

Next, we focus on the case
$\frac{\Pr[N_1\geq 2]}{\Pr[N_1\geq 1]} \neq \frac{\Pr[N_2\geq 2]}{\Pr[N_2\geq 1]} \neq \frac{\Pr[M\geq 2]}{\Pr[M \geq 1]}.$ 
By  FME, the region in~\eqref{eq:innerbound} can be equivalently written as follows in terms of $R_1$ and $R_2$, 
\begin{subequations}
\begin{align}
&\text{bounds~\eqref{outer-bound.A} to~\eqref{outer-bound.D}}, 
\\& R_1 + R_2 \leq c_1 + c_2,  
    \label{inner-condition1}
\\& \text{bound in~\eqref{eq:FMEcase4} next},
    \label{inner-condition2}
\end{align}
where~\eqref{inner-condition2} has the following expression,  depending on the case:
\label{eq:FMEform} 
\end{subequations}

\begin{subequations}
\begin{enumerate}
\item $\xi_1 = \frac{c_1}{a_1}> \xi_2 = \frac{c_2}{a_2}$, $\xi_3 = \frac{b_1}{c_1}> \xi_4 = \frac{b_ 2}{c_2}$:
\begin{align}
& \frac{b_1c_2}{b_1c_2 - c_1b_2}R_1 + (1+\frac{c_1c_2}{b_1c_2 - c_1b_2})R_2 \leq  \notag
\\
&c_1 + c_2+ c_1c_2\frac{b_1+b_2}{b_1c_2 - c_1b_2}. 
\label{case1.condition2}
\end{align}

\item $\xi_1 = \frac{c_2}{a_2}> \xi_2 = \frac{c_1}{a_1}$, $\xi_3 = \frac{b_1}{c_1}> \xi_4 = \frac{b_2}{c_2}$:
\begin{align}
&(\frac{c_1}{a_1c_2 -c_1a_2} + \frac{b_1}{b_1c_2 - c_1 b_2} )R_1 + 
(\frac{c_1}{b_1c_2 -c_1b_2} + 
\notag \\ &
\frac{a_1}{a_1c_2 - c_1 a_2} )R_2 \leq 
1 + c_1\frac{a_1+a_2}{a_1c_2-c_1a_2} + c_1\frac{b_1+b_2}{b_1c_2-c_1b_2}. 
\label{case2.condition2}
\end{align}

\item  $\xi_1 = \frac{c_2}{a_2}> \xi_2 = \frac{c_1}{a_1}$, $\xi_3 = \frac{b_2}{c_2}> \xi_4 = \frac{b_ 1}{c_1}$:
\begin{align}
&(1+\frac{c_1c_2}{a_1c_2 - a_2c_1}) R_1 + \frac{a_1c_2}{a_1 c_2 - a_2c_1} R_2 \leq c_1 + c_2  \notag
\\
& +c_1c_2\frac{a_1+ a_2}{a_1 c_2 -a_2c_1}. 
\label{case3.condition2}
\end{align} 

\item $\xi_1 = \frac{c_1}{a_1}> \xi_2 = \frac{c_2}{a_2}$, $\xi_3 = \frac{b_2}{c_2}> \xi_4 = \frac{b_ 1}{c_1}$:
\begin{align}
&(1+\frac{c_1c_2}{a_2c_1 - a_1c_2}) R_1 + (\frac{c_2}{b_2} + \frac{a_2c_1}{a_2 c_1 - a_1c_2} )R_2 \leq   \notag
\\
& c_1 + c_2 + \frac{c_2b_1}{b_2} + c_1c_2\frac{a_1+ a_2}{a_2 c_1 -a_1c_2}. 
\label{case4.condition2}
\end{align}

\end{enumerate}
\label{eq:FMEcase4}
\end{subequations}

First, consider the condition (C1) that one of bound.B and bound.C in the outer bound is redundant.
To give the explicit expressions of C1,
we define the convex polygonal region surrounded by (bound.1, bound.A, bound.C, bound.D, bound.2) as $\mathcal{R}_{\text{noB}}$, and
       the convex polygonal region surrounded by (bound.1, bound.B, bound.2) as $\mathcal{R}_{\text{onlyB}}$. Similarly, define the convex polygonal region surrounded by (bound.1, bound.A, bound.B, bound.D, bound.2) as $\mathcal{R}_{\text{noC}}$, and
       the convex polygonal region surrounded by (bound.1, bound.C, bound.2) as $\mathcal{R}_{\text{onlyC}}$.
       
The condition of bound.B in~\eqref{eq:outerbound} redundant is that $\mathcal{R}_{\text{noB}}$ is a subset of $\mathcal{R}_{\text{onlyB}}$. 
 Likewise,  bound.C in~\eqref{eq:outerbound} is redundant if $\mathcal{R}_{\text{noC}}$ is a subset of $\mathcal{R}_{\text{onlyC}}$. 
Also, $\mathcal{R}_{\text{noB}}$  is the convex-hull of $[(0,0),$ $(R_{1}^{1A},0),$ $(R_{1}^{AC}, R_{2}^{AC}),$ $(R_{1}^{CD}, R_{2}^{CD}),$ $(0, $ $R_{2}^{D2})]$; $\mathcal{R}_{\text{noC}}$  is the convex-hull of  $[(0,0),$ $(R_{1}^{1A}, 0),$ $(R_{1}^{AB}, R_{2}^{AB}), (R_{1}^{BD}, R_{2}^{BD}),$ $(0,$ $R_{2}^{D2})]$, where
$R_1^{1A} = \mathbb{E}[N_{1}], \ \ R_2^{D2} = \mathbb{E}[N_{2}], 
R_1^{AC} = \frac{\xi_1 \mathbb{E}[N_{1}] - \xi_3 \mathbb{E}[M]}{\xi_1-\xi_3},  
R_2^{AC} = \xi_1 \xi_3 \frac{\mathbb{E}[M] - \mathbb{E}[N_{1}]}{\xi_1-\xi_3}, 
R_1^{CD} = \frac{\xi_3 \mathbb{E}[M] - \mathbb{E}[N_2]}{\xi_3-\xi_4},  
R_2^{CD} = \frac{\xi_3 \mathbb{E}[N_2] - \xi_3 \xi_4 \mathbb{E}[M]}{\xi_3- \xi_4},  
R_1^{AB} = \frac{\xi_1 \mathbb{E}[N_1] - \mathbb{E}[M]}{\xi_1-\xi_2},  
R_2^{AB} = \frac{\xi_1 \mathbb{E}[M] - \xi_1 \xi_2 \mathbb{E}[N_1]}{\xi_1- \xi_2},  
R_1^{BD} = \frac{\mathbb{E}[M] - \mathbb{E}[N_{2}]}{\xi_2- \xi_4}, 
R_2^{BD} = \frac{\xi_2 \mathbb{E}[N_{2}] - \xi_4 \mathbb{E}[M]}{\xi_2- \xi_4}.$ 
Hence, we need all the corner points of $\mathcal{R}_{\text{noB}}$ inside the region of $\mathcal{R}_{\text{onlyB}}$ to make bound.B inactive.
It is obvious that the corner points on the $R_1$- and $R_2$-axis of $\mathcal{R}_{\text{onlyB}}$ are outside $\mathcal{R}_{\text{noB}}$. So we further need 
\begin{align}
\begin{cases}
\xi_2 R_1^{AC} + R_2^{AC} \leq \mathbb{E}[M]\\
\xi_2 R_1^{CD} + R_2^{CD} \leq \mathbb{E}[M].
\end{cases} \label{bound.B-redundant-cond}
\end{align} 
Plugging the coordinates of $(R_1^{AC}, R_2^{AC})$ and $(R_1^{CD}, R_2^{CD})$ into~\eqref{bound.B-redundant-cond}, we obtain 
\begin{align}
\begin{cases}
\frac{\xi_1 (\xi_2- \xi_3)\mathbb{E}[N_1]}{\xi_1(1-\xi_3) + \xi_3(\xi_2-1)} \leq \mathbb{E}[M] 
\\
\mathbb{E}[M] \leq \frac{(\xi_2-\xi_3)\mathbb{E}[N_2]}{\xi_4(1-\xi_3) + \xi_3(\xi_2-1)}. \label{bound.B-redundant}
\end{cases}
\end{align}
Similarly,  bound.C in~\eqref{eq:outerbound} is redundant if
$ R_{1,AB} + \frac{R_{2,AB}}{\xi_3} \leq \mathbb{E}[M]$ and 
$ R_{1,BD} +\frac{ R_{2,BD}}{\xi_3} \leq \mathbb{E}[M], $
which gives
$\frac{(\xi_2-\xi_3)\mathbb{E}[N_2]}{\xi_4(1-\xi_3) + \xi_3(\xi_2-1)} \leq \mathbb{E}[M] $ and 
$ \mathbb{E}[M] \leq \frac{\xi_1(\xi_2-\xi_3)\mathbb{E}[N_1]}{\xi_1(1-\xi_3) + \xi_3(\xi_2-1)}.$ 
 Furthermore, concerning condition (C2) and taking bound.B redundant as an example, after some simple algebra, we list all possible conditions of~\eqref{bound.B-redundant} as follows. 
\begin{enumerate}
\item $\xi_1 = \frac{c_1}{a_1} > \xi_2 = \frac{c_2}{a_2}$; $\xi_3 = \frac{b_1}{c_1} > \xi_4 = \frac{b_2}{c_2}$: 
$ b_1(c_1 + c_2) \leq a_2b_1+c_1^{2} $ 
and 
$ a_2b_1+c_2^{2} \leq a_2(c_1 + c_2).$ 

\item $\xi_1 = \frac{c_2}{a_2} > \xi_2 = \frac{c_1}{a_1}$; $\xi_3 = \frac{b_1}{c_1} > \xi_4 = \frac{b_2}{c_2}$: 
$b_1(c_1 + c_2) \leq a_1b_1+c_1c_2$
and 
$a_1b_1+c_1c_2 \leq a_1(c_1 + c_2).$

\item $\xi_1 = \frac{c_1}{a_1} > \xi_2 = \frac{c_2}{a_2}$; $\xi_3 = \frac{b_2}{c_2} > \xi_4 = \frac{b_1}{c_1}$: 
$ b_2(c_1 + c_2) \leq a_2b_2+c_1c_2$ 
and 
$ a_2b_2+c_1c_2 \leq a_2(c_1 + c_2).$ 

\item $\xi_1 = \frac{c_2}{a_2} > \xi_2 = \frac{c_1}{a_1}$; $\xi_3 = \frac{b_2}{c_2} > \xi_4 = \frac{b_1}{c_1}$:
$ b_2(c_1 + c_2) \leq a_1b_2+c_2^{2}$  
and 
$ a_1b_2+c_1^{2} \leq a_1(c_1 + c_2).$  
\end{enumerate}


 When bound.B is redundant and also $\xi_1=\frac{c_1}{a_1}\geq \xi_2 =\frac{c_2}{a_2}; \ \xi_3=\frac{b_1}{c_1} \geq \xi_4 = \frac{b_2}{c_2}$ holds,  
 we can simplify the corner points of bound.A $\&$ bound.C and bound.C $\&$ bound.D as follows
 \begin{subequations}
\begin{align}
R_{1,AC} 
&=  \frac{c_1^2(a_1+a_2) -a_1b_1 (c_1 + c_2)}{c_1^2 - a_1b_1}, \label{R1-ac}
\\
R_{2,AC} 
&= \frac{b_1c_1[(c_1+c_2) - (a_1 + a_2)]}{c_1^2 - a_1b_1}, \label{R2-ac}
\\
R_{1,CD} 
& = c_2, \ 
R_{2,CD} 
= b_1.\label{R1-cd}
\end{align}
Plugging~\eqref{R1-R2-ac-cd} into~\eqref{eq:FMEform}, one can verify that all the inequalities in~\eqref{eq:FMEform} are satisfied.
 \label{R1-R2-ac-cd}
\end{subequations}
Similarly, when bound.C is redundant and also $\xi_1=\frac{c_1}{a_1}\geq \xi_2=\frac{c_2}{a_2}; \ \xi_3=\frac{b_1}{c_1} \geq \xi_4 = \frac{b_2}{c_2}$ holds,  
we compute the coordinates of the corner points, then plug the values into~\eqref{eq:FMEform}  and obtain that~\eqref{eq:FMEform} holds as well. 
 
 
Considering all possible cases in~\eqref{eq:FMEform}, we conclude that the scheme in Theorem~\ref{thm:Prototype} is optimal  
when (C1) either bound.B or bound.C in~\eqref{eq:outerbound} is redundant, and (C2) $\xi_1 = \frac{c_1}{a_1}> \xi_2 = \frac{c_2}{a_2}$, $\xi_3 = \frac{b_1}{c_1}> \xi_4 = \frac{b_ 2}{c_2}$ which is $\frac{\Pr[N_1\geq 2]}{\Pr[N_1\geq 1]} \geq \frac{\Pr[M\geq 2]}{\Pr[M\geq 1]} \geq \frac{\Pr[N_2\geq 2]}{\Pr[N_2\geq 1]}$ or  $\xi_1 = \frac{c_2}{a_2}> \xi_2 = \frac{c_1}{a_1}$, $\xi_3 = \frac{b_2}{c_2}> \xi_4 = \frac{b_ 1}{c_1}$ which is $\frac{\Pr[N_2\geq 2]}{\Pr[N_2\geq 1]} \geq \frac{\Pr[M\geq 2]}{\Pr[M\geq 1]} \geq \frac{\Pr[N_1\geq 2]}{\Pr[N_1\geq 1]}$. 

\end{appendices}

\bibliographystyle{IEEEtran}
\bibliography{LPEBC-journal-v9-SL}

\begin{thebibliography}{10}
\providecommand{\url}[1]{#1}
\csname url@samestyle\endcsname
\providecommand{\newblock}{\relax}
\providecommand{\bibinfo}[2]{#2}
\providecommand{\BIBentrySTDinterwordspacing}{\spaceskip=0pt\relax}
\providecommand{\BIBentryALTinterwordstretchfactor}{4}
\providecommand{\BIBentryALTinterwordspacing}{\spaceskip=\fontdimen2\font plus
\BIBentryALTinterwordstretchfactor\fontdimen3\font minus
  \fontdimen4\font\relax}
\providecommand{\BIBforeignlanguage}[2]{{%
\expandafter\ifx\csname l@#1\endcsname\relax
\typeout{** WARNING: IEEEtran.bst: No hyphenation pattern has been}%
\typeout{** loaded for the language `#1'. Using the pattern for}%
\typeout{** the default language instead.}%
\else
\language=\csname l@#1\endcsname
\fi
#2}}
\providecommand{\BIBdecl}{\relax}
\BIBdecl

\bibitem{FadingBC}
D.~N.~C. Tse and R.~D. Yates, ``Fading broadcast channels with state
  information at the receivers,'' \emph{IEEE Transactions on Information
  Theory}, vol.~58, no.~6, pp. 3453--3471, June 2012.

\bibitem{book:ElGamal-Kim}
A.~{El Gamal} and Y.-H. Kim, \emph{Network Information Theory}.\hskip 1em plus
  0.5em minus 0.4em\relax UK: Cambridge University Press, 2011.

\bibitem{Cover:2006}
T.~Cover and J.~Thomas, \emph{Elements of Information Theory}, 2nd~ed.\hskip
  1em plus 0.5em minus 0.4em\relax New York:Wiley, 2006.

\bibitem{on-the-capacity-of-1-to-k}
C.~Wang, ``On the capacity of 1-to-$k$ broadcast packet erasure channels with
  channel output feedback,'' \emph{IEEE Transactions on Information Theory},
  vol.~58, no.~2, pp. 931--956, Feb 2012.

\bibitem{multiuser-BEC-with-feedback}
M.~Gatzianas, L.~Georgiadis, and L.~Tassiulas, ``Multiuser broadcast erasure
  channel with feedback--capacity and algorithms,'' \emph{IEEE Transactions on
  Information Theory}, vol.~59, no.~9, pp. 5779--5804, Sept 2013.

\bibitem{Ozarow:1984}
L.~Ozarow and S.~Leung-Yan-Cheong, ``An achievable region and outer bound for
  the gaussian broadcast channel with feedback,'' \emph{IEEE Transactions on
  Information Theory}, vol.~30, no.~7, pp. 667--671, Jul. 1984.

\bibitem{4655434Bhaskaran}
S.~R. Bhaskaran, ``Gaussian broadcast channel with feedback,'' \emph{IEEE
  Transactions on Information Theory}, vol.~54, no.~11, pp. 5252--5257, Nov
  2008.

\bibitem{ElGamal-BC}
A.~E. Gamal, ``The feedback capacity of degraded broadcast channels
  (corresp.),'' \emph{IEEE Transactions on Information Theory}, vol.~24, no.~3,
  pp. 379--381, May 1978.

\bibitem{two-user-Gaussian-fading-BC}
A.~{Jafarian} and S.~{Vishwanath}, ``{The Two-User Gaussian Fading Broadcast
  Channel},'' \emph{arXiv e-prints}, p. arXiv:1102.3216, Feb 2011.

\bibitem{achievable-throughput-of-a-multiantenna-Gaussian-BC}
G.~{Caire} and S.~{Shamai}, ``On the achievable throughput of a multiantenna
  gaussian broadcast channel,'' \emph{IEEE Transactions on Information Theory},
  vol.~49, no.~7, pp. 1691--1706, July 2003.

\bibitem{dynamic-power-allocation}
M.~J. Neely, E.~Modiano, and C.~E. Rohrs, ``Resource allocation and cross-layer
  control in wireless networks,'' \emph{Found. Trends Netw.}, 2006.

\bibitem{Dimitriou}
I.~Dimitriou and N.~Pappas, ``Stable throughput and delay analysis of a random
  access network with queue-aware transmission,'' \emph{IEEE Transactions on
  Wireless Communications}, vol.~17, no.~5, pp. 3170--3184, 2018.

\bibitem{stable-scheduling}
A.~{Eryilmaz}, R.~{Srikant}, and J.~R. {Perkins}, ``Stable scheduling policies
  for broadcast channels,'' in \emph{Proceedings IEEE International Symposium
  on Information Theory,}, 2002, pp. 382--.

\bibitem{fundamental-stability}
C.~{Zhou} and G.~{Wunder}, ``A fundamental characterization of stability in
  broadcast queueing systems,'' in \emph{2009 IEEE International Symposium on
  Information Theory}, 2009, pp. 1418--1422.

\bibitem{stability-degraded-bc}
K.~K. {Sesha Sayee} and U.~{Mukherji}, ``Stability of scheduled message
  communication over degraded broadcast channels,'' in \emph{2006 IEEE
  International Symposium on Information Theory}, 2006, pp. 2764--2768.

\bibitem{duality-and-stability-regions}
V.~R. Cadambe and S.~A. Jafar, ``Duality and stability regions of multi-rate
  broadcast and multiple access networks,'' in \emph{IEEE International
  Symposium on Information Theory}, July 2008, pp. 762--766.

\bibitem{BroadcastStability}
Y.~E. {Sagduyu} and A.~{Ephremides}, ``On broadcast stability of queue-based
  dynamic network coding over erasure channels,'' \emph{IEEE Transactions on
  Information Theory}, vol.~55, no.~12, pp. 5463--5478, 2009.

\bibitem{Sagduyu-unusual}
Y.~E. Sagduyu, L.~Georgiadis, L.~Tassiulas, and A.~Ephremides, ``Capacity and
  stable throughput regions for the broadcast erasure channel with feedback: An
  unusual union,'' \emph{IEEE Transactions on Information Theory}, vol.~59,
  no.~5, pp. 2841--2862, May 2013.

\bibitem{Ephremides}
A.~Ephremides and B.~Hajek, ``Information theory and communication networks: an
  unconsummated union,'' \emph{IEEE Transactions on Information Theory},
  vol.~44, no.~6, pp. 2416--2434, 1998.

\bibitem{Siyao-Li}
S.~Li, D.~Tuninetti, and N.~Devroye, ``On the capacity region of the layered
  packet erasure broadcast channel with feedback,'' in \emph{ICC 2019 - 2019
  IEEE International Conference on Communications (ICC)}, 2019, pp. 1--6.

\bibitem{XOR-in-the-air}
S.~{Katti}, H.~{Rahul}, W.~{Hu}, D.~{Katabi}, M.~{Medard}, and J.~{Crowcroft},
  ``Xors in the air: Practical wireless network coding,'' \emph{IEEE/ACM
  Transactions on Networking}, vol.~16, no.~3, pp. 497--510, June 2008.

\bibitem{Exact-decoding-probability}
O.~{Trullols-Cruces}, J.~M. {Barcelo-Ordinas}, and M.~{Fiore}, ``Exact decoding
  probability under random linear network coding,'' \emph{IEEE Communications
  Letters}, vol.~15, no.~1, pp. 67--69, January 2011.

\bibitem{on-the-equivalence-of-shannon-capacity-and-stable-capacity}
H.~{Yao}, T.~{Ho}, and M.~{Effros}, ``On the equivalence of shannon capacity
  and stable capacity in networks with memoryless channels,'' in \emph{2011
  IEEE International Symposium on Information Theory Proceedings}, July 2011,
  pp. 503--507.

\bibitem{throughput-capacity-stability-regions}
J.~Luo and A.~Ephremides, ``On the throughput, capacity, and stability regions
  of random multiple access,'' \emph{IEEE Trans. Inf. Theor.}, vol.~52, no.~6,
  pp. 2593--2607, June 2006.

\bibitem{Geng-Quan}
Q.~{Geng}, H.~T. {Do}, R.~{Wu}, M.~{Yuan}, Y.~{Li}, and W.~{Ding}, ``{On the
  Capacity Region of Broadcast Packet Erasure Relay Networks With Feedback},''
  \emph{arXiv e-prints}, p. arXiv:1312.1727, Dec 2013.

\bibitem{GeometricChernoff}
B.~{Doerr}, ``{Probabilistic Tools for the Analysis of Randomized Optimization
  Heuristics},'' \emph{arXiv e-prints}, p. arXiv:1801.06733, Jan 2018.

\bibitem{book:probabilityAGraduate}
A.~Gut, \emph{Probability: A Graduate Course.}\hskip 1em plus 0.5em minus
  0.4em\relax Springer, 2005.

\bibitem{wolff_1989}
R.~W. Wolff, \emph{Stochastic modeling and the theory of queues}.\hskip 1em
  plus 0.5em minus 0.4em\relax Prentice-Hall, 1989.

\end{thebibliography}

\end{document}